\documentclass[12pt]{article}
\usepackage{amsmath}
\usepackage{graphicx}
\usepackage{enumerate}
\usepackage{natbib}
 \usepackage{float}
\usepackage{url} % not crucial - just used below for the URL 
\newcommand{\blind}{1}

\usepackage{placeins}
\usepackage{graphicx, color, url}
\usepackage{dsfont}
\usepackage{amssymb}
\usepackage{amsfonts}
\usepackage{amsbsy}
\usepackage{setspace}
\usepackage{bm}
\usepackage{textcomp}
\usepackage{booktabs}
\usepackage{rotating}
\usepackage{enumitem}
\usepackage{amsmath,amsthm,nccmath}
\usepackage{mathtools}
\usepackage{bbm}
\usepackage{soul}
\usepackage[ruled,linesnumbered,noend]{algorithm2e}
\usepackage{multirow}
\usepackage{bm}
\usepackage{threeparttable}
\usepackage{threeparttablex}
\usepackage{longtable}
\usepackage{lscape}
\usepackage{parskip}
\let\oldnl\nl% Store \nl in \oldnl
\newcommand{\nonl}{\renewcommand{\nl}{\let\nl\oldnl}}% Remove line number for one line

\usepackage[dvipsnames]{xcolor}

\usepackage{setspace}
  \usepackage[hidelinks]{hyperref}
\makeatletter
\newcommand{\oset}[2]{%
  {\mathop{#2}\limits^{\vbox to 1.75\ex@{\kern-\tw@\ex@
   \hbox{\scriptsize #1}\vss}}}}
\makeatother
\usepackage{bbm}
\newcommand{\I}[1]{\mathbbm{1}\left[#1\right]}
\newcommand{\E}[1]{\mathbb{E}\left[#1\right]}

\newcommand{\indep}{\perp \!\!\! \perp} %tianmin add

\usepackage{caption}
\usepackage{subcaption}

\newtheorem{theorem}{Theorem}

\newtheorem{proposition}{Proposition}

\newtheorem{lemma}{Lemma}

\begin{document}

\def\spacingset#1{\renewcommand{\baselinestretch}%
{#1}\small\normalsize} \spacingset{1}

%%%%%%%%%%%%%%%%%%%%%%%%%%%%%%%%%%%%%%%%%%%%%%%%%%%%%%%%%%%%%%%%%%%%%%%%%%%%%%

\if1\blind
{
  \title{\bf Searching for local associations while controlling the false discovery rate}
\author{
Paula Gablenz \\  Department of Statistics\\
Stanford University, California, USA \\[0.5em] 
Matteo Sesia \\
Departments of Data Sciences and Operations, and of Computer Science\\ University of Southern California, California, USA\\[0.5em]
Tianshu Sun \\
Cheung Kong Graduate School of Business, Beijing, China \\[0.5em] and \\[0.5em]
Chiara Sabatti \\  Departments of Biomedical Data Science, and of Statistics\\
Stanford University, California, USA \\
}
  \maketitle
} \fi

\if0\blind
{
  \bigskip
  \bigskip
  \bigskip
  \begin{center}
    {\LARGE\bf Searching for local associations while controlling the false discovery rate}
\end{center}
  \medskip
} \fi

\bigskip
\begin{abstract}
We introduce local conditional hypotheses that express how the relation between explanatory variables and outcomes changes across different contexts, described by covariates. By expanding upon the model-X knockoff filter, we show how to adaptively discover these local  associations, all while controlling the false discovery rate. Our enhanced inferences can help explain sample heterogeneity and uncover interactions, making better use of the capabilities offered by modern machine learning models. Specifically, our method is able to leverage any model for the identification of data-driven hypotheses pertaining to different contexts. Then, it rigorously test these hypotheses without succumbing to selection bias. Importantly, our approach is efficient and does not require sample splitting. We demonstrate the effectiveness of our method through  numerical experiments and by studying the genetic architecture of Waist-Hip-Ratio across different sexes in the UKBiobank.

%%% Local Variables:
%%% mode: latex
%%% TeX-master: "main_jasa"
%%% End:

 \end{abstract}

\noindent%

\vfill

\newpage
\spacingset{1} % DON'T change the spacing!

\section{Introduction} \label{sec:intro}

\subsection{Motivation}\label{sec:preview}

As data sets become more detailed and extensive, the methods and models used for analysis must similarly advance in sophistication.
A compelling example comes from precision medicine, where the goal is to tailor disease prevention and treatment to individual differences in genes, environments, and lifestyles, departing from a one-size-fits-all approach.
 
Anna Karenina's opening line---``All happy families are alike; each unhappy family is unhappy in its own way''---is often cited to describe the heterogeneity of disease. 
Increasingly, we understand that similar symptoms leading to the same diagnosis can arise from distinct disruptions in biological pathways.
Identifying the underlying causes of these symptoms may be the key to achieving the most effective and safest treatments. Tolstoy's quote is especially fitting, as it refers not to an ``unhappy individual'' but to an ``unhappy family''---suggesting that making broad generalizations requires finding a homogeneous group of subjects. This highlights one of the greatest challenges in handling heterogeneous data: identifying subgroups within which a consistent mechanism operates. This paper presents a specific formulation of this problem and a corresponding solution.

Specifically, we focus on settings where we have an outcome $Y$, a large number $p$ of possible explanatory (or ``treatment'') variables $X$, and a set of covariates $Z$ that capture essential individual features which may help identify relatively homogeneous groups within an otherwise heterogeneous population.
In other words, we say that these covariates $Z$ define distinct ``environments'' across which the observations may differ significantly.
The goal is to identify the subset of $X$ variables that carry unique information on $Y$, while accounting for the possibility that these key variables may vary across environments.

This situation is quite common in genomics.
For instance, in genome-wide association studies (GWAS), the goal is to identify which of the $p$ single nucleotide polymorphisms (SNPs) in $X$ predict a trait $Y$, while controlling for covariates $Z$ such as sex, age, medication use, diet, or healthcare access.
Here, different combinations of $Z$ values create diverse environments that may reveal unique genetic associations relevant to each subgroup.

\subsection{Main Contributions and Outline}

Our first contribution is to introduce a highly expressive family of {\em local} conditional independence hypotheses, designed to rigorously address, within a versatile non-parametric statistical framework, the challenge of identifying meaningful associations that may be specific to certain subgroups of a heterogeneous population. 
This extends the classical notion of {\em population-wide} conditional independence, as studied for example by \citet{candes2018}, which can be viewed as a more restrictive ``one-size-fits-all'' version of our approach. 

The second contribution is an extension of the model-X knockoff filter \citep{candes2018}, originally developed for testing population-wide hypotheses, to enable testing local hypotheses while controlling the false discovery rate (FDR). 
We call our method the {\em adaptive Local Knockoff Filter (aLKF)}. 
A key innovation of the aLKF is its ability to test not only pre-defined hypotheses but also {\em data-driven local hypotheses}, automatically identified by powerful machine learning models as likely to be informative for the data at hand.

This method efficiently partitions the information in a dataset, allowing both learning and testing \citep{Tian2018}. 
Thus, it can be seen as a ``data-thinning'' approach \citep{Jacob2024}, specifically designed for knockoff inference and taking advantage of the unique capacity of knockoffs to partially ``mask'' data \citep{spector2024}.

After outlining the general framework of the aLKF, which is very flexible, we describe specific implementations, including an approach tailored to large-scale GWAS data. The effectiveness of our approach is then demonstrated 
through numerical experiments with simulated data, and 
with an analysis of the genetic basis of waist-hip ratio and its variation between women and men in the UK Biobank data \citep{bycroft2018}.

 \subsection{Related Work} \label{sec:related}

 The analysis of heterogeneous data extends beyond the focus of this paper and has attracted interest for various reasons. While a comprehensive review is beyond the scope of this work, it is worth noting that heterogeneous data analysis can help mitigate collinearity issues \citep{peterson2019genome}, improve the robustness of predictive models to distribution shifts \citep{berg2019reduced}, and help identify causal associations \citep{peters2016causal}.

This paper focuses on identifying subgroups of a heterogeneous population within which there are relatively homogeneous conditional associations.
There are many algorithms and models that can discover subgroups exhibiting coherent behaviors \citep{novak2009supervised}, detect interactions \citep{bien2013lasso,wen2014bayesian}, or estimate heterogeneous effects in regression and causal inference 
\citep{athey2016recursive}; see also \citet{caron2022estimating} for a review.
Even if such methods are not designed to obtain statistical inferences with finite-sample guarantees, they can be leveraged in the ``learning'' phase of our method.

This work is also inspired by the literature on explainability in machine learning \citep{molnar2022}, where a central challenge is enhancing the interpretability of opaque predictive models by identifying ``important" variables.
There are many definitions of variable importance, from descriptive measures capturing a model's input-output relation to inferential measures offering insights into the underlying population.
Given the nonlinear nature of many machine learning models, assessments of variable importance are often context-dependent. 
Despite limitations in current approaches \citep{ghassemi2021}, this research highlights the value of ``local" importance measures in capturing data heterogeneity. 

A few prior studies have addressed heterogeneity in knockoff-based analyses \citep{barber2015controlling}, particularly within the model-X framework \citep{candes2018} used in this paper. \citet{Gimenez2019} introduced a notion of local variable importance expressed as statistical hypotheses testable via the knockoff filter. Their approach formulates inference in terms of conditional associations within neighborhoods in the space of $X$, with each neighborhood corresponding to distinct hypotheses tested under FDR control.
To enable efficient inference, these neighborhoods are defined (possibly adaptively) using both original features and synthetic knockoffs, which can complicate interpretability. Additionally, \citet{Gimenez2019} note that testing all local hypotheses within fixed neighborhoods may significantly reduce power due to smaller effective sample sizes.

In contrast, we use knockoffs only as analytical tools, without influencing the definition or interpretation of the hypotheses. Further, our method adjusts the ``locality" of hypotheses in a data-driven manner, striking an effective balance between power and specificity.

Finally, our work is inspired by \citet{li2021searching}, which extends the knockoff filter to analyze heterogeneous data. While their definition of local conditional hypotheses informs our approach, we address a different problem. \citet{li2021searching} assumes each observation belongs to a predetermined environment—for instance, in a GWAS, individuals may be categorized by self-reported race/ethnicity, with these labels representing distinct ``environments."

Our approach differs in two key ways. First, we do not assume fixed environments for individuals. Instead, we allow each variable to ``interact" with different covariates, creating subgroup-specific associations. For example, one SNP might interact with sex, defining male and female environments, while another interacts with age, defining younger and older environments. Second, unlike \citet{li2021searching}, we do not assume the relevant environments for each variable are known in advance.
 
\section{Methods} \label{sec:methods}

\subsection{Review of Global Conditional Hypotheses} \label{sec:cond-hyp}

Consider $n$ observations of a vector of random variables $(X,Y,Z).$ 
For each individual $i \in [n] = \{1,\ldots,n\}$, ${X^{i} = (X^{i}_1, \ldots, X^{i}_p) \in \mathbb{R}^p}$ describes $p$ explanatory variables, $Y^{i} \in \mathbb{R}$ is the outcome of interest (either numerical or categorical), and $\smash{Z^{i} = (Z^{i}_1, \ldots, Z^{i}_m) \in \mathcal{Z} \subseteq \mathbb{R}^m}$ represents $m$ covariates. 
The distinction between explanatory variables and covariates reflects our inferential question: we want to discover which of the $X_j$ for $j \in [p]$ is ``important'' for $Y$, while $Z$ is used to account for known possible confounders. An example of $Z$ in the GWAS setting is sex: SNP $X_j$ could have an effect only for females, but not for males. We consider the genetic underpinning of Waist-Hip-Ratio across sexes in Section~\ref{sec:real-data-genetic}.

One way of formalizing this problem, introduced in \citet{candes2018}, is to test under FDR control the following family of {\em global} conditional independence hypotheses:
\begin{align} \label{eq:null-hyp}
 \mathcal{H}_{j} : Y \indep X_j \mid X_{-j}, Z \qquad j \in [p],
\end{align}
where $X_{-j}$ indicates all the variables in $X$ except $X_j$.
The term ``global'' is adopted in this paper to emphasize these hypotheses describe relations between variables across the entire population, as opposed to the more specific {\em local} hypotheses that we will introduce later.
Appendix A1 provides a brief review of the {\em global} knockoff filter, designed to test~\eqref{eq:null-hyp} while controlling the FDR.
Testing~\eqref{eq:null-hyp} makes us focus on the unique information that $X_j$ carries on $Y$: the resulting discoveries are both precise and informative. 
In the GWAS context, it makes it possible to identify a collection of non-redundant SNPs that capture the genetic signature of a trait, minimizing spurious associations (variants that are merely ``guilty by association'') and focusing on more likely causal mechanisms.
(See Appendix A1 for connections of~\eqref{eq:null-hyp} to parametric inference in generalized linear models and causal inference.)

The non-parametric nature of~\eqref{eq:null-hyp} brings two advantages.
First, this type of hypothesis can be interpreted without relying on assumptions whose alignment with reality may be difficult to validate, since it emphasizes testing general relationships between variables rather than estimating parameters within a model. 
Second, this non-parametric approach offers some robustness to heterogeneity, as explained below. 

Consider a population with $p=3$ and $m=2$, with $Z_1=1$ for half of the individuals (e.g., females) and $Z_1=0$ for the others (e.g., males), while $Z_2$ is a continuous variable (e.g., Body Mass Index, BMI).
Suppose the $X_i$'s are independent standard normal and $Y \sim \mathcal{N}(Z_1X_1\beta_1+\I{Z_2>t_u}X_2\beta_2, 1)$, for some coefficient $(\beta_1,\beta_2) \in \mathbb{R}$ and threshold $t_u$, which identifies overweight individuals. 
The hypotheses in~\eqref{eq:null-hyp} are well defined, with $\mathcal{H}_{1},\mathcal{H}_{2}$ true if and only if (respectively) $\beta_1=0$, $\beta_2=0$, and $\mathcal{H}_{3}$ true. 
Thus, testing~\eqref{eq:null-hyp} is already useful for distinguishing the importance of \( X_1, X_2 \) in predicting \( Y \) from the irrelevance of \( X_3 \).
However, this approach does not inform about how the importance of \( X_1, X_2 \) varies with changes in the value of \( Z_1, Z_2 \).
In other words, testing~\eqref{eq:null-hyp} cannot tell us that $X_1$ is only {\em locally} important among females, and $X_2$ is only {\em locally} important for individuals with elevated BMI.

As a first step towards capturing heterogeneity, \citet{li2021searching} extended the framework of \citet{candes2018} to test a family of {\em environment-level} hypotheses defined as
\begin{align} \label{eq:null-hyp-env}
 \mathcal{H}^{(e)}_{j} : Y^{(e)} \indep X^{(e)}_j \mid X^{(e)}_{-j}, Z^{(e)},
\end{align}
simultaneously for different environments $e \in \{1,2,\ldots,E\}$. 
Here, each environment corresponds to a distinct distribution over $(X,Y,Z)$, which may be interpreted as representing a fixed and {\em pre-defined} subgroup.
This approach, for example, allows testing the importance for a trait of each SNP across distinct ancestries, like African or European.

However, \citet{li2021searching} do not address our problem: their approach lacks flexibility to test hypotheses corresponding to different subgroups for different variables and to enable the discovery of relevant subgroups adaptively.
We thus present a novel approach to overcome these challenges, beginning with the introduction of a convenient non-parametric framework for capturing data heterogeneity and a new family of {\em local} conditional hypotheses.

\subsection{A Model-X Framework for Heterogeneous Data} \label{sec:individual-hyp}

Let us denote as $\mathbf{X} \in \mathbb{R}^{n \times p}$, $\mathbf{Y} \in \mathbb{R}^{n \times 1}$, and $\mathbf{Z} \in \mathbb{R}^{n \times m}$ the data matrices collecting the observations for all $n$ individuals. 
To capture heterogeneity without an explicit parametric model, we assume that the joint distribution of the data can be written as:
\begin{align} \label{eq:indep-model}
 P(\mathbf{X}, \mathbf{Y}, \mathbf{Z})
 & = \prod_{i=1}^{n} p_{X,Z}(X^i, Z^i) \cdot p^{i}_{Y \mid X, Z}(Y^i \mid X^i, Z^i),
\end{align}
where each \( p^{i}_{Y \mid X, Z} \) denotes the (unknown) conditional distribution of the outcome \( Y^i \) for the \( i \)-th individual, given that individual’s variables and covariates. 
This distribution may vary across individuals, who are assumed to be independent. To simplify notation, we assume that \( p_{X,Z} \) remains constant, though this assumption could be relaxed. The concept of an individual-specific distribution \( p^{i}_{Y \mid X, Z} \) is abstract and provides limited opportunities for learning, as each individual essentially forms their own category. We use this framework to capture unknown heterogeneity, but in practice, learning will be conducted at the subgroup level by aggregating information from ``similar" individuals, as it will soon become clear.

Given the distribution \eqref{eq:indep-model} one may consider testing a family of {\em individual-level} conditional hypotheses defined as:
\begin{align} \label{eq:null-hyp-ind}
 \mathcal{H}^i_{j} : Y^i \indep X^i_j \mid X^i_{-j}, Z^i, \qquad i \in [n], j \in [p].
\end{align}
Intuitively, this asks whether the distribution \( p^{i}_{Y \mid X, Z} \) for the \( i \)-th individual in~\eqref{eq:indep-model} depends on \( X_j \), representing an extreme departure from the global hypotheses of \citet{candes2018}.
While these hypotheses are theoretically appealing because they allow us to think of conditional associations on a per-person basis, they are impractical to test within any non-parametric framework, as each hypothesis applies only to a single data point.

\subsection{Local Conditional Hypotheses} \label{sec:loc-hyp}

To gather strength from multiple data points, we introduce a new family of \emph{local} conditional hypotheses, offering a pragmatic middle ground between the highly flexible but impractical individual hypotheses in~\eqref{eq:null-hyp-ind} and the more rigid environment-level hypotheses in~\eqref{eq:null-hyp-env}. 
Intuitively, the idea is to define flexible environments specific to each variable \( j \in [p] \), grouping together individuals who are relatively homogeneous in the conditional association between \( Y \) and \( X_j \). By testing conditional hypotheses within these local environments, we can pool information from similar individuals, gaining power to make discoveries.

To make this idea precise, we use a collection of {\em partition}
 functions $\nu_j : \mathcal{Z} \mapsto [L_j]$ that, for each $j \in [p]$, use the covariate space $\mathcal{Z}$ to identify $L_j$ disjoint neighborhoods $\ell \in [L_j]$, each defining a sub-population of relevance for variable $X_j$. In the example from Section~\ref{sec:cond-hyp}, a useful set of partition functions is $\nu_1(Z) = 1+ \I{Z_1=1}$ (where $\I{\cdot}$ is the indicator function identifying the female ($Z_1=1$) and male ($Z_1=0$) subgroups across which $X_1$ has different impacts), $\nu_2(Z) = 1+\I{Z_2> t_l}+\I{Z_2>t_u}$ (which identifies the subgroups corresponding to ``underweight'', ``healthy weight'' and ``overweight'' subjects), and $\nu_3(Z) = 1$ (a single trivial neighborhood encompassing the whole sample, since $X_3$ is unimportant for all individuals). This example is illustrated in Table A1 in Appendix A2.

For a given collection of partition functions $\nu = (\nu_1,\ldots,\nu_p)$, we are interested in testing the family of {\em local} conditional hypotheses defined as:
\begin{align} \label{eq:null-hyp-loc}
 \mathcal{H}_{j,\ell}(\nu) = \bigcap_{i \in [n] \,:\, \nu_j(Z^i) = \ell} \mathcal{H}^i_{j}, \qquad j \in [p], \ell \in [L_j],
\end{align}
where $\mathcal{H}^i_{j}$ is the {\em individual-level} hypothesis defined in~\eqref{eq:null-hyp-ind}.
In words, $\mathcal{H}_{j,\ell}(\nu)$ states that $Y$ does not depend on $X_j$ conditional on $X_{-j}$ for any of the individuals in the group defined by $\nu_j(Z) = \ell$.
Considering~\eqref{eq:null-hyp-loc} for all $j \in [p]$ and $\ell \in [L_j]$ gives a multiple testing problem with a total of $L = \sum_{j=1}^pL_j$ hypotheses, if the $\nu_j$ are fixed. In the following, we will denote the subset of true null local hypotheses as $\mathcal{H}_0(\nu) = \{ \mathcal{H}_{j,\ell}(\nu) : j \in [p], \ell \in [L_j], \mathcal{H}_{j,\ell}(\nu) \text{ is true}\}$.
To simplify notation, rather than referring to the number of neighborhoods $L_j$ that vary across different variables, we will often use $L_{\max} =\max_{j}(L_j)$ with the understanding that this is an upper bound for the number of neighborhoods.

Let us take a moment to clarify the interpretation the local hypotheses and how they differ from the population-wide conditional hypotheses in~\eqref{eq:null-hyp} and the environment-level hypotheses in~\eqref{eq:null-hyp-env}.
If $L=1$, \eqref{eq:null-hyp-loc} essentially has the same interpretation as~\eqref{eq:null-hyp}.
In general, however, a rejection of~\eqref{eq:null-hyp-loc} is more informative than one of~\eqref{eq:null-hyp} because it pinpoints a specific subgroup of individuals displaying a significant association.
Further, the connection with the approach of \citet{li2021searching} can be seen by noting that $\mathcal{H}_{j,\ell}(\nu)$ in \eqref{eq:null-hyp-loc} can be interpreted as being equivalent to an environment-level hypothesis
\begin{align} \label{eq:null-hyp-loc-env}
 \bar{\mathcal{H}}_{j,\ell}(\nu) : Y^{(j,\ell)} \indep X^{(j,\ell)}_j \mid X^{(j,\ell)}_{-j}, Z^{(j,\ell)}\;\;\;\; j \in [p], \ell \in [L_j],
\end{align}
 where $\smash{(X^{(j,\ell)},Y^{(j,\ell)},Z^{(j,\ell)})}$ denotes a random sample from an ``environment'' where the distribution of $(X,Y,Z)$ is conditioned on $\nu_j(Z) = \ell$. 
Thus, rejecting $ \mathcal{H}_{j,\ell}(\nu)$ helps pinpoint the effect of $X_j$ to the sub-population with $\nu_j(Z) = \ell$. 
This highlights the higher flexibility of \eqref{eq:null-hyp-loc} relative to~\eqref{eq:null-hyp-env}: it allows us to focus on different environments for different variables.

In the toy example in Section~\ref{sec:cond-hyp}, \( \mathcal{H}^{i}_{1} \) being true if and only if the i-th individual is male ($Z^i_1=0$) reflects that the local importance of \( X_1 \) for \( Y \) depends on \( Z_1 \). In this sense, the hypotheses in~\eqref{eq:null-hyp-loc} provide a non-parametric extension of the concept of interaction between \( X_2 \) and \( Z_1 \), moving beyond the common but narrower definition as a ``departure from additivity in a linear model on a selected scale of measurement'' \citep{wang2011statistical}.

This being said, two elements distinguish testing \eqref{eq:null-hyp-loc} from the more standard search for statistical interactions within a generalized linear model.
 First, in our non parametric framework, there is no notion of ``effect size'': a variable is either conditionally independent of the outcome, under $\smash{P^{(j,\ell)}_{X,Y,Z}}$, or it is not. Consider a modification of our running example that sets $Y \sim \mathcal{N}((Z_1+1)X_1\beta_1+\I{Z_2>t_u}X_2\beta_2, 1)$.
Now $X_1$ is important for the entire population, just with a different effect size for the female ($Z_1=1)$ and male ($Z_1=0$) subgroups.
In this case, both $\mathcal{H}_{1,1}$ and $\mathcal{H}_{1,2}$ are true. The hypotheses~\eqref{eq:null-hyp-loc} then can only describe what are sometimes called ``sufficient cause interactions'' \citep{tutorial}: situations where the outcome occurs when both the ``interacting" exposures are present. 
Interestingly, these type of interactions often lend themselves to a robust, ``mechanistic'' interpretation, less dependent of a specific scale and linear model.
Second, failing to reject $\mathcal{H}_{1,1}$ is not equivalent to establishing it is true. 
In interpreting the results of tests for $\mathcal{H}_{1,1}$ and $\mathcal{H}_{1,2}$, we need to keep in mind the asymmetry of the hypothesis testing framework. 

Two considerations are important in deciding which hypotheses to test. First, typically the sample size available to test hypotheses in~\eqref{eq:null-hyp-loc} is lower than that for \eqref{eq:null-hyp}. Moreover, the hypotheses in~\eqref{eq:null-hyp-loc}
are truly more informative than~\eqref{eq:null-hyp} only if the partition function $\nu_{j}$ identifies neighborhoods which are relatively homogeneous in the associations between $X_j$ and $Y$. 
Since this information is typically not available a priori, the selection of $\nu$ should generally be data-driven. 
As mentioned in Section~\ref{sec:intro}, there exist many algorithms that could be applied to learn a meaningful partition function $\nu$ from the data.
However, one must be careful that allowing the hypotheses to be random may create the risk of selection bias in the subsequent tests.
In particular, we will see that if the same data utilized to select the subgroups were re-used naively to test the hypotheses in~\eqref{eq:null-hyp-loc}, the type-I errors might be inflated \citep{wager2018estimation}.
Sample splitting could circumvent this issue, but it is wasteful. 
This paper presents an efficient solution that overcomes the limitations of both aforementioned benchmarks.

To present our method, we begin in the next section by outlining our key modeling assumptions, including the availability of suitable knockoff variables. In Section~\ref{sec:knockoffs-testing}, we introduce a knockoff filter to test local hypotheses \( \mathcal{H}_{j,\ell}(\nu) \) defined by a fixed partition function \( \nu \) while controlling the FDR. Then, in Section~\ref{sec:adaptive}, we extend this method to accommodate a data-driven partition function \( \hat{\nu} \), ensuring FDR control for the resulting \emph{random} hypotheses \( \mathcal{H}_{j,\ell}(\hat{\nu}) \) conditional on the choice of $\hat{\nu}$.

\subsection{Modeling Assumptions and Knockoffs} \label{sec:knockoffs-model}

In general, Equation~\eqref{eq:indep-model} can be factored as $P(\mathbf{X}, \mathbf{Y}, \mathbf{Z}) =\prod_{i=1}^{n} p_{Z}(Z^i) \cdot p_{X \mid Z}(X^i \mid Z^i) \cdot p^{i}_{Y \mid X, Z}(Y^i \mid X^i, Z^i)$.
Following a {\em model-X} approach \citep{candes2018}, we treat $\mathbf{Z}$ as fixed and $P_{X \mid Z}$ as known, while allowing $p^{i}_{Y \mid X, Z}$ to be arbitrary and unknown for all $i \in [n]$.

The assumption that $P_{X \mid Z}$ is known may not be justified in every application, but its validity is testable and is well-suited for the analysis of data from randomized experiments \citep{nair2023randomization} or GWAS \citep{sesia2019}. The existing knockoffs literature discusses robustness questions in detail \citep{barber2020robust}. \citet{sesia2021false} specifically address robustness in GWAS settings where $Z$ is ancestry.

The key ingredient of our method are the knockoff variables, which we denote as $\tilde{X}$.
As prescribed by \citet{candes2018}, these are synthetic variables constructed by the statistician, using knowledge of $P_{X \mid Z}$, as a function of $X$ and $Z$ without looking at $Y$, so that $\tilde{X} \indep Y \mid X,Z$.
Each observation of $X$ is assigned a corresponding knockoff $\tilde{X}$, with $\tilde{\mathbf{X}} \in \mathbb{R}^{n \times p}$ denoting the corresponding data matrix of knockoff variables for $\mathbf{X}$.

Knockoffs are constructed to be exchangeable with $X$ conditional on $Z$.
That is, if $[ \mathbf{X}, \tilde{\mathbf{X}} ] \in \mathbb{R}^{n \times 2p}$ is the matrix obtained by concatenating $\mathbf{X}$ with $\tilde{\mathbf{X}} \in \mathbb{R}^{n \times p}$ and, for any $j \in [p]$, $[ \mathbf{X}, \smash{\tilde{\mathbf{X}}]_{\mathrm{swap}(j)}}$ is obtained by swapping the $j$-th columns of $\mathbf{X}$ and ${\tilde{\mathbf{X}}}$, then
\begin{align} \label{eq:knock_cond_1}
 \big[ \mathbf{X}, \tilde{\mathbf{X}} \big]_{\mathrm{swap}(j)} \mid \mathbf{Z}
 \; \overset{d}{=} \; \big[ \mathbf{X}, \tilde{\mathbf{X}} \big] \mid \mathbf{Z}, \qquad \forall j \in [p],
\end{align}
where the symbol $\overset{d}{=}$ denotes equality in distribution.
Thus, the only meaningful difference between $X_j$ and $\tilde{X}_j$ may be the lack of conditional association of the latter with~$Y$.

Constructing knockoffs that satisfy~\eqref{eq:knock_cond_1} not only requires knowledge of $P_{X \mid Z}$ but it can also be computationally expensive in general. 
However, the problem is well-studied \citep{candes2018,gimenez2019knockoffs,bates2020metropolized}, with practical and effective solutions available for GWAS data \citep{chu2024second, sesia2019,sesia2020multi,sesia2021false}.
Therefore, in this paper we assume the availability of suitable knockoffs for the data at hand, using this as a starting point for developing our methodology to test local conditional independence hypotheses. In our GWAS analyses specifically, we follow the approach of \citet{sesia2021false} to generate valid knockoffs, leveraging hidden Markov models.

\subsection{A Knockoff Filter for Fixed Local Hypotheses} \label{sec:knockoffs-testing}

We now present an extension of the knockoff filter \citep{candes2018} that utilizes the augmented data $\mathcal{D} = (\smash{[\mathbf{X}, \tilde{\mathbf{X}}],\mathbf{Y},\mathbf{Z}})$ to simultaneously test all local conditional independence hypotheses in~\eqref{eq:null-hyp-loc}, for a {\em fixed} partition function $\nu$, while controlling the FDR.

The first step is to compute a vector of importance scores for all $p$ variables and knockoffs, with separate scores for each region of the covariate space determined by $ \nu = (\nu_1,\ldots,\nu_p)$.
We will denote this vector of scores as $ \smash{[\mathbf{T},\tilde{\mathbf{T}}] \in \mathbb{R}^{2{L}}} $, where $ L = \sum_{j=1}^{p} {L}_j(\nu) $ and $ {L}_j(\nu) = \max_{z \in \mathcal{Z}} \nu_j(z) \in \mathbb{N} $ is the number of disjoint subgroups induced by $ \nu_j $, for any $j \in [p]$. 
More precisely, $ \mathbf{T} $ (respectively, $ \tilde{\mathbf{T}} $) is the concatenation of $ p $ sub-vectors $ (T_{j,\ell})_{\ell \in [L_j]} $ (respectively, $ (\tilde{T}_{j,\ell})_{\ell \in [L_j]} $) for $ j \in [p] $, where each element indexed by $(j,\ell)$ quantifies the importance of $ X_j $ (respectively, $ \tilde{X}_j $) in predicting $ Y $ within the subgroup defined by $ \nu_j(z) = \ell $, according to a suitable predictive model for $ Y $ given $ (X, \tilde{X}, Z) $.

A key requirement in the knockoff framework is that ${[\mathbf{T},\tilde{\mathbf{T}}]}$ satisfy a specific symmetry condition. This, coupled with the exchangeability of knockoffs, guarantees that, $T_{j,\ell}$ and $\tilde{T}_{j,\ell}$ corresponding to a null hypotheses, have the same distribution---a property that results in FDR control.
Specifically, ${[\mathbf{T},\tilde{\mathbf{T}}]}$ must be computed in such a way that swapping a variable $X_j$ with its knockoff $\tilde{X}_j$ within any subgroup $\ell \in [L_j]$ would have the sole effect of swapping the corresponding scores in that subgroup, namely $T_{j,\ell}$ and $\tilde{T}_{j,\ell}$.

This symmetry is easy to achieve when testing global hypotheses. For example, one can define importance scores using the absolute value of the estimated coefficient for each variable in a Lasso model. The case of local hypotheses, however, where the observations are partitioned in different ways for each variable, achieving this symmetry with sensitive measures of importance is not trivial (see Appendix Figure A2). To achieve this we resort to a data augmentation and masking ideas.

We start by creating a cloaked version of the data.
Let $\mathbf{V} \in \{0,1\}^{n \times p}$ be i.i.d.~Bernoulli(0.5) random variables, representing a random noise matrix generated by the analyst independent of everything else.
Then, let $[\mathbf{X},\tilde{\mathbf{X}}]_{\mathrm{swap}(\mathbf{V})} \in \mathbb{R}^{n \times 2p}$ denote the concatenation of $\mathbf{X}$ and $\tilde{\mathbf{X}}$, with the $i$-th observation of $X_j$ swapped with its knockoff if and only if $V_{ij} = 1$.
Combining $[\mathbf{X},\tilde{\mathbf{X}}]_{\mathrm{swap}(\mathbf{V})}$ with $\mathbf{Y}$ and $\mathbf{Z}$ gives us a {\em cloaked} data set $\tilde{\mathcal{D}}(\mathbf{V}) = (\smash{[\mathbf{X}, \tilde{\mathbf{X}}]_{\mathrm{swap}(\mathbf{V})}, \mathbf{Y},\mathbf{Z}})$, where the identities of the original variables and knockoffs are hidden.
Although $\tilde{\mathcal{D}}(\mathbf{V})$ (deliberately) contains additional noise compared to $\mathcal{D}$, it still carries valuable information about the relation between $X$ and $Y$ because each column of the matrix $\smash{[\mathbf{X}, \tilde{\mathbf{X}}]_{\mathrm{swap}(\mathbf{V})}}$ contains approximately $n/2$ real observations.

As will soon become evident, the cloaked dataset $\tilde{\mathcal{D}}(\mathbf{V})$ is useful for \textit{augmenting} the data in $\mathcal{D}$, as it can be safely reused across multiple stages of our analysis. In contrast, the remaining information, specifically the true identities of the variables and knockoffs not captured by $\tilde{\mathcal{D}}(\mathbf{V})$, must be used with great care because it can only be accessed once---to compute the test statistics for the corresponding local hypotheses.

Without loss of generality, the importance scores ${[\mathbf{T},\tilde{\mathbf{T}}]}$ can be written as the output of a (possibly randomized) function $\boldsymbol{\tau}$ taking as input both $\mathcal{D}$ and $\tilde{\mathcal{D}}(\mathbf{V})$:
\begin{align} \label{eq:def-tau}
 & [\mathbf{T},\tilde{\mathbf{T}}] = \boldsymbol{\tau} \big( \mathcal{D}, \tilde{\mathcal{D}}(\mathbf{V}) \big)
  = \big[ \bm{t} \big( \mathcal{D}, \tilde{\mathcal{D}}(\mathbf{V}) \big), \tilde{\bm{t}} \big( \mathcal{D}, \tilde{\mathcal{D}}(\mathbf{V}) \big) \big],
\end{align}
where $\bm{t}$ (resp.~$\tilde{\bm{t}}$) are defined in terms of $\boldsymbol{\tau}$, as its first (resp.~last) ${\hat{L}}$ elements.
This notation is useful to highlight the distinct roles played by the original dataset \( \mathcal{D} \) and the cloaked dataset \( \tilde{\mathcal{D}}(\mathbf{V}) \) in computing importance scores. 
Specifically, \( \tilde{\mathcal{D}}(\mathbf{V}) \) can be used freely, while FDR control requires strict constraints on how \( \boldsymbol{\tau} \) uses the data in \( \mathcal{D} \), as explained next.

For any $\mathcal{S} \subseteq [\smash{L}]$, whose elements uniquely identify pairs $(j,\ell)$ for $j \in [p]$ and $\smash{ \ell \in[{L}_j]}$, let $\smash{[\mathbf{X}, \tilde{\mathbf{X}}]_{\mathrm{swap}(\mathcal{S})}}$ be the matrix obtained from $\smash{[\mathbf{X}, \tilde{\mathbf{X}}]}$ after swapping the sub-column $\smash{\mathbf{X}^{(j,\ell)}_j}$, which contains all observations of $X_j$ in the subgroup $(j,\ell)$, with the corresponding knockoffs, for all $(j,\ell) \in \mathcal{S}$.
Similarly, define $\mathcal{D}_{\mathrm{swap}(\mathcal{S})} = ( [\mathbf{X}, \tilde{\mathbf{X}}]_{\mathrm{swap}(\mathcal{S})}, \mathbf{Y}, \mathbf{Z})$.
Then, our importance scores must be computed in such a way that swapping a variable with its knockoff within any subgroup in $\mathcal{D}$ would have the sole effect of swapping the corresponding importance scores for that subgroup. 
Formally, this means that $\boldsymbol{\tau}$ must satisfy:
\begin{align} \label{eq:swap-tau}
 \boldsymbol{\tau}\big( \mathcal{D}_{\mathrm{swap}(\mathcal{S})}, \tilde{\mathcal{D}}(\mathbf{V}) \big)
 =
 \big[
 \bm{t}\big( \mathcal{D}, \tilde{\mathcal{D}}(\mathbf{V}) \big),
 \tilde{\bm{t}}\big( \mathcal{D}, \tilde{\mathcal{D}}(\mathbf{V}) \big) \big]_{\mathrm{swap}(\mathcal{S})}.
\end{align}

In truth, it is sufficient for~\eqref{eq:swap-tau} to hold in distribution if $\boldsymbol{\tau}$ itself is randomized (e.g., it involves cross-validation to tune some hyper-parameters). However, we can safely ignore this detail here to avoid complicating the notation unnecessarily.

One way to achieve~\eqref{eq:swap-tau} is to compute each pair $(T_{j,\ell}, \tilde{T}_{j,\ell})$ using an appropriate measure of the importance of $(X_j, \tilde{X}_j)$ within a {\em local} model $\hat{\psi}^{(j,\ell)}$ predicting $Y$ given $(X, \tilde{X}, Z)$. This model can be trained using the cloaked dataset $\tilde{\mathcal{D}}(\mathbf{V})$ and a smaller {\em local} dataset $\mathcal{D}^{(j,\ell)} = (\mathbf{X}_j, \tilde{\mathbf{X}}_j, \mathbf{Y}, \mathbf{Z})^{(j,\ell)}$, including only observations $(X_j, \tilde{X}_j, Y, Z)$ for individuals $i \in [n]$ with $\nu_j(Z^i) = \ell$. Figure A3 provides a schematic visualization of this workflow.
Then, to guarantee~\eqref{eq:swap-tau}, it suffices to compute $(T_{j,\ell}, \tilde{T}_{j,\ell})$ in such a way that swapping $X_j$ and $\tilde{X}_j$ in $\mathcal{D}^{(j,\ell)}$ results in $T_{j,\ell}$ being swapped with $\tilde{T}_{j,\ell}$; see Proposition A3 in Appendix A3.

Concretely, in this paper, we implement $\hat{\psi}^{(j,\ell)}$ using a regularized generalized linear model (e.g., the lasso), trained with $\mathbf{Y}^{(j,\ell)}$ as the outcome and $[\mathbf{X}_j, \tilde{\mathbf{X}}_j, [\mathbf{X}, \tilde{\mathbf{X}}]_{\mathrm{swap}(\mathbf{V}, -j)}, \mathbf{Z}]^{(j,\ell)}$ as the predictor matrix, where
$[\mathbf{X}, \tilde{\mathbf{X}}]_{\mathrm{swap}(\mathbf{V}, -j)}$ denotes the matrix $[\mathbf{X}, \tilde{\mathbf{X}}]_{\mathrm{swap}(\mathbf{V})}$ without the $j$-th and $(p+j)$-th columns.
The additional observations contained in the cloaked dataset $\tilde{\mathcal{D}}(\mathbf{V})$ are incorporated into this model through the use of appropriate regularization weights, as explained in Appendix A2. 
The scores $T_{j,\ell}$ and $\tilde{T}_{j,\ell}$ are then defined as the absolute values of the estimated regression coefficients for the first two (scaled) variables.

While it is convenient to think about the lasso, the main idea behind our approach is applicable with any model. The key to achieving~\eqref{eq:swap-tau} is that the scores $(T_{j,\ell},\tilde{T}_{j,\ell})$ are computed by a local model that sees the other variables (other than $X_j$ and $\tilde{X}_j$) and the observations in other subgroups (other than $(j,\ell)$) only through the lenses of the swapped data in $\tilde{\mathcal{D}}(\mathbf{V})$.
However, revealing the true values of $X_j$ and $\tilde{X}_j$ within the environment $(j,\ell)$ when fitting the corresponding local model is essential to achieve non-trivial power---otherwise, it would be impossible to tell important variables apart from knockoffs.

Further implementation details on the computation of local importance scores, as well as additional computational shortcuts that enable the efficient application of this method to very large-scale GWAS data, are provided in Appendix A2.

After computing the importance scores $[\mathbf{T},\tilde{\mathbf{T}}]$, we assemble test statistics $\smash{ \mathbf{W} \in \mathbb{R}^{{\hat{L}}}}$ for all local hypotheses as in \citet{candes2018}, by computing $\smash{ W_{j,\ell} = T_{j,\ell} - \tilde{T}_{j,\ell} }$ for each pair $(j,\ell)$.
Finally, we vectorize $\smash{ \mathbf{W}}$ and apply the selective SeqStep+ procedure of \citet{barber2015controlling} (i.e., the final component of the standard knockoff filter), which computes an adaptive significance threshold determining which hypotheses are rejected.
Algorithm~\ref{alg:sskf_second_phase} summarizes this method, which is guaranteed to control the FDR.

\begin{algorithm}[!htb]
\caption{Local Knockoff Filter for fixed hypotheses}
\label{alg:sskf_second_phase}
\KwIn{Knockoff-augmented data set $\mathcal{D} = ([\mathbf{X},\tilde{\mathbf{X}}], \mathbf{Y}, \mathbf{Z})$; Random noise matrix $\mathbf{V}$; \newline 
 Partition function $\nu = (\nu_1, \ldots, \nu_p)$; Nominal FDR level $\alpha \in (0,1)$\;}

Compute the partition width $L = \sum_{j=1}^{p} L_j$, where $L_j = \max_{z \in \mathcal{Z}} \nu_j(z)$\;
Define local hypotheses $\mathcal{H}_{j,\ell}(\nu)$ as in~\eqref{eq:null-hyp-loc}, for $j \in [p]$ and $\ell \in [L_j]$\;
Assemble the cloaked data matrix $\tilde{\mathcal{D}}(\mathbf{V}) = ([\mathbf{X},\tilde{\mathbf{X}}]_{\mathrm{swap}(\mathbf{V})}, \mathbf{Y}, \mathbf{Z})$\;
\For {$j \in [p]$} {
  \For {$\ell \in [L_j]$} {
   Define the local dataset $\mathcal{D}^{(j,\ell)} = (\mathbf{X}_j, \tilde{\mathbf{X}}_j, \mathbf{Y}, \mathbf{Z})^{(j,\ell)}$\;
   Compute importance scores $T_{j,\ell}$ and $\tilde{T}_{j,\ell}$ using the data in $\mathcal{D}^{(j,\ell)}$ and $\tilde{\mathcal{D}}(\mathbf{V})$, in such a way that~\eqref{eq:swap-tau} holds; see Appendix A2.1 for implementation details\;
    Compute the test statistic $W_{j,\ell} = T_{j,\ell} - \tilde{T}_{j,\ell}$\;
  }
}
Vectorize the test statistic: $\mathbf{W} = (W_{1,1}, \ldots, W_{1,L_1}, W_{2,1}, \ldots, W_{p,1}, \ldots, W_{p,L_p})$\;
Apply Selective SeqStep+ \citep{barber2015controlling} at level $\alpha$ to the vector $\mathbf{W}$\;

\KwOut{List of rejected hypotheses $\hat{\mathcal{R}} \subseteq \{\mathcal{H}_{j,\ell}(\nu)\}_{j \in [p], \ell \in [L_j]}$.}
\end{algorithm}

\begin{theorem} \label{thm:coin-flip-fixed}
Algorithm~\ref{alg:sskf_second_phase} applied to a data set $\mathcal{D} = (\smash{[\mathbf{X}, \tilde{\mathbf{X}}],\mathbf{Y},\mathbf{Z}})$, where $\tilde{\mathbf{X}}$ are valid knockoffs for $\mathbf{X}$, and a fixed partition function $\nu$, controls the FDR at level $\alpha$ for the local hypotheses $\mathcal{H}_{j,\ell}(\nu)$ defined in~\eqref{eq:null-hyp-loc}.
That is, $\mathbb{E}[|\hat{\mathcal{R}} \cap \mathcal{H}_0(\nu)| / \max(1, |\hat{\mathcal{R}}|) ] \leq \alpha$
\end{theorem}

\subsection{A Knockoff Filter for Data-Driven Local Hypotheses}\label{sec:adaptive}

We extend our method to test data-driven local hypotheses~\eqref{eq:null-hyp-loc}.
The idea is that one can {\em learn} an informative partition function $\hat{\nu} = (\hat{\nu}_1, \ldots, \hat{\nu}_p)$ using {\em any algorithm} and then test the corresponding local hypotheses~\eqref{eq:null-hyp-loc} while controlling the FDR, as long as the learning step only uses the {\em cloaked} dataset $\tilde{\mathcal{D}}(\mathbf{V}) = (\smash{[\mathbf{X}, \tilde{\mathbf{X}}]_{\mathrm{swap}(\mathbf{V})}, \mathbf{Y},\mathbf{Z}})$.
In other words, for each $j \in [p]$, we can use any random partition function $\smash{\hat{\nu}_j: {\mathcal{Z}} \mapsto {\mathbb{N}}}$ whose parametrization may depend on $\tilde{\mathcal{D}}(\mathbf{V})$; i.e., $\smash{ \hat{\nu}_j(z; \tilde{\mathcal{D}}(\mathbf{V})) \in [\hat{L}_j] }$ for all $z \in \mathcal{Z}$ and some $\hat{L}_j \in \mathbb{N}_+$.

This two-step framework for adaptively selecting and testing local hypotheses is outlined by Algorithm~\ref{alg:sskf} and visualized schematically in Figure A3 in Appendix A2.
We call this approach {\em Adaptive Local Knockoff Filter} (aLKF).

\begin{algorithm}[!htb]
\caption{Adaptive Local Knockoff Filter (aLKF)}
\label{alg:sskf}
\KwIn{Knockoff-augmented data set $\mathcal{D} = ([\mathbf{X},\tilde{\mathbf{X}}], \mathbf{Y}, \mathbf{Z})$; FDR level $\alpha \in (0,1)$\;}
Randomly generate $\mathbf{V} \in \{0,1\}^{n \times p}$, a matrix of i.i.d.~Bernoulli noise\;
Assemble the cloaked data matrix $\tilde{\mathcal{D}}(\mathbf{V}) = ([\mathbf{X},\tilde{\mathbf{X}}]_{\mathrm{swap}(\mathbf{V})}, \mathbf{Y}, \mathbf{Z})$\;
Looking only at $\tilde{\mathcal{D}}(\mathbf{V})$, learn a data-driven partition function $\hat{\nu}=(\hat{\nu}_1,\ldots,\hat{\nu}_p)$, with total width $ \hat{L} = \sum_{j=1}^{p} \hat{L}_j $; e.g., see Appendix A2.2 for implementation details\;
Apply Algorithm~\ref{alg:sskf_second_phase} using the partition function $\hat{\nu}$\;
\KwOut{List of rejected hypotheses $\hat{\mathcal{R}} \subseteq \{\mathcal{H}_{j,\ell}(\hat{\nu})\}_{j \in [p], \ell \in [\hat{L}_j]}$.}
\end{algorithm}

aLKF provably controls the FDR for~\eqref{eq:null-hyp-loc} conditional on the selected set of hypotheses, a nontrivial result that does not directly follow from prior work. While the proof strategy draws on \citet{candes2018} and \citet{li2021searching}, the data-dependent nature of our hypotheses adds complexity and could introduce selection bias. However, the random noise introduced into the cloaked dataset $\tilde{\mathcal{D}}(\mathbf{V})$ protects us from this issue.

\vspace{0.5em}
\begin{theorem} \label{thm:coin-flip}
Algorithm~\ref{alg:sskf} applied to a knockoff-augmented data set $\mathcal{D} = (\smash{[\mathbf{X}, \tilde{\mathbf{X}}],\mathbf{Y},\mathbf{Z}})$, where $\tilde{\mathbf{X}}$ are valid knockoffs for $\mathbf{X}$, controls the FDR at level $\alpha$ for the family of local hypotheses $\mathcal{H}_{j,\ell}(\hat{\nu})$ in~\eqref{eq:null-hyp-loc} conditional on $\smash{\hat{\nu}}$.
That is, $\mathbb{E}[|\hat{\mathcal{R}} \cap \mathcal{H}_0(\hat{\nu})| / \max(1, |\hat{\mathcal{R}}|) \mid \hat{\nu} ] \leq \alpha$.
\end{theorem}
\vspace{0.5em}

A relatively simple yet effective implementation of aLKF uses sparse generalized linear models with interactions \citep{bien2013lasso} to learn informative partition functions. While the details of this implementation are deferred to Appendix A2.2, the underlying idea is straightforward, particularly in the case of a single binary covariate $Z_1 \in \{0,1\}$. 

For example, we can fit the lasso to predict $\mathbf{Y}$ using $\smash{[\mathbf{X}, \tilde{\mathbf{X}}]_{\mathrm{swap}(\mathbf{V})}}$ and $\smash{\mathbf{Z}_1}$, including all possible pairwise interactions between $\mathbf{Z}_1$ and $\smash{[\mathbf{X}, \tilde{\mathbf{X}}]_{\mathrm{swap}(\mathbf{V})}}$. For each $j \in [p]$, the partition function $\smash{\hat{\nu}_j}$ then divides the samples into two subgroups, corresponding to $\nu_j(0)=1$ and $\nu_j(1)=1$, if the model selects at least one interaction term for either the $j$-th or $(j+p)$-th variables in $\smash{[\mathbf{X}, \tilde{\mathbf{X}}]_{\mathrm{swap}(\mathbf{V})}}$. Otherwise, $\smash{\hat{\nu}_j(z)}=1$ for all $z$ values.
In general, when dealing with $m$ covariates, this approach is applied using a hyper-parameter $G_{\max} \in \mathbb{N}$ controlling the maximum number of interactions that may be selected per variable.
The choice of $G_{\max}$ does not affect the theoretical validity of our method, but it mediates a trade-off: larger values may reveal more specific (and thus potentially more homogeneous) subgroups, but can also reduce power to reject the associated local hypotheses due to the smaller effective sample size within each subgroup. In this paper, we fix $G_{\max} = 1$ or $G_{\max} = 2$, which achieve a reasonable balance and yield informative results. Optimally tuning $G_{\max}$ for a specific dataset is an interesting direction for future research.

As detailed in Appendix A2.2, this implementation can be extended to handle multiple binary covariates as well as non-binary ones. For non-binary covariates, we apply discretization, as the ability to detect meaningful subgroups and reject the corresponding null hypotheses generally depends on having a sufficient number of individuals in each subgroup.
Additionally, Appendix A2.3 describes a specialized implementation with computational shortcuts designed for GWAS-like settings, characterized by a very large number $p$ of variables and a modest number $m$ of covariates.

\section{Simulations} \label{sec:empirical}

We demonstrate the performance of aLKF through two simulated settings, one real GWAS data analysis, and a hybrid semi-simulated setting based on data from a randomized experiment studying the effectiveness of blood donation incentives \citep{sun2019mobile}.
These are designed to capture the most practically relevant scenarios. In the first simulation (Section~\ref{sec:experiments}), the challenge is to identify important variables, when their effects are highly localized, and enviroment is described by a large number of covariates. The second simulation (Section~\ref{subsec:simulation-genetic}), reflects GWAS-like conditions with a multiplicity challenge that comes from the high number of explanatory variables ($p > n$) and fewer environments. It serves as a stepping stone to the real GWAS data analysis presented in Section~\ref{sec:real-data-genetic}. 
The semi-simulated analysis, where both $m$ and $p$ are small, is presented in Appendix A4.

\subsection{Simulation: Finding Highly Localized Important Variables} \label{sec:experiments}

We generate synthetic data with $p=20$ binary variables and $m=80$ covariates, where the first 20 covariates are binary and the remaining 60 are continuous in $[0,1]$. Details on the joint distribution of these variables are in Appendix A4. 

The outcome is generated using a ``causal'' linear model with heterogeneous effects and homoscedastic standard Gaussian noise. 
The expected outcome for the $i$-th individual is $ \E{Y^{i} \mid X^{i}, Z^{i}} = \sum_{j=1}^{20} X^{i}_j \beta_j^{i} + \sum_{j=1}^{80} Z^{i}_j \gamma_j$,
where $\smash{\beta_j^{i} = \bar{\beta}_j Z_{l_{j,1}}^{i} Z_{l_{j,2}}^{i}}$ for some $l_{j,1}, l_{j,2} \in [20]$. 
In words, the effects of $X_j$ are subgroup-specific, with $\smash{\beta_j^{i}}$ determined by a parameter $\smash{\bar{\beta}_j}$ and the interaction of two randomly selected binary covariates $Z_{l_{j,1}}, Z_{l_{j,2}}$. 
For $j \in [10]$, $\smash{\bar{\beta}_j = 0}$, while $\smash{\bar{\beta}_j \in \{\pm 4\}}$ for $j \in \{11, \ldots, 20\}$, with independent random signs.
Thus, if $X_j$ has an effect on $Y$, it has it only in the subgroup where $Z_{l_{j,1}} = 1$ and $Z_{l_{j,2}} = 1$, comprising approximately 1/4 of individuals when $l_{j,1} \neq l_{j,2}$. The indices $l_{j,1}, l_{j,2}$ are sampled with replacement from $[20]$, independently for each~$j$.
The covariate main effects, captured by $\gamma$, are homogeneous. 
The first 40 covariates have no effects: $\gamma_1 = \ldots = \gamma_{40} = 0$. Among the remaining 40 covariates, half (randomly chosen) have effects with absolute value 4 and independent random signs, while the others have no effects; i.e., $\gamma_{41}, \ldots, \gamma_{80} \in \{0, \pm 4\}$.

Based on this simulation setup, valid knockoffs can be sampled independently conditional on the covariates from a Bernoulli distribution; see Appendix A4 for further details.

Given a sample of $n$ independent observations, our goal is to identify which variables are non-null within which subgroups, aiming for both power (many rejections) and precision (informative subgroups).
We apply our method ({\em aLKF}) at the nominal FDR level $\alpha = 0.1$, using the lasso-based approach for importance scores (Appendix A2.1) and learning data-driven partitions via a lasso model with interactions (Appendix 2.2), allowing at most two interactions per variable ($G_{\max} = 2$).
We compare {\em aLKF} against three benchmarks that are applicable within the knockoffs framework: (1) {\em Global-KF}, the knockoff filter of \cite{candes2018}, which controls the FDR but tests only population-wide hypotheses; (2) {\em Split-LKF}, a less powerful version of {\em aLKF} that splits the data, using half to learn the partition function and the other half to test local hypotheses, sacrificing power due to reduced sample size but maintaining FDR control; (3) {\em Naive-LKF}, an invalid version of {\em aLKF} that uses all data twice---to learn the partition and to test local hypotheses---without cloaking, failing to control the FDR due to selection bias.

For each method, we report the false discovery proportion (FDP), the power, and a ``homogeneity'' metric that captures the informativeness of their findings, averaged over 100 independent experiments. These metrics are defined as follows.

The FDP is defined intuitively from~\eqref{eq:null-hyp-loc}: it is the proportion of discoveries for which the variable \( X_j \) has no causal effect on \( Y \) for any individuals within the reported subgroup (which is always equal to the whole sample for the global approach), as determined by the true causal model. Ideally, we aim to keep the average FDP, or FDR, below 10\%.

Power is defined as the proportion of variables with a non-zero causal effect that are correctly identified in at least one subgroup, even if the identified subgroup does not precisely match the true subgroup with non-zero effects, provided the rejected variable has an effect in the identified group. Ideally, power should be as close to one as possible.

The homogeneity metric is defined as the proportion of individuals within a reported subgroup for whom a discovered variable $X_j$ has a non-zero coefficient in the true causal model. Ideally, homogeneity should be close to one, indicating that the identified subgroups are highly homogeneous in their conditional association between $Y$ and $X_j$.

\begin{figure}[!htb]
  \centering
  \includegraphics[height=4cm]{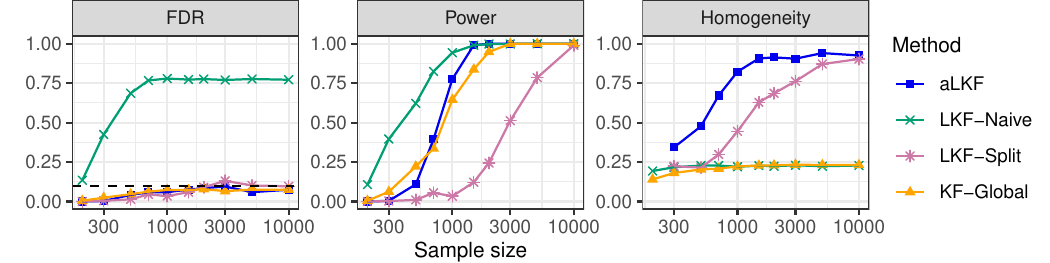}
  \caption{Performance of the adaptive Local Knockoff Filter ({\em aLKF)} and benchmark methods on synthetic data. The informativeness of the discoveries is quantified by the homogeneity of the corresponding subgroups (higher is better). The nominal FDR level is 0.1. }
  \label{fig:experiment-heterogeneous-1}
\end{figure}

Figure~\ref{fig:experiment-heterogeneous-1} reports the performances of all methods over 100 independent experiments.
These results validate our method and demonstrate its advantages relative to the three benchmarks. 
In summary, the aLKF discoveries are more informative than those of {\em Global-KF}, more numerous than those of {\em Split-LKF}, and more reliable than those of {\em Naive-LKF}.

{\em Global-KF} is effective at discovering global associations and controls the FDR. However, it cannot identify the subgroups where significant variables have non-zero effects. This limitation is evident in Figure~\ref{fig:experiment-heterogeneous-1}, where the homogeneity remains relatively unchanged as the sample size increases.
In contrast, aLKF uncovers increasingly informative discoveries as the dataset grows, ultimately enabling precise identification of subgroups with non-zero effects. It avoids the selection bias that inflates the FDR in {\em Naive-LKF} and overcomes the inefficiency of {\em Split-LKF}, which loses power due to unnecessary sample splitting.

\subsection{Simulation: Searching for Gene-Environment Interactions}\label{subsec:simulation-genetic}

The following experiments simulate a search for gene-environment interactions in GWAS data, using real chromosome 21 genotypes from unrelated white non-British individuals in the UK Biobank. This subset is chosen to reduce computational cost, but the results of the following experiments can be expected to generalize to British individuals as well as non-European individuals, as the validity of our method depends only on having valid knockoffs, which are available for both populations \citep{sesia2021false}.
Working with real genetic variants introduces additional complexity due to strong collinearity among nearby variants on the same chromosome—a phenomenon known as linkage disequilibrium (LD). Because of LD, identifying the precise causal variant is often infeasible; instead, associations are typically detected at the level of broader genomic regions comprising multiple correlated SNPs. Following \citet{sesia2021false}, we partition the genome into contiguous groups of SNPs in strong LD, which serve as the units of inference in place of individual variants. Intuitively, our method tests whether each group of SNPs is conditionally independent of the outcome within a given environment, conditional on all other SNPs and covariates. The formal definition of the local hypotheses for groups of SNPs is provided in Appendix A2.3.4. In the following analysis the median SNP group width of 3 kilobases (kb)—a scale that captures strong linkage disequilibrium while allowing variation in signal strength.

Because our hypotheses target groups of SNPs, we can use more powerful knockoffs that satisfy a weaker version of the exchangeability condition in Equation~\eqref{eq:knock_cond_1}, which now only needs to hold under \textit{group-wise} swapping. Using knockoffs designed for individual SNPs would reduce power unnecessarily. This issue has been addressed by \citet{sesia2021false}, who developed an algorithm for generating valid knockoffs for SNP groups in GWAS data using hidden Markov models. We adopt their knockoffs in our analysis. 

There are $14,733$ individuals in total and $p = 8,832$ variants. We randomly sample 30\% of the individuals ($n = 4,420$) to imitate a more typical high-dimensional GWAS setting.
We simulate two independent binary random covariates $Z_1$ and $Z_2$, independently across individuals from a Bernoulli($0.5$) distribution. 
The results of additional experiments using only one covariate can be found in Appendix A6.
A continuous outcome $Y$ is simulated based on a linear model: $Y^i = \beta_0 X^i + (\beta_1 X^i) * \mathbf{1}(Z_1^i = 0) + \epsilon^i$, with $\beta_0,\beta_1 \in \mathbb{R}^p$, where $\epsilon^i \sim N(0, 1)$ independently for each $i \in [n]$.
Thus, $Z_1$ is used to create local effects, among the subgroup with $Z_1=0$, while $Z_2$ is unimportant.
We randomly choose 1\% of the genetic variants to be non-null, in the sense that they have a non-zero coefficient either globally (in $\beta_0$) or locally (in $\beta_1$).
Additional results obtained using different proportions of non-null variants (0.05\% and 2\%) can be found in Appendix A6. 
Each non-null variant is randomly assigned as either a global or a local effect according to the desired proportions. 

The nonzero elements in $\beta_0$ and $\beta_1$ are initialized with absolute values equal to $a / \sqrt{n}$, where the signal amplitude $a \in \mathbb{R}^{+}$ is varied as a control parameter, with independent random signs. Additionally, to mitigate disparities in the detectability of global versus local effects, each element in $\beta_0$ and $\beta_1$ is re-scaled by a factor equal to the square root of proportion of individuals in the affected subgroup. For global effects, this adjustment factor is 1, while for local effects it is approximately $\sqrt{2}$.

Our goal is to identify which groups of SNPs are non-null within which subgroups of individuals, aiming for both high power and precise localization. We apply {\em aLKF} as described in Section~\ref{sec:experiments}, with modifications to account for SNP grouping as detailed in Appendix A2.3.4, and we constrain the data-driven partition function to select at most one interaction per variable ($G_{\max} = 1$). In addition to the three benchmarks from Section~\ref{sec:experiments}—{\em Global-KF}, {\em Split-LKF}, and {\em Naive-LKF}—we introduce a fourth, {\em Fixed-LKF}, a non-adaptive version of LKF described below.
All methods in this comparison use the same SNP group definitions and the same group knockoffs from \citet{sesia2021false}.

{\em Fixed-LKF} applies Algorithm~\ref{alg:sskf_second_phase} using a fixed partition function with four local environments for each variable: $(Z_1, Z_2) \in \{0,1\}^2$. 
This does not utilize data cloaking and computes importance scores within each local environment using only the data from that environment, similar to applying {\em Global-KF} to approximately one quarter of the observations.
While {\em Fixed-LKF} tests hypotheses similar to those of {\em aLKF} and controls the FDR, it cannot adaptively identify informative partitions. Consequently, it may lose power, due to reduced effective sample sizes, and produce less informative discoveries, making it more difficult to distinguish truly local effects from those that are likely global.

For each method, we report the FDP, power, and ``homogeneity'' of the discoveries, as in Section~\ref{sec:experiments}. However, to gain additional insight, we evaluate two separate power metrics: {\em global power} and {\em local power}.
The {\em global power} measures the proportion of variables with a global non-zero effect that are correctly identified in at least one subgroup. The {\em local power} measures the proportion of variables with a local non-zero effect that are correctly identified in at least one subgroup. 
As in Section~\ref{sec:experiments}, these power definitions are agnostic to whether the identified subgroup exactly matches the subgroup with true causal effects.

\begin{figure}[!htb]
  \centering
  \includegraphics[width=14cm]{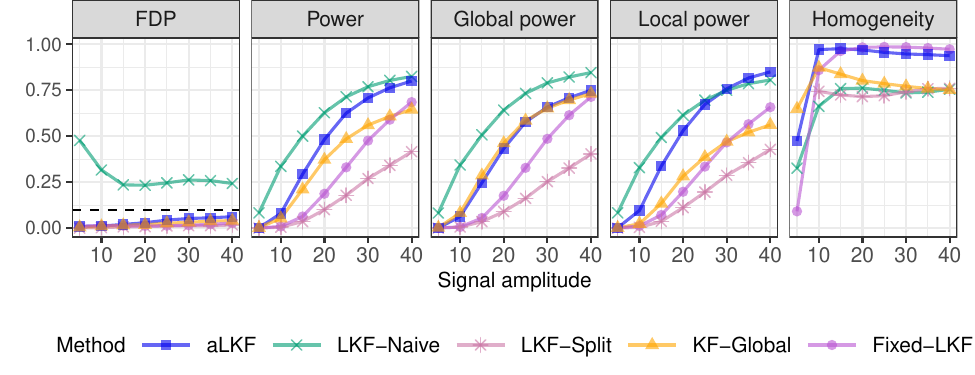}
  \caption{Performance of the adaptive Local Knockoff Filter ({\em aLKF)} and benchmark methods on real genotype data with a simulated outcome, as a function of the signal amplitude. Half of the important genetic variables have global causal effects, while the other half have local effects. Other details are as in Figure~\ref{fig:experiment-heterogeneous-1}.}
  \label{fig:experiment-genetic-2}
\end{figure}

Figure~\ref{fig:experiment-genetic-2} presents the results of these experiments as a function of the signal amplitude parameter $a$, fixing the proportion of causal effects that are local equal to 50\%. 
The results show that {\em aLKF} achieves the highest overall power and the highest power for local effects among all methods with valid FDR control. 
It has comparable power for global effects to {\em Global-KF} but demonstrates significantly higher power for local effects. Both {\em Split-LKF} and {\em Fixed-LKF} have lower power than {\em aLKF}.
The informativeness of the discoveries, measured by homogeneity, is highest for {\em aLKF} and {\em Fixed-LKF}. Note that {\em Fixed-LKF} always achieves a homogeneity of 1 by definition for all correct rejections, whereas {\em aLKF} can also reject local effects for the entire population. As expected, {\em Naive-LKF} fails to control the FDR.

Appendix A6 provides additional details, including the heritability associated with the signal amplitudes in Figure~\ref{fig:experiment-genetic-2}, and results from experiments conducted under similar setups, with qualitatively similar conclusions. These include experiments varying the proportions of local and global effects or the sparsity levels (0.05\% and 2\%), and independent analyses performed with {\em Global-KF} in each of the four subgroups defined by two binary covariates.

\section{Genetic Underpinning of WHR Across Sexes}\label{sec:real-data-genetic}

We study the genetic basis of Waist-Hip Ratio (WHR) across sexes using the UK Biobank data \citep{bycroft2018}. Understanding sex-specific effects could help inform more precise and personal medical decisions, for example by refining genetic risk scores. Previous studies have examined sex-specific effects for WHR or WHR adjusted for BMI, identifying between 7 and 44 sex-specific loci \citep{heid2010meta, randall2013sex, winkler2015influence, shungin2015new, funkhouser2020deciphering}. Given the relatively few sex-specific loci discovered in prior research, we anticipate a high proportion of global effects.

We work with $n=346,119$ unrelated British individuals for whom we have genotypes ($p=591,513$ variants) and measures on sex (46.35\% male), age, height, weight, BMI, waist and hip circumference. While we focus on common variants and unrelated individuals, our method is more generally applicable in all settings where valid knockoffs are available, including for the analysis of rare variants and related individuals \citep{sesia2021false}. We define WHR = waist circumference / hip circumference. 
After the construction of knockoff variables with the approach of \citet{sesia2021false} and the assembly of cloaked data, the analysis described below takes approximately 15 hours on a computing cluster. We refer to Appendix A5 for additional computational details.

We apply {\em aLKF} using the implementation described in Appendix A2.3, designed for large-scale GWAS data with a very large number $p$ of explanatory variables. This leverages lasso models and two key computational shortcuts: pre-screening and batch computation of importance scores. 
The lasso models adjust for the following covariates: age, age squared, the first five genetic principal components, and sex (if applicable). The target FDR level is $\alpha = 0.1$. 
As in the simulation in Section~\ref{subsec:simulation-genetic}, and detailed in Appendices A2.3.4 and A5, inference is performed at the level of SNP groups, with a median group width of 81 kilobases (kb). In real data, this resolution offers a practical balance between power and precision of the tested hypotheses \citep{sesia2020multi, sesia2021false, gablenz2025catch}.

To learn an informative partition function, after pre-screening, we fit a lasso model including all possible pairwise interactions between the pre-screened SNPs and sex; see Appendix A2.3.4 for further details. This leads to the definition of 15,001 SNP-group-level hypotheses for the full sample and 175 hypotheses each for males and females.

As shown in Table~\ref{tab:num_rejections_whr_ukb}, aLKF yields 541 rejections, of which 57 are sex-specific (46 for females and 11 for males). Of the 46 female-specific rejections, 41 are unique to females, while 5 are shared with males. Overall, the number of local rejections for females or males slightly exceeds those reported in prior studies \citep{heid2010meta, randall2013sex, funkhouser2020deciphering, winkler2015influence}, while remaining within a comparable range. The larger number of loci identified for females aligns with previous findings, which observed multiple variants with significant effects in females but smaller or no effects in males \citep{heid2010meta, randall2013sex, winkler2015influence, funkhouser2020deciphering}.

\begin{table}[!htbp]
\centering
\caption{Summary of local and global discoveries for WHR on the UK Biobank GWAS data set, obtained with different methods at the 10\% FDR level.}
\label{tab:num_rejections_whr_ukb}
\begin{tabular}{lcccc}
\toprule
\multirow{2}{*}{Local environment} & \multicolumn{4}{c}{Number of discoveries} \\ 
\cmidrule(lr){2-5}
                                    & {\em aLKF} & {\em Global-KF} & {\em Fixed-LKF} & {\em Separate-KF} \\ 
\midrule
All individuals                     & 484        & 595 & NA & NA\\ 
Female                               & 46         & NA & 257 & 271\\ 
Male                                 & 11         & NA & 82 & 63 \\ 
\midrule
Total                      & 541 & 595 & 339 & 334\\ 
\bottomrule
\end{tabular}
\end{table}

\begin{figure}[!htb]
  \centering
  \includegraphics[width=16cm]{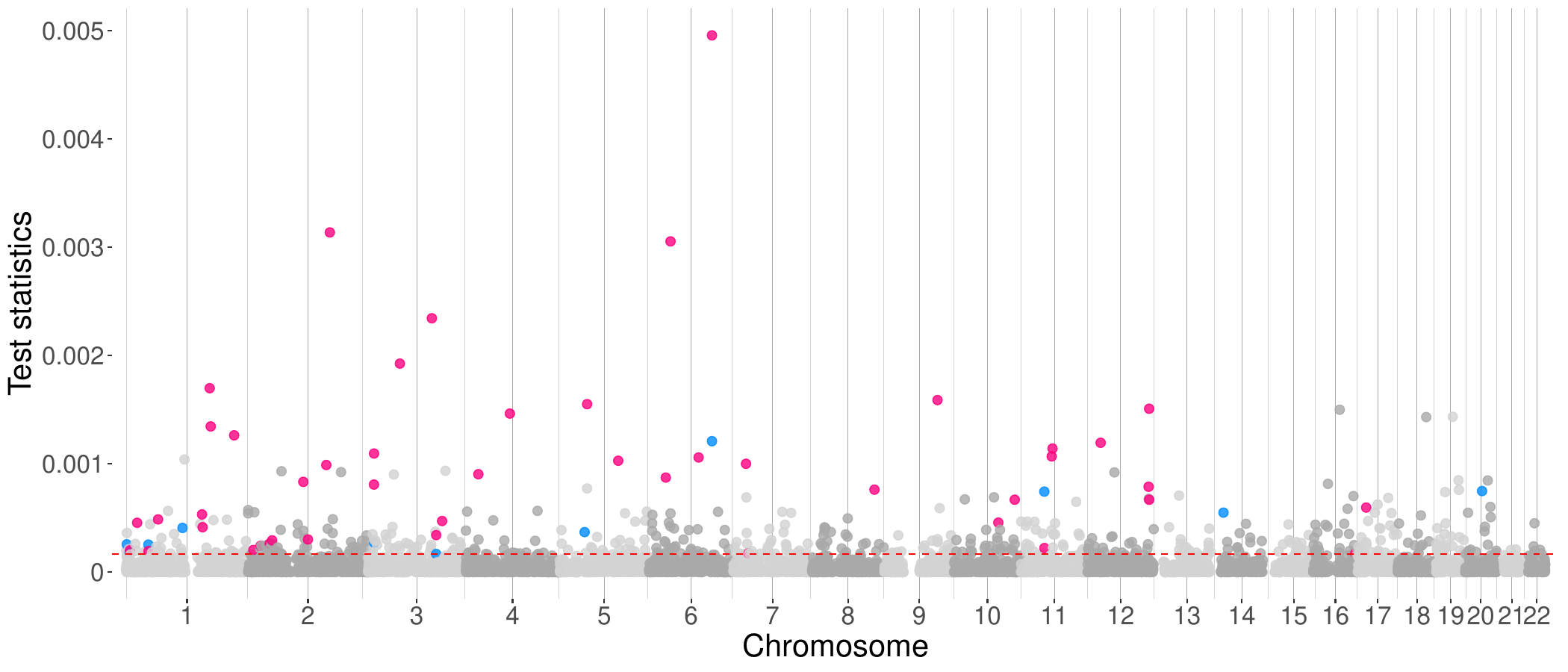}\\[-0.5em]
  \caption{Manhattan plot of {\em aLKF} test statistics for the analysis of WHR using the UK Biobank GWAS data. Each dot represents a local hypothesis for either the entire sample or a sex-specific subgroup. The x-axis indicates the genomic location of the leading SNP within each rejected region, labeled by chromosome, while the y-axis shows the corresponding test statistic. The dashed red line marks the rejection threshold for FDR control at 10\%. 
Discoveries specific to females are shown in pink, and those specific to males are in blue.
}
  \label{fig:ukb-manhattan}
\end{figure}

The Manhattan plot in Figure~\ref{fig:ukb-manhattan} highlights the discoveries made by {\em aLKF}, with female-specific findings shown in pink and male-specific findings in blue. Notably, the largest scores are associated with local hypotheses, and the scores for rejected local hypotheses are distributed across the full range, rather than being concentrated at the lower end.

Table A8 in Appendix A7 lists the 57 sex-specific discoveries reported by {\em aLKF}, along with validation results. Of these, 28 regions show evidence of sex differences based on prior studies \citep{heid2010meta, randall2013sex, winkler2015influence, funkhouser2020deciphering}. These studies also discuss possible biological mechanisms: the top (local) hit is on the RSPO3 gene, which impacts body fat distribution and regulates adipose cell biology. Another top (local) finding on chromosome 6 is the gene VEGFA, which is presumed to act as a mediator of adipogenesis. Additionally, 26 regions have been associated with WHR or a closely related phenotype, WHR adjusted for BMI \citep{sollis2023nhgri, costanzo2023cardiovascular}, but the discovery of sex differences is novel. Three regions have been weakly associated with WHR or WHR adjusted for BMI in earlier studies \citep{costanzo2023cardiovascular}.

Table~\ref{tab:num_rejections_whr_ukb} also contrasts the results of {\em aLKF} with three benchmark approaches. {\em Global-KF} identifies slightly more global discoveries (595 genetic regions), consistent with the experiments in Figure A6 (Appendix A6), which show higher power for {\em Global-KF} when most causal effects are global. However, while {\em Global-KF} captures 37 of the 57 sex-specific loci identified by {\em aLKF}, it misses 20, including 6 previously reported sex-specific effects (see Table A8 in Appendix A7). Moreover, {\em Global-KF} does not distinguish between sex-specific and global effects, reducing its informativeness in this context.

The {\em Fixed-KF} approach described in Section~\ref{subsec:simulation-genetic} results in 339 total rejections. Applying {\em Separate-KF} (the global knockoff filter) separately to the male and female datasets (185,694 observations for females, 160,425 for males) yields 271 rejections for females and 63 for males, for a combined total of 334. These findings underscore that {\em aLKF} is a powerful method for testing local hypotheses, consistent with earlier numerical experiments.

Finally, the Venn diagram in Figure A15 (Appendix A7 compares the unique loci discovered by {\em aLKF}, {\em Global-KF}, and {\em Fixed-KF}. These approaches agree on 215 loci, while {\em aLKF} and {\em Global-KF} share 277 additional loci. Table A7 (Appendix A7) further compares the partition-specific rejections between {\em aLKF} and {\em Fixed-KF}.

\section{Discussion} \label{sec:discussion}

We introduced conditional hypotheses to explore dependencies between explanatory variables and an outcome within contexts defined by other covariates. 
Our inferential method uses knockoffs to create a ``cloaked" version of the data, enabling the identification of heterogeneity while preserving the ability to test local hypotheses after unmasking the knockoffs and original variables. An application to the interaction between sex and DNA variation in the genetic architecture of WHR highlights the potential of this approach.

Applying the LKF necessitates generating knockoffs based on the conditional distribution of $X$ given $Z$, assuming this distribution is known. In our study, this was straightforward since sex is independent of genotypes from non-sexual chromosomes. However, in other contexts, obtaining conditionally valid knockoffs may be more complicated. In such cases, approximate second-order knockoffs may be beneficial \citep{chu2024second}.

The data analysis and simulations in this paper all reflect the results of one knockoff generation, with associated randomness. 
Future research could investigate how to use multiple knockoffs to improve stability \citep{he2021genome}, an how to integrate the de-randomization approach of \citet{ren2024derandomised}, which leverages e-values \citep{wang2022false}.

The first step of our procedure consists in identifying the relevant ``neighborhoods'' in which to test for association between $Y$ and each of the $X_j$. 
In the examples considered in this paper, we found that the lasso with interactions can be useful in defining a partition $\nu_j(Z)$. We also noted that different penalization levels for main effects and interactions might be appropriate to navigate the trade-off between the precision of the tested hypotheses with the power to reject them, linked to the size of the subgroup defined by $\nu_j(Z)$ (see Appendix A2.3.1). A better balance between these might be achieved with a resolution-adaptive procedure as the one described in \citet{gablenz2025catch}, leveraging e-values.

The local conditional hypotheses introduced in this paper can be thought as one non parametric definition of ``interaction'' between explanatory variable $X_j$ and covariates $Z$ in influencing outcome $Y$. 
Extending this idea, one could explore interactions between different explanatory variables $X_j$ and $X_k$.
While some elements of our proposed framework can be adapted for this purpose, careful consideration is needed to properly index and label hypotheses to ensure interpretability. We leave this extension for future work.

Appendix A8, instead, presents an extension of our method designed to detect variables with a robust local association across different subgroups, and demonstrates its practical usefulness in a ``transfer learning'' setting inspired by \citet{li2022transfer}. 

\if1\blind
A software implementation of our methods is available at \url{https://github.com/msesia/i-modelx}, along with the code needed to reproduce our simulations and data analysis.
\else
A software implementation of our methods is available at [masked url], along with the code needed to reproduce our simulations and data analysis.
\fi

\if1\blind
\section*{Acknowledgments}
P.~G.~and C.~S.~were supported by NSF grant DMS 2210392.  M.~S.~was supported by NSF grant DMS 2210637. P.~G.~was supported by a Ric Weiland fellowship. This research has been conducted using data from UK Biobank (Applications 27837 and 74923). 

\else

\fi

%%% Local Variables:
%%% mode: latex
%%% TeX-master: "main_jasa"
%%% End:

\bibliographystyle{Chicago}

\bibliography{bibliography.bib}

%%%%%%%%%%%%%%%%%%%%%%%%%%%%%%%%%%%%%%%%%%%%%%%%%%%%%%%%%%%%%%%%%%%%%%%%%%%%%%
\newpage

%\spacingset{1.9} % DON'T change the spacing!
\spacingset{1}

\appendix
% Special numbering for appendix (Use "S" instead of "A" in Supplement)
\renewcommand{\thesection}{A\arabic{section}}
\renewcommand{\theequation}{A\arabic{equation}}
\renewcommand{\thetheorem}{A\arabic{theorem}}
\renewcommand{\theproposition}{A\arabic{proposition}}
\renewcommand{\thelemma}{A\arabic{lemma}}
\renewcommand{\thetable}{A\arabic{table}}
\renewcommand{\thefigure}{A\arabic{figure}}
\renewcommand{\thealgocf}{A\arabic{algocf}}
\setcounter{figure}{0}
\setcounter{table}{0}
\setcounter{proposition}{0}
\setcounter{theorem}{0}
\setcounter{lemma}{0}
\setcounter{equation}{0}
\setcounter{algocf}{0}

\section{Review of Relevant Technical Background} \label{app:review}

\subsection{The Interpretation of Global Conditional Hypotheses} 

The model-X problem studied by \citet{candes2018} can be thought of as testing, for all $j \in [p]$, whether $X_j$ is associated with $Y$ given $Z$ and all variables excluding $X_j$ (i.e., $X_{-j}$); that is, whether the null hypothesis defined in~\eqref{eq:null-hyp} holds true: $\mathcal{H}_{0,j} : Y \indep X_j \mid X_{-j}, Z$.
For example, if one assumed a generalized linear model for $Y \mid X,Z$ (which we do not do here), then the hypothesis in~\eqref{eq:null-hyp} would reduce, under relatively mild assumptions on $P_{X \mid Z}$, to stating that the linear coefficient for $X_j$ is zero~\citep{candes2018}.

More generally, without any parametric model for the distribution of the outcome, a rejection of~\eqref{eq:null-hyp} can be interpreted as a discovery that $X_j$ has some effect on $Y$ conditional on the other observed variables, for at least some of the individuals in the population. (Note that we are not assuming the individuals are identically distributed.)
These discoveries can be generally useful to screen variables in high-dimensional data analyses \citep{candes2018}, to help prioritize follow-up studies \citep{sesia2020multi}, and in some cases even to make approximate causal inferences \citep{li2021searching}.
In fact, under some additional assumptions such as the absence of unmeasured confounders, a rejection of~\eqref{eq:null-hyp} can sometimes be rigorously interpreted as stating that $X_j$ causes $Y$ \citep{bates2020}. 
 
\subsection{The Connection to Causal Inference} \label{app:causal-inference}

\begin{figure}[H]
  \centering
  \includegraphics[]{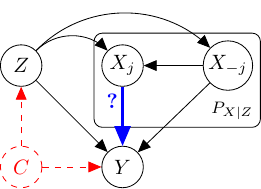}
  \caption{Graphical representation of a non-parametric causal model linking the treatment ($X$), the outcome ($Y$), the measured covariates ($Z$), and possibly also other unmeasured covariate ($C$). A typical goal in this setting would be to test whether a particular treatment $X_j$ has any causal effect on the outcome. The joint distribution of $X \mid Z, C$ is assumed to be known and may depend only on $Z$, so that $X \indep C \mid Z$.
}
  \label{fig:dag}
\end{figure}

To establish an explicit connection between conditional independence testing and causal inference, it would be necessary to introduce two additional assumptions not made in this paper. 
The first additional assumption would be that the variables $X$ can be interpreted as {\em randomized treatments} that may depend on the observed covariates $Z$ but are independent of anything else.
The second additional assumption would be that the joint distribution of $(X,Y,Z)$ takes the form of a structural equation model \citep{bollen2014structural, pearl_2009} in which $Z$ and $X$ may cause $Y$, but not the other way around, as visualized by the directed acyclic graph shown in Figure~\ref{fig:dag}. 

It is easy to see this setup implicitly rules out confounding in a test of conditional independence between $X$ and $Z$, because the treatments are conditionally independent of any other unmeasured covariate $C$.
Therefore, to test whether $X_j$ has a causal effect on $Y$, it suffices to test whether the conditional independence hypothesis in~\eqref{eq:null-hyp}.
\begin{proposition}[From~\cite{bates2020}] \label{prop:causal-reduction}
Let $C$ be any unmeasured covariate. If $X \indep C \mid Z$, a valid test of the conditional independence null hypothesis $\mathcal{H}_{0,j}$ in \eqref{eq:null-hyp} is also a valid test of the stronger null hypothesis
\begin{align} \label{eq:causal-hyp}
  \mathcal{H}_{0,j}^* : Y \indep X_j \mid Z, X_{-j}, C.
\end{align}
\end{proposition}
\begin{proof}
The proof is analogous to that of Proposition~1 in~\cite{bates2020}.
Suppose $\mathcal{H}_{0,j}^* : Y \indep X_j \mid Z, X_{-j},C$ is true.
Then, it follows from $C \indep X_{j} \mid Z, X_{-j}$ that $(Y,C) \indep X_{j} \mid Z, X_{-j}$. Therefore, $\mathcal{H}_{0,j}$ must also be true.
\end{proof}

\subsection{The Global Knockoff Filter} \label{app:review-gkf}

We briefly review here the {\em global} knockoff filter method developed by \citet{candes2018}, which is designed to test the family of global conditional independence hypotheses defined in~\eqref{eq:null-hyp} while controlling the FDR.

Using a knockoff-augmented data set $\mathcal{D} = (\smash{[\mathbf{X}, \tilde{\mathbf{X}}],\mathbf{Y},\mathbf{Z}})$, with variables $\mathbf{X}$, knockoffs $\tilde{\mathbf{X}}$, outcomes $\mathbf{Y}$, and covariates $\mathbf{Z}$, the global knockoff filter fits a predictive model for $Y$ given $X,\tilde{X}$ and $Z$.
This model is utilized compute importance measures $T_j$ and $\tilde{T}_j$ for each $X_j$ and $\tilde{X}_j$.
Any model can be employed for this purpose, as long as swapping $X_j$ with $\tilde{X}_j$ only results in $T_j$ being swapped with $\tilde{T}_j$.
Formally, we may write this property as:
\begin{align} \label{eq:global-symmetry}
  T_{j}\left( \mathbf{X}_j, \tilde{\mathbf{X}}_j, \mathbf{X}_{-j}, \tilde{\mathbf{X}}_{-j}, \mathbf{Y}, \mathbf{Z}\right) 
  & = \tilde{T}_{j}\left( \tilde{\mathbf{X}}_j, \mathbf{X}_j, \mathbf{X}_{-j}, \tilde{\mathbf{X}}_{-j}, \mathbf{Y}, \mathbf{Z}\right),
\end{align}
which may be seen as a {\em global} counterpart of our {\em local symmetry} property defined in~\eqref{eq:local-symmetry}.

A typical choice is to fit a (generalized) lasso model \citep{tibshirani1996regression} and define the importance measures as the absolute values of the scaled regression coefficients for $X$ and $\tilde{X}$, after tuning the regularization via cross-validation.

After computing $T_j$ and $\tilde{T}_j$ for all $j \in p$, these importance scores are combined pairwise into anti-symmetric statistics $W_j$; i.e., $W_j = T_j - \smash{\tilde{T}_j}$.
This ensures that the signs of the $W_j$ are mutually independent coin flips for all $j$ corresponding to a true $\mathcal{H}_{0,j}$, while larger and positive values provide evidence against the null.
Letting $\epsilon \in \{-1,+1\}^p$ be an independent random vector such that $\epsilon_j = +1$ if $\mathcal{H}_{0,j}$ is false and $\mathbb{P}[\epsilon_j = +1 ] = 1/2$ otherwise, then it can be proved that $W$ satisfies the {\em flip-sign} property:
\begin{align} \label{eq:flip-sign}
  W \mid \mathbf{Z} \overset{d}{=} W \odot \epsilon \mid \mathbf{Z},
\end{align}
where $\odot$ indicates element-wise multiplication.
Intuitively, the sign of $W_j$ gives rise to a conservative binary p-value $p_j$ for $\mathcal{H}_{0,j}$.
That is, $p_j = 1/2$ if $W_j>0$ and $p_j = 1$ otherwise, as long as $\mathcal{H}_{0,j}$ is true.

Finally, a rejection rule for~\eqref{eq:null-hyp} can be obtained by applying a sequential testing procedure to these p-values, in the order defined by the absolute values of $W$.
As each of these p-values contains a single bit of information, an appropriate sequential testing procedure is the selective SeqStep+ test of~\cite{barber2015controlling}, which can compute an adaptive threshold for $W_j$ controlling the FDR below any desired threshold.

Although inspired by it, the local knockoff filter method developed in this paper is designed to address a problem that differs significantly from the one targeted by the global knockoff filter of \citet{candes2018}. 
First, our method tests the family of local conditional hypotheses defined in~\eqref{eq:null-hyp-loc}, as opposed to the global hypotheses in~\eqref{eq:null-hyp}. Second, it treats these hypotheses as random, relying on a data-driven partition function $\hat{\nu}$, rather than treating them as fixed a priori. 
Technically, the differences in operation lie in how the vector of test statistics $W$ is defined and computed. Once $W$ is obtained, both approaches apply the selective SeqStep+ test of~\cite{barber2015controlling} in a similar manner.

 \clearpage

\section{Method Implementation} \label{app:implementation}

We now describe in more details concrete implementations of our method.
We start with an approach to compute powerful importance scores for testing local hypotheses with Algorithm~\ref{alg:sskf_second_phase}.
We then introduce a versatile algorithm to learn informative local environments using generalized linear models with interactions.
Finally, we describe a specialized implementation of our method that is designed for settings similar to GWAS, where there is a very large number $p$ of possible explanatory variables and a modest number $m$ of covariates.

To guide the reader through the numerous details, we will rely on graphical illustrations. Because of space considerations, these will focus on two explanatory variables only. We will illustrate two different partition functions for the variables $X_1$ and $X_2$, carrying through the toy example we have used in the text. We convenience we visualize the local conditional hypotheses described in the example in the following Table~\ref{table:partition-example}.

\begin{table}[!htb]
  \caption{Example of partitions and corresponding local hypotheses. For the first variable, $\nu_1$ partitions the observations in two subgroups (male and female), while $\nu_2$ relative to the second variable partitions the sample in three subgroups (``underweight'', ``healthy weight'', and ``overweight''.)} \vspace{0.5em}
  \label{table:partition-example}
    \centering
     \begin{tabular}{ccccl}
        \toprule
      \multirow{2}{*}{Variable} & \multirow{2}{*}{Covariates} & \multicolumn{2}{c}{Partition} & \multirow{2}{5cm}{\centering Hypotheses to test}\\
      \cmidrule(l{2pt}r{2pt}){3-4} 
       &  & Label & Definition &  \\
      \midrule
      \multirow{2}{*}{$X_1$} & \multirow{2}{*}{\{$Z_1=\text{Sex}$\}} & (1) & $Z_1 =0$ & $X_1$ is cond. ind. of $Y$ if $Z_1=0$ (male) \\
       &  & (2) & $Z_1 = 1$  & $X_1$ is cond. ind. of $Y$ if $Z_1=1$ (females) \\
       
      \hline
      \multirow{3}{*}{$X_2$} & \multirow{3}{*}{\{$Z_3=\text{BMI}$\}} & (1) & $Z_2 \leq t_{l}$  & $X_2$  is cond. ind. of  $Y$ if $Z_2\leq t_{l}$\\
       &  & (2) & $t_{l}\leq Z_2<t_{u}$  & $X_2$  is cond. ind. of  $Y$ if $t_{l} < Z_2 \leq t_{u}$ \\
       &  & (3) & $Z_2 > t_u$  & $X_2$  is cond. ind. of  $Y$ if $Z_2 >t_u $ \\[0.25em]
       \hline
     $X_3$ & $\emptyset$ & 1 & All individuals  & $X_3$   is cond. ind. of  $Y$ \\
        \bottomrule

    \end{tabular}

\end{table}

\begin{figure}[!htb]
  \centering
 (a) \hspace*{8cm} (b) 
  
  \includegraphics[width=0.95\textwidth]{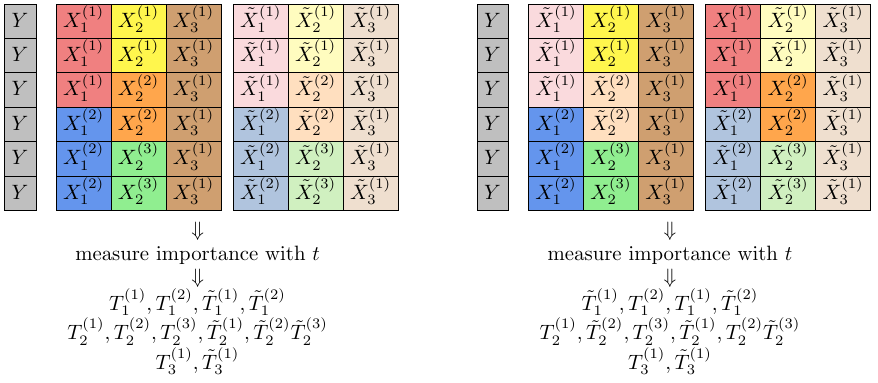}
  \caption{Illustration of the symmetry requirement for measures of importance. Each column corresponds to one variable: we have three columns for the original variables, and three columns for their knockoffs. Each row corresponds to one observation. Colors are used to indicate the partition of observations corresponding to the tested local hypotheses (we have two subgroups for variable 1, three subgroups for variable 2, and one group for variable 3). Original observations are in a more saturated shade, while corresponding knockoffs are lighter. The (b) panel represents a possible swapping of $X_j$ with its knockoff $\tilde{X}_j$ within subgroups $\ell \in [L_j]$ and indicates how the measures of importance need to correspond to those in panel (a), with the swapping of the corresponding scores.}
  \label{fig:symmetry}
\end{figure}

Figure \ref{fig:symmetry} illustrates the symmetry requirements corresponding to the collection of these 6 hypotheses. 

\subsection{Computing Powerful Local Importance Scores} \label{sec:powerful-statistics}

We describe here in more detail a specific implementation of Algorithm~\ref{alg:sskf_second_phase} that uses regularized generalized linear models to compute powerful local importance scores satisfying~\eqref{eq:swap-tau}. While this is not the only possible approach, it is one that we have found to perform well in practice, particularly when the dataset size is moderate. Additionally, it serves to illustrate the flexibility of our local knockoff filter method.

\begin{sidewaysfigure}
  \centering
  \includegraphics[width=24cm]{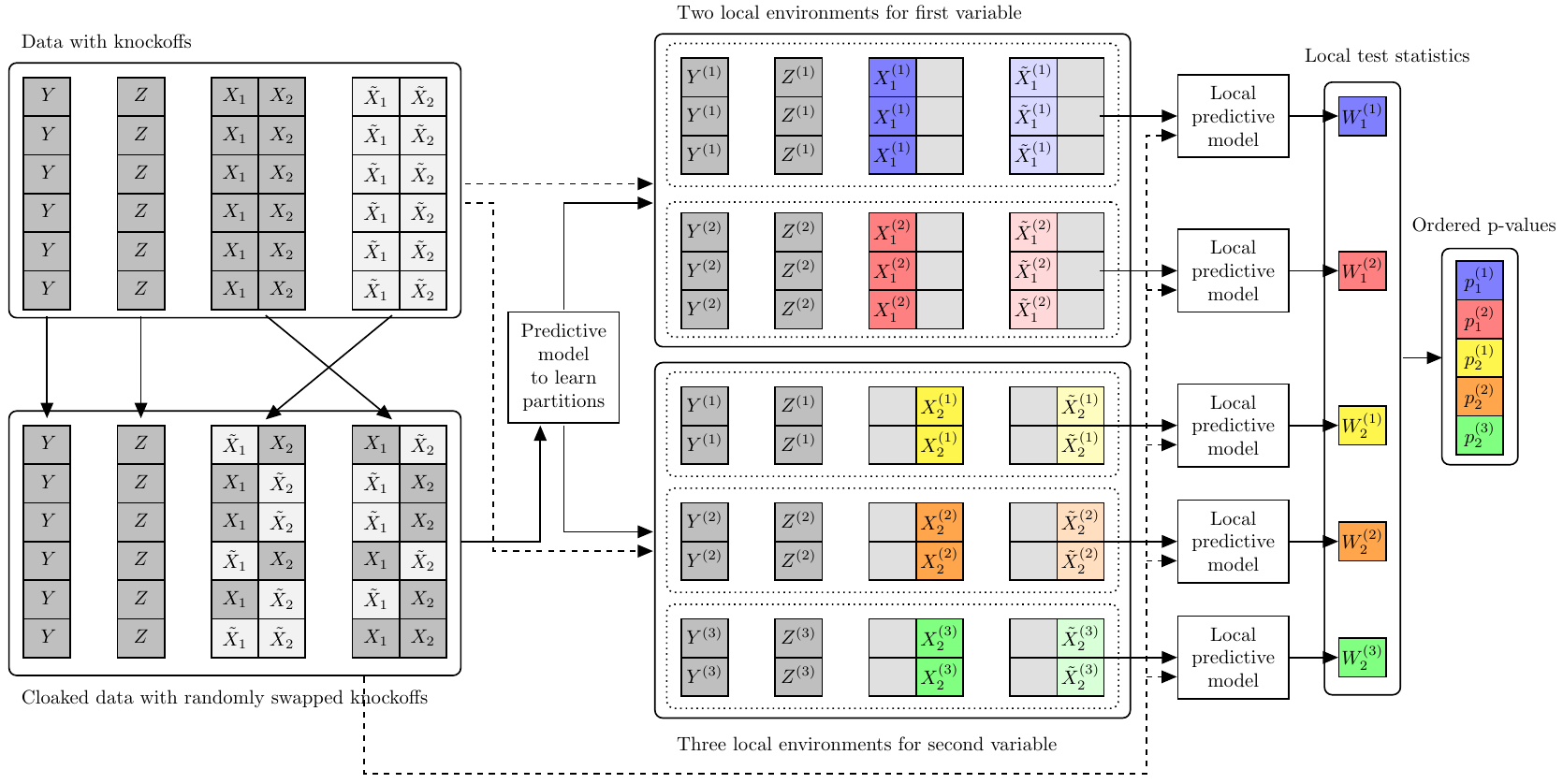}
  \caption{Schematic of the Adaptive Local Knockoff Filter. The relevant variables and associated statistics for different local environments are shown in different colors. The original variables are shown in darker shades and the knockoffs in lighter shades. We describe each step in Section~\ref*{sec:knockoffs-testing} in detail. In summary, we start by constructing the \textit{cloaked} dataset, which involves hiding the identities of the original variables ($X_1, X_2$) and their knockoffs ($\widetilde{X}_1$, $\widetilde{X}_2$). This cloaked dataset can be safely used to fit a predictive model to learn the partitioning of the data. In this toy example, we find two local environments for $X_1$ and three local environments for $X_2$. Next, we fit a single local predictive model within each partition of the data, revealing the true identities of the relevant variables in each local model (for example, we would reveal the true identities of $X_1$ and $\widetilde{X}_1$ in the partitions for the first variable). This allows to obtain a local test statistic for each variable and each partition, resulting in five local test statistics in total (two for $X_1$ and three for $X_2$). These local test statistics are then passed through the knockoff filter \citep{candes2018}.}
  \label{fig:diagram}
\end{sidewaysfigure}

Recall the high-level ideas of the procedure described in Section~\ref*{sec:knockoffs-testing}, also visualized schematically in Figure~\ref{fig:diagram}.
For each pair $(j,\ell)$, we extract the scores $(T_{j,\ell},\tilde{T}_{j,\ell})$ from a local predictive model $\hat{\psi}^{(j,\ell)}$ trained using information from the cloaked dataset $\tilde{\mathcal{D}}(\mathbf{V})$ and the {\em local} dataset $\mathcal{D}^{(j,\ell)} = (\mathbf{X}_j, \tilde{\mathbf{X}}_j, \mathbf{Y}, \mathbf{Z})^{(j,\ell)}$, which includes only observations $(X_j, \tilde{X}_j, Y, Z)$ for individuals $i \in [n]$ with $\nu_j(Z^i) = \ell$.
We now outline this implementation in more detail.

First, a sparse generalized linear model (e.g., the lasso) $\hat{\psi}^{(0)}$ is trained to predict $\mathbf{Y}$ given the data in $\tilde{\mathcal{D}}(\mathbf{V})$, tuning the regularization via cross-validation.
The absolute values of the (scaled) regression coefficients from $\hat{\psi}^{(0)}$ will serve as ``prior'' importance measures, $\smash{T_j^{\text{prior}}}$ and $\smash{\tilde{T}_j^{\text{prior}}}$, for all variables and knockoffs indexed by $j \in [p]$.
These are combined pairwise into a weight $\pi_j$ for each $j$; for example, as $\smash{ \pi_j = \zeta(T^{\text{prior}}_j+\tilde{T}^{\text{prior}}_j) }$, where $\zeta : \mathbb{R}_+ \mapsto \mathbb{R}_+$ is a positive and decreasing function such as $\zeta(t) = 1/(0.05+t)$. Larger values of $\pi_j$ suggest the $j$-th variable is more likely to have a significant effect among some individuals. Similar weights can also be defined for the covariates.

Then, separately for each pair $(j,\ell)$, a {\em local} model $\hat{\psi}^{(j,\ell)}$ is trained with $\mathbf{Y}^{(j,\ell)}$ as the outcome and $[\mathbf{X}_j, \tilde{\mathbf{X}}_j, [\mathbf{X}, \tilde{\mathbf{X}}]_{\mathrm{swap}(\mathbf{V}_{-j})}, \mathbf{Z}]^{(j,\ell)}$ as the predictor matrix, where $[\mathbf{X}, \tilde{\mathbf{X}}]_{\mathrm{swap}(\mathbf{V}_{-j})}$ is defined as a narrower version of the cloaked data set $\tilde{\mathcal{D}}(\mathbf{V})$, in which the $j$-th and $(p+j)$-th columns are omitted.
Omitting the $j$-th and $(p+j)$-th columns of $[\mathbf{X}, \tilde{\mathbf{X}}]$ is intuitive since the true variable $\mathbf{X}_j$ and its knockoff $\tilde{\mathbf{X}}_j$ are already included in this local model.

To leverage any additional (non-local) information that may be contained in the cloaked data set $\tilde{\mathcal{D}}(\mathbf{V})$, each sparse local model $\hat{\psi}^{(j,\ell)}$ is trained with feature-specific regularization that depends on the prior weights $\pi$ extracted from $\hat{\psi}^{(0)}$.
Specifically, for each $l \in [p]$, the penalty for the $l$-th and $(l+p)$-th variables in $\hat{\psi}^{(j,\ell)}$ is $\smash{\lambda^{(j,\ell)}_l = \lambda^{(j,\ell)} (1-\xi^{(j,\ell)}) + \xi^{(j,\ell)} \pi_l}$, where $\lambda^{(j,\ell)}>0$ and $\xi^{(j,\ell)} \in [0,1]$ are two hyper-parameters to be tuned by cross-validation.

If $\smash{\xi^{(j,\ell)} = 0}$, this reduces to a local model that only depends on the data in subgroup $(j,\ell)$. But, in general, larger values of $\xi^{(j,\ell)}$ can make our approach more powerful, as we gather strength from the data in all subgroups. In fact, null variables will tend to receive smaller values of $\pi$ and will thus be less likely to be incorrectly selected by the final model, thereby reducing the noise in the test statistics.
Of course, the weights $\pi$ may not always be informative, hence why $\xi^{(j,\ell)}$ is tuned by cross-validation. For that purpose, an expensive two-dimensional grid search can be avoided by tuning first $\lambda^{(j,\ell)}>0$ and then $\xi^{(j,\ell)}$.

Finally, the importance scores for the $j$-th variable and knockoff in subgroup $\ell$ are respectively defined as the absolute values of the (scaled) regression coefficients for $X_j$ and $\tilde{X}_j$ in the local model $\hat{\psi}^{(j,\ell)}$.
This procedure is summarized in Algorithm~\ref{alg:sskf_local_scores}.
The following result states that this leads to importance scores satisfying the equivariance property in~\eqref{eq:swap-tau}.

\begin{algorithm}[!htb]
\caption{Computing powerful lasso-based local importance scores}
\label{alg:sskf_local_scores}
\KwIn{Data $\mathcal{D} = ([\mathbf{X},\tilde{\mathbf{X}}], \mathbf{Y}, \mathbf{Z})$;
  Cloaked data set $\tilde{\mathcal{D}}(\mathbf{V}) = ([\mathbf{X},\tilde{\mathbf{X}}]_{\mathrm{swap}(\mathbf{V})}, \mathbf{Y}, \mathbf{Z})$, Partition function $\nu = (\nu_1, \ldots, \nu_p)$;\newline
  Positive and decreasing function $\zeta : \mathbb{R}_+ \mapsto \mathbb{R}_+$;\newline
  Finite lists of possible hyper-parameter values $\Lambda \subseteq \mathbb{R}_+$ and $\Xi \subseteq \mathbb{R}_+$\;
}

Fit a sparse generalized linear model $\hat{\psi}^{(0)}$ to predict $Y$, using the data in $\tilde{\mathcal{D}}(\mathbf{V})$\;
\For {$j \in [p]$} {
  Define $\smash{T_j^{\text{prior}}}$ and $\smash{\tilde{T}_j^{\text{prior}}}$ as the absolute values of the (scaled) regression coefficients in $\hat{\psi}^{(0)}$ for the $j$-th and $(j+p)$-th variables in $[\mathbf{X},\tilde{\mathbf{X}}]_{\mathrm{swap}(\mathbf{V})}$, respectively\;
  Compute $\pi_j = \zeta(T^{\text{prior}}_j+\tilde{T}^{\text{prior}}_j)$.
}
Compute the partition width $L = \sum_{j=1}^{p} L_j$, where $L_j = \max_{z \in \mathcal{Z}} \nu_j(z)$\;
\For {$j \in [p]$} {
  \For {$\ell \in [L_j]$} {
  Define the local data set $\tilde{\mathcal{D}}^{(j,\ell)}(\mathbf{V}) = ([\mathbf{X}, \tilde{\mathbf{X}}]_{\mathrm{swap}(\mathbf{V}_{-j})}, \mathbf{Y}, \mathbf{Z})^{(j,\ell)}$\;
%      Define the local data set $\mathcal{D}^{(j,\ell)} = ([\mathbf{X}, \tilde{\mathbf{X}}], \mathbf{Y}, \mathbf{Z})^{(j,\ell)}$\;
    \For {$(\lambda, \xi) \in \Lambda \times \Xi$ } {
      For each $l \in [p]$, define the weight $\smash{\lambda_l = \lambda (1-\xi) + \xi \pi_l}$\;
      Fit a sparse generalized linear model $\hat{\psi}^{(j,\ell)}(\lambda, \xi)$ to predict $Y$ using the data in a cross-validated subset of $\tilde{\mathcal{D}}^{(j,\ell)}(\mathbf{V})$, with weights $(\lambda_1,\ldots,\lambda_p,\lambda_1,\ldots,\lambda_p, 0, \ldots, 0)$ for $([\mathbf{X}, \tilde{\mathbf{X}}]_{\mathrm{swap}(\mathbf{V}_{-j})}, \mathbf{Z})$\;
    }
  }
  Select the hyper-parameters $(\hat{\lambda}^{(j,\ell)}, \hat{\xi}^{(j,\ell)})$ that minimize the cross-validated test prediction error for $Y$ based on the data in $\tilde{\mathcal{D}}^{(j,\ell)}(\mathbf{V})$\;
  For each $l \in [p]$, define the weight $\smash{\hat{\lambda}^{(j,\ell)}_l = \hat{\lambda}^{(j,\ell)} (1-\hat{\xi}^{(j,\ell)}) + \hat{\xi}^{(j,\ell)} \pi_l}$\;
  Fit a sparse generalized linear model $\hat{\psi}^{(j,\ell)}$ for $Y$ using all data in $\tilde{\mathcal{D}}^{(j,\ell)}(\mathbf{V})$, with weights $(\hat{\lambda}^{(j,\ell)}_1,\ldots,\hat{\lambda}^{(j,\ell)}_p,\hat{\lambda}^{(j,\ell)}_1,\ldots,\hat{\lambda}^{(j,\ell)}_p, 0, \ldots, 0)$ for $([\mathbf{X}, \tilde{\mathbf{X}}]_{\mathrm{swap}(\mathbf{V}_{-j})}, \mathbf{Z})$\;
    Define $\smash{T_{j,\ell}}$ and $\smash{\tilde{T}_{j,\ell}}$ as the absolute values of the (scaled) regression coefficients in $\hat{\psi}^{(j,\ell)}$ for $X_j$ and $\tilde{X}_j$, respectively\;
}
\KwOut{List of importance scores $(T_{j,\ell},\smash{\tilde{T}_{j,\ell}})$ for all $j \in [p]$ and $\ell \in [L_j]$.}
\end{algorithm}

\begin{proposition} \label{prop:sskf_local_scores}
The local importance scores $[\mathbf{T},\tilde{\mathbf{T}}]$ output by Algorithm~\ref{alg:sskf_local_scores} satisfy the equivariance property defined in~\eqref{eq:swap-tau}.
\end{proposition}
\begin{proof}
This is an immediate corollary of Proposition~\ref{prop:sskf_local_scores-simple}, stated in Appendix~\ref{app:proofs}.
\end{proof}

\begin{sidewaysfigure}
  \centering
  \includegraphics[width=23cm]{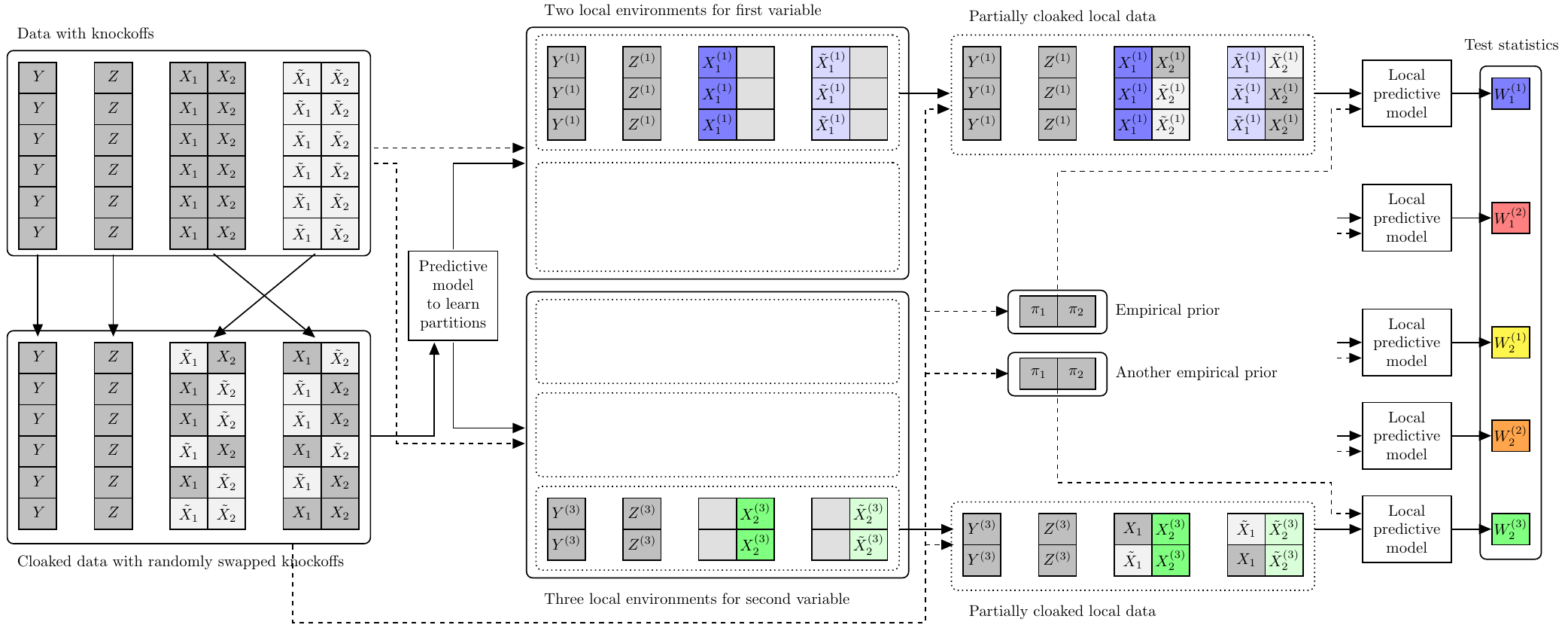}
  \caption{Schematic of Adaptive Local Knockoff Filter, focusing on the computation of the statistics for the first (blue) and fifth (green) local environments. 
    The test statistics are based on a simple empirical prior for the two variables that is learned from the cloaked data set with randomly swapped knockoffs. Other details are as in Figure~\ref{fig:diagram}.}
  \label{fig:diagram_2}
\end{sidewaysfigure}

A natural extension of this implementation would involve estimating different prior weights $\pi$ in different local environments, as sketched by Figure~\ref{fig:diagram_2}. This more sophisticated implementation still satisfies~\eqref{eq:swap-tau} as long as the models utilized to compute the priors only look at the cloaked data in $\tilde{\mathcal{D}}(\mathbf{V})$.

\FloatBarrier

\subsection{Learning Local Environments Using Interaction Models} \label{sec:learn-interactions}

This section describes a practical approach for learning knockoff-invariant partition functions that allow testing informative local conditional hypotheses with our method.
This approach is based on parametric regression models with interaction terms.

Suppose that we suspect some of the variables denoted as $X$ may interact with certain covariates $Z$, akin to the toy example from Section~\ref*{sec:loc-hyp}.
For simplicity, we focus on binary covariates here, although one could also similarly handle more general categorical covariates, or even continuous-valued covariates, through standard discretization techniques.

In this scenario, the analyst first generates a matrix $\mathbf{V}$ of i.i.d.~Bernoulli random variables, independent of the data, and assembles $\tilde{\mathcal{D}}(\mathbf{V})$.
Then, the analyst proceeds by fitting a sparse generalized linear model (i.e., the lasso) to predict $\mathbf{Y}$ given $\smash{[\mathbf{X}, \tilde{\mathbf{X}}]_{\mathrm{swap}(\mathbf{V})}}$ and $\smash{\mathbf{Z}}$, including all possible pairwise interactions between $\mathbf{Z}$ and $\smash{[\mathbf{X}, \tilde{\mathbf{X}}]_{\mathrm{swap}(\mathbf{V})}}$.
Several algorithms exist for fitting such models \citep{bien2013lasso,lim2015learning,tibshirani2020pliable}, making this approach practical and effective.

After tuning the lasso regularization via cross-validation, let $\hat{\beta}_j$ and $\hat{\beta}_{j+p}$ indicate the estimated main effects for the (scaled) variables in the $j$-th and $(j+p)$-th columns of $[\mathbf{X}, \widetilde{\mathbf{X}}]_{\text{swap}(\mathbf{V})}$, respectively.
Since $X_j$ and $\widetilde{X}_j$ are cloaked, in this model there is no meaningful distinction between the original variables and the knockoffs.
Let also $\hat{\gamma}_{l,j}$ and $\hat{\gamma}_{l,j+p}$ indicate the corresponding interaction coefficients involving $Z_l$ and the variables in the $j$-th and $(j+p)$-th columns of $[\mathbf{X}, \widetilde{\mathbf{X}}]_{\text{swap}(\mathbf{V})}$, respectively, for all $l \in [m]$ and $j \in [p]$.

Given an upper bound $G_{\max}$ on the number of interactions per variable (e.g., $G_{\max}=2$), for all $j \in [p]$ let $\smash{\hat{\mathcal{I}}_j \subset [m]}$ indicate the subset of $G_{\max}$ covariates with the strongest (nonzero) interactions involving the randomly swapped versions of $X_j$ or $\tilde{X}_j$.
Here, we understand that any ties are broken at random and $\smash{|\hat{\mathcal{I}}_j| < G_{\max}}$ if the number of covariates with nonzero interaction coefficients is too small.
In other words, we compute
\begin{align*}
  \hat{\mathcal{I}}_j := \big\{ l \in [m] : |\hat{\gamma}_{l,j}|+|\hat{\gamma}_{l,j+p}| > 0, \sum_{l' \neq l} \I{|\hat{\gamma}_{l,j}|+|\hat{\gamma}_{l,j+p}| > |\hat{\gamma}_{l',j}|+|\hat{\gamma}_{l',j+p}|} > m-G_{\max} \big\}.
\end{align*}
This model thus links each pair $(X_j,\tilde{X}_j)$ to at most $G_{\max}$ covariates.
The corresponding functions $\smash{\hat{\nu}_j}$ take values in $\smash{\{1,\ldots,2^{|\hat{\mathcal{I}}_j|}\}}$, which indexes all possible configurations of the covariates in $\hat{\mathcal{I}}_j$.
Therefore, we may write $|\hat{L}_j| = 2^{|\hat{\mathcal{I}}_j|}$ for each $j \in [p]$.

For example, in the special case where $\hat{\mathcal{I}}_j = \emptyset$, meaning that no interaction effects are selected for the pair $(X_j, \tilde{X}_j)$, this procedure yields $\nu_j(z) = 1$ for all possible $z \in \mathbb{R}^m$. In other words, in this case, the partition function $\hat{\nu}_j$ defines a single local environment for the $j$-th variable, comprising the full sample.
By contrast, if $|\hat{\mathcal{I}}_j| = \{l\}$ for some $l \in [m]$, the partition function $\hat{\nu}_j$ will define two local environments for the $j$-th variable: one corresponding to $Z_l = 0$ and the other to $Z_l = 1$.

%Each $\smash{\hat{\nu}_j}$ is a mapping from the Cartesian space $\smash{\{0,1\}^{|\hat{\mathcal{I}}_j|}}$ to $\smash{\{1,\ldots,2^{|\hat{\mathcal{I}}_j|}}\}$, indexing all possible configurations of the covariates in $\hat{\mathcal{I}}_j$.

Table~\ref{table:interactions-example} helps visualize this approach by highlighting a thought experiment involving 3 explanatory variables and 3 binary covariates.
In this toy example, we imagine that the lasso selects $Z_1, Z_2$ as potentially interacting with $X_1$, and $Z_3$ for $X_2$.
No interactions with $X_3$ are detected.
Thus, the population is partitioned into 4 subgroups for $X_1$, 2 subgroups for $X_2$, and 1 trivial subgroup for $X_3$, yielding a total of 7 data-driven hypotheses.

% The interpretation of the findings reported by the adaptive local knockoff filter is that $X_1$ is associated with $Y$ for individuals with $Z_2=1$, while $X_2$ is associated with $Y$ for individuals with $Z_3=1$, and $X_3$ is associated with $Y$ at the aggregate level. No evidence of an association between $X_1$ and $Y$ is detected among individuals with $Z_2=0$, and no evidence of association between $X_2$ and $Y$ is detected among individuals with $Z_3=0$.

\begin{table}[!htb]
  \caption{Subgroups selected by fitting a lasso model with interactions, and corresponding interpretations of the possible discoveries obtained with our method, in a toy example with 3 variables and 3 binary covariates. The term ``influences'' is utilized loosely here to indicate a (possibly non-causal) significant conditional association.} \vspace{0.5em}
  \label{table:interactions-example}
    \centering
    \begin{tabular}{ccccc}
        \toprule
      \multirow{2}{*}{Variable} & \multirow{2}{*}{Covariates} & \multicolumn{2}{c}{Partition} & \multirow{2}{5cm}{\centering Interpretation of the findings (upon rejection)}\\
      \cmidrule(l{2pt}r{2pt}){3-4}
       &  & Label & Definition &  \\
      \midrule
      \multirow{4}{*}{$X_1$} & \multirow{4}{*}{\{$Z_1,Z_2$\}} & 1 & $Z_1 = 0, Z_2 = 0$ & $X_1$ influences $Y$ if $Z_1=0$ and $Z_2=0$\\
       &  & 2 & $Z_1 = 0, Z_2 = 1$  & $X_1$ influences $Y$ if $Z_1=0$ and $Z_2=1$\\
       &  & 3 & $Z_1 = 1, Z_2 = 0$  & $X_1$ influences $Y$ if $Z_1=1$ and $Z_2=0$ \\
       &  & 4 & $Z_1 = 1, Z_2 = 1$  & $X_1$ influences $Y$ if $Z_1=1$ and $Z_2=1$\\
      \hline
      \multirow{2}{*}{$X_2$} & \multirow{2}{*}{\{$Z_3$\}} & 1 & $Z_3 = 0$  & $X_2$ influences $Y$ if $Z_3=0$\\
       &  & 2 & $Z_3 = 1$  & $X_2$ influences $Y$ if $Z_3=1$ \\[0.25em]
      \hline
      $X_3$ & $\emptyset$ & 1 & All individuals  & $X_3$ influences $Y$ \\
        \bottomrule
    \end{tabular}
\end{table}

Note that the parameter $G_{\max}$ generally controls an important trade-off between the power of our method and the interpretability of the findings. A larger value of $G_{\max}$ tends to lead to more specific hypotheses corresponding to smaller subgroups, but it also makes it more difficult to reject those hypotheses by reducing the effective sample size available to compute the local importance scores by a factor of up to $2^{G_{\max}}$, in the case of binary covariates.
Concretely, all demonstrations presented in this paper utilize $G_{\max} \in \{1,2\}$, but the optimal value of $G_{\max}$ may generally depend on the data.

%Of course, our lasso approach is only one among countless possible ways to define sample partitions satisfying Definition~\ref{def:knockoff-invariant-function}.
%Alternatively, one could take advantage of more sophisticated sample partitioning techniques, which may for example be based on Bayesian models \citep{alaa2017bayesian}, random forests \citep{wager2018estimation}, or neural networks \citep{yao2018representation}.

\subsection{Implementing aLKF for GWAS Data} \label{subsec:adaptation-gwas}

One of the motivations behind the methods developed in this paper is to study how genetic variants influencing a phenotype may vary across environments, often referred to as ``gene-environment interactions''. Consequently, it is important to describe a version of the aLKF that can be applied effectively and without excessive computational costs to the analysis of genetic data. Key characteristics to consider are that the number of potential explanatory variables, $p$, typically ranges from hundreds of thousands to millions, while the number of measured covariates, $m$, is much smaller, often on the order of tens.

These characteristics make it impractical to implement our method as described in the previous sections, creating a need for a specialized implementation.
This implementation departs from the approach described so far in three main aspects.

\begin{itemize}
    \item \textbf{Pre-screening variables.} A lasso model fitted to the cloaked data identifies a subset of promising variables ($p_{\text{screen}} \ll p$) by selecting variables with non-zero lasso coefficients for either version of the cloaked original and knockoff SNPs.
    
    \item \textbf{Learning local environments.} A lasso model with interaction terms is applied to the pre-screened data to define data-driven partition functions. To prioritize global associations and avoid unnecessary sample splits, the penalty for interaction terms is scaled by a hyperparameter $c \in (0,1]$ (e.g., $c = 0.25$).
    
    \item \textbf{Batch computation of importance scores.} Variables with identical partition functions are grouped into batches. A single predictive model is fitted per batch, reducing the number of models needed, especially when $m \ll p_{\text{screen}}$.
\end{itemize}

Algorithms~\ref{alg:sskf_second_phase_genetic} and~\ref{alg:sskf_genetic} summarize this specialized implementation of our method, with additional details provided in Appendices~\ref{subsec:adaptation-gwas-1}--\ref{subsec:adaptation-gwas-3}.
Furthermore, Appendix~\ref{subsec:adaptation-gwas-LD} describes how our method can be applied after partitioning the genome into groups of SNPs.
This accounts for strong linkage disequilibrium (LD) among nearby variants on the same chromosome, enabling a {\em multi-resolution} analysis \citep{sesia2020multi,sesia2021false}.

\begin{algorithm}[!htb]
\caption{Local Knockoff Filter for GWAS data, with fixed hypotheses}
\label{alg:sskf_second_phase_genetic}
\KwIn{Pre-screened variables $\mathbf{X}$ and knockoffs $\tilde{\mathbf{X}}$, outcomes $\mathbf{Y}$, covariates $\mathbf{Z}$; \newline
  Random noise matrix $\mathbf{V}$; Partition function $\nu = (\nu_1, \ldots, \nu_{p_\text{screen}})$; \newline
  Nominal FDR level $\alpha \in (0,1)$\;
}

Compute the partition width $L = \sum_{j=1}^{p_{\text{screen}}} L_j$, where $L_j = \max_{z \in \mathcal{Z}} \nu_j(z)$\;
Define local hypotheses $\mathcal{H}_{j,\ell}(\nu)$ as in~\eqref{eq:null-hyp-loc}, for $j \in [p_{\text{screen}}]$ and $\ell \in [L_j]$\;
Partition all $j \in [p_{\text{screen}}]$ into batches $\mathcal{Q}$ such that  $\hat{\nu}_j(\mathbf{Z}) = \hat{\nu}_k(\mathbf{Z})$ $\forall$ $j, k \in Q$\;
Assemble the cloaked data matrix $\tilde{\mathcal{D}}(\mathbf{V}) = ([\mathbf{X},\tilde{\mathbf{X}}]_{\mathrm{swap}(\mathbf{V})}, \mathbf{Y}, \mathbf{Z})$\;
\For {$Q \in \mathcal{Q}$} {
  \For {$\ell \in [L_Q]$} {
    Fit a lasso model of $\mathbf{Y}^{(Q,\ell)}$ on $[[\mathbf{X}_j, \tilde{\mathbf{X}}_j]_{j \in Q}, [\mathbf{X}, \tilde{\mathbf{X}}]_{\mathrm{swap}(\mathbf{V},-Q)}, \mathbf{Z}]^{(Q,\ell)}$\;
    \For {$j \in Q$} {
      Define $T_{j,\ell} = |\hat{\beta}_{j, \ell}|$ and $\tilde{T}_{j,\ell} = |\hat{\beta}_{j + |Q|, \ell}|$, where $\hat{\beta}$ are the lasso coefficients\;
      Compute the test statistic $W_{j,\ell} = T_{j,\ell} - \tilde{T}_{j,\ell}$ for all $j \in \mathcal{Q}$\;
    }
  }
}
Vectorize the test statistics $W_{j, \ell}$ for all $j \in [p_{\text{screen}}], \ell \in [L_j]$ into $\mathbf{W}$\;
Apply Selective SeqStep+ \citep{barber2015controlling} at level $\alpha$ to the vector $\mathbf{W}$\;

\KwOut{List of rejected hypotheses $\hat{\mathcal{R}} \subseteq \{\mathcal{H}_{j,\ell}(\nu)\}_{j \in [p_{\text{screen}}], \ell \in [L_j]}$.}
\end{algorithm}

\begin{algorithm}[!htb]
\caption{Adaptive Local Knockoff Filter for GWAS data}
\label{alg:sskf_genetic}
\KwIn{Knockoff-augmented data set $\mathcal{D} = ([\mathbf{X},\tilde{\mathbf{X}}], \mathbf{Y}, \mathbf{Z})$; FDR level $\alpha \in (0,1)$\;}
Randomly generate $\mathbf{V} \in \{0,1\}^{n \times p}$, a matrix of i.i.d.~Bernoulli noise\;
Assemble the cloaked data matrix $\tilde{\mathcal{D}}(\mathbf{V}) = ([\mathbf{X},\tilde{\mathbf{X}}]_{\mathrm{swap}(\mathbf{V})}, \mathbf{Y}, \mathbf{Z})$\;
Obtain pre-screened variants $[\mathbf{X}, \mathbf{\widetilde{X}}]_{\text{(swap(\textbf{V}), screen)}}$ by fitting a lasso model for the outcome $\mathbf{Y}$ given $([\mathbf{X}, \tilde{\mathbf{X}}]_{\mathrm{swap}(\mathbf{V})}, \mathbf{Z})$; see Appendix~\ref{subsec:adaptation-gwas-1}\;
Learn a partition function $\hat{\nu}$, with width $ \hat{L} = \sum_{j\in [p_{\text{screen}}]} \hat{L}_j $, by fitting a lasso model of $\mathbf{Y}$ on $[[\mathbf{X}, \mathbf{\widetilde{X}}]_{\text{(swap(\textbf{V}), screen)}}, \mathbf{Z}]$, including interactions; see Appendix~\ref{subsec:adaptation-gwas-2}\;
Apply Algorithm~\ref{alg:sskf_second_phase_genetic} using the partition function $\hat{\nu}$\;
\KwOut{List of rejected hypotheses $\hat{\mathcal{R}} \subseteq \{\mathcal{H}_{j,\ell}\}_{j \in [p_{\text{screen}}], \ell \in [\hat{L}_j]}$.}
\end{algorithm}

\subsubsection{Pre-Screening of Promising Variables Using Cloaked Data} \label{subsec:adaptation-gwas-1}

To reduce the number of local hypotheses that need to be considered, while maintaining control over the FDR for the selected hypotheses, we begin by identifying a smaller set of promising explanatory variables, $p_{\text{screen}} < p$ (e.g., with $p_{\text{screen}}$ on the order of tens of thousands), through a pre-screening step based on the cloaked data in $\tilde{\mathcal{D}}(\mathbf{V})$. The goal of this step is to reduce the dimensionality of the problem while retaining variables that are likely to be relevant.

To pre-screen relevant variables, we fit a lasso model of $\mathbf{Y}$ on $[[\mathbf{X}, \widetilde{\mathbf{X}}]_{\text{swap}(\mathbf{V})}, \mathbf{Z}]$, tuning the regularization parameter via cross validation. Note that computationally efficient software packages \citep{prive2018efficient, prive2019efficient} make this task feasible even for large-scale GWAS data \citep{sesia2020multi, sesia2021false}, as this model only needs to be fitted once and does not include interaction terms.

Let $\hat{\beta}_j$ and $\hat{\beta}_{j+p}$ indicate the estimated lasso coefficients for the (scaled) variables in the $j$-th and $(j+p)$-th columns of $[\mathbf{X}, \widetilde{\mathbf{X}}]_{\text{swap}(\mathbf{V})}$, respectively.
Since $X_j$ and $\widetilde{X}_j$ are cloaked, in this model there is no meaningful distinction between the original variables and the knockoffs.
Therefore, the pre-screening selects both columns containing the randomly swapped values of $X_j$ and $\widetilde{X}_{j + p}$, for all $j \in [p]$ such that $|\hat{\beta}_j| + |\hat{\beta}_{j+p}| > 0$.

To simplify the notation in the following, we will assume, by convention and without loss of generality, that the indices $j \in [p]$ corresponding to the $p_{\text{screen}}$ original variables that passed the screening step are given by $[p_{\text{screen}}] = \{1, \ldots, p_{\text{screen}}\}$. Furthermore, we will denote by $[\mathbf{X}, \widetilde{\mathbf{X}}]_{\text{swap}(\mathbf{V}), \text{screen}}$ the data matrix containing the $p_{\text{screen}}$ (cloaked) pairs of variables and knockoffs that passed the screening step.

\subsubsection{Learning Local Environments Using Pre-Screened Data} \label{subsec:adaptation-gwas-2}
 
\paragraph{Fitting a lasso model with interaction terms.}
The second step of our method focuses on learning a potentially interesting partition function $\hat{\nu}$ by fitting a lasso model with interaction terms, similar to the approach described in Appendix~\ref{sec:learn-interactions}.
Again, for simplicity we are assuming here that the covariates are binary, as in Appendix~\ref{sec:learn-interactions}.
The key difference is that, instead of using the full cloaked dataset $\tilde{\mathcal{D}}(\mathbf{V}) = (\smash{[\mathbf{X}, \tilde{\mathbf{X}}]_{\mathrm{swap}(\mathbf{V})}, \mathbf{Y}, \mathbf{Z}})$, now we use its smaller pre-screened version, $\tilde{\mathcal{D}}_{\text{screen}}(\mathbf{V}) = (\smash{[\mathbf{X}, \tilde{\mathbf{X}}]_{\mathrm{swap}(\mathbf{V}), \text{screen}}, \mathbf{Y}, \mathbf{Z}})$.

In particular, we fit a lasso model to predict $\mathbf{Y}$ given $[[\mathbf{X}, \mathbf{\widetilde{X}}]_{\text{(swap(\textbf{V}), screen)}},\mathbf{Z}]$, after augmenting the data matrix with all possible pairwise interactions between $\mathbf{Z}$ and $[\mathbf{X}, \mathbf{\widetilde{X}}]_{\text{(swap(\textbf{V}), screen)}}$, and tuning the regularization parameter via cross validation.
Again, let $\hat{\beta}_j$ and $\hat{\beta}_{j+p}$ indicate the estimated main effects for the $j$-th and $(p+j)$-th variables in $[\mathbf{X}, \mathbf{\widetilde{X}}]_{\text{(swap(\textbf{V}), screen)}}$, for any $j \in [p_{\text{screen}}]$.
 Further, let $\hat{\gamma}_{l, j}$ and $\hat{\gamma}_{l, j + p}$ indicate the corresponding interaction coefficients involving $Z_l$, for all $l \in [m]$ and $j \in [p_{\text{screen}}]$.

Finally, we define the data-driven partition function $\hat{\nu}$ with the same approach described in Appendix~\ref{sec:learn-interactions}.
In particular, the interaction effect for $Z_l$ with $l \in [m]$ is selected whenever $|\hat{\gamma}_{l, j}| +  |\hat{\gamma}_{l, j + p}| > 0$ and the main effect if $|\hat{\gamma}_{l, j}| +  |\hat{\gamma}_{l, j + p}| = 0$ but $|\hat{\beta}_j| + |\hat{\beta}_{j+p}| > 0$. As in Appendix~\ref{sec:learn-interactions}, we consider only the $G_{\max}$ covariates with the strongest (nonzero) interactions involving $X_j$ or $\tilde{X}_j$, where $G_{\max}$ is a fixed hyper-parameter.

\paragraph{Encouraging the selection of coarser partitions.}
Large-scale GWAS data often exhibit two additional characteristics that must be considered when applying the aLKF method in order to maximize power.
First, in the analysis of polygenic traits, there are typically many discoveries to be made. The global knockoff filter of \citet{candes2018} already demonstrates high power, often identifying hundreds or thousands of significant associations between genetic variants and the phenotype \citep{sesia2020multi, sesia2021false}. 
Second, not all genetic variants are likely to have subgroup-specific effects. For many significant SNPs, it may be sufficient to detect associations at the population level. Only a relatively small fraction of global associations are expected to correspond to effects specific to particular subgroups. In other words, not all significant genetic effects are expected to involve interactions with covariates.

These considerations highlight the need to ensure that the aLKF method does not unnecessarily reduce power to detect global associations relative to existing methods. A key concern is that if the lasso model with interactions described above may select too many spurious interaction terms, the data-driven partition function $\hat{\nu}$ may unnecessarily split the sample into overly small subgroups. This would reduce the sample size available for computing local importance scores and potentially diminish power.

To address this issue, the lasso model with interactions must be calibrated more conservatively by increasing the practical threshold for selecting interaction terms, thus creating an incentive to select coarser sample partitions. 
In practice, this is implemented by adjusting the lasso penalty applied to main effects. Specifically, the penalty for main effects in the interaction model described above is multiplied by a factor $c \in (0, 1]$, with smaller values of $c$ effectively increasing the cost of selecting interaction terms.
While the choice of $c$ involves some arbitrariness, our numerical experiments have shown that $c = 0.25$—which emphasizes global effects—works well in practice across a variety of scenarios, involving synthetic data with different proportions of true local vs.~global effects.

\subsubsection{Batch Computation of Local Importance Scores} \label{subsec:adaptation-gwas-3}

Recall that typical GWAS datasets are characterized by a very large number, $p$, of explanatory variables and a much smaller number, $m$, of relevant covariates. Even after pre-screening, we generally expect $p_{\text{screen}}$ to remain much larger than $m$. This imbalance creates a natural opportunity to streamline the computation of the importance scores.

Rather than re-fitting a local predictive model for each variable $j \in [p_{\text{screen}}]$ and subgroup $\ell \in \hat{L}_j$, we can perform ``batch computation'':
the key intuition is that it is possible to utilize the same local predictive models for different variables assigned to identical partition functions.
When $m$ is much smaller than $p_{\text{screen}}$, this shortcut significantly reduces the number of local models that need to be fitted compared to Algorithm~\ref{alg:sskf_local_scores}.

To illustrate this process, consider a simplified setting with $p_{\text{screen}} = 4$ explanatory variables, post-screening, and a single binary covariate, $Z_1 \in \{0,1\}$. Suppose the detected local environments are as described in Table~\ref{table:interactions-example-bulk-lasso} below.
Based on the subgroups identified in Table~\ref{table:interactions-example-bulk-lasso}, Algorithm~\ref{alg:sskf_second_phase} would require fitting six distinct predictive models. In contrast, the batch processing shortcut described below reduces this to just three local models: one for the entire sample, one for the subgroup defined by $Z_1 = 0$ and one for the subgroup defined by $Z_1 = 1$.

\begin{table}[!htb]
  \caption{Partition functions learnt by fitting a lasso model with interactions in an imaginary toy example with 4 variables and a single binary covariate.} \vspace{0.5em}
  \label{table:interactions-example-bulk-lasso}
    \centering
    \begin{tabular}{ccccc}
        \toprule
      \multirow{2}{*}{Variable} & \multirow{2}{*}{Covariates} & \multicolumn{2}{c}{Partition}\\
      \cmidrule(l{2pt}r{2pt}){3-4}
       &  & Label & Definition &  \\
      \midrule
        $X_1$ & $\emptyset$ & 1 & All individuals   \\[0.25em]
      \hline
     \multirow{2}{*}{$X_2$} & \multirow{2}{*}{\{$Z_1$\}} & 1 & $Z_1 = 0$ \\
       &  & 2 & $Z_1 = 1$  \\[0.25em]
      \hline
      \multirow{2}{*}{$X_3$} & \multirow{2}{*}{\{$Z_1$\}} & 1 & $Z_1 = 0$  \\
       &  & 2 & $Z_1 = 1$ \\[0.25em]
      \hline
      $X_4$ & $\emptyset$ & 1 & All individuals \\[0.25em]
        \bottomrule
    \end{tabular}
\end{table}

Our ``batch computation'' shortcut can be generally described as follows.

\paragraph{Defining the batches.}
First, we partition $[p_{\text{screen}}]$ into disjoint ``batches'' $\mathcal{Q}$, such that $[p_{\text{screen}}] = \cup_{Q \in \mathcal{Q}} Q$ and all variables within each batch have the same partition function: $\hat{\nu}_j(\mathbf{Z}) = \hat{\nu}_k(\mathbf{Z})$ for all $j, k \in Q$, for all $Q \in \mathcal{Q}$.
Let this joint partition be denoted by $\hat{\nu}_Q$, with $\hat{L}_Q(\nu)$ representing the number of disjoint batchs induced by $\nu_Q$, and let the associated covariates be denoted by $\mathbf{Z}_Q \subseteq \mathbf{Z}$.

In the toy example of Table~\ref{table:interactions-example-bulk-lasso}, the partition function shared by $X_2$ and $X_3$ divides the sample into two subgroups corresponding to $Z_1 = 0$ and $Z_1 = 1$, while the partition function for $X_1$ and $X_4$ does not partition the sample. Based on this, we would assign $X_1$ and $X_4$ to the first batch and $X_2$ and $X_3$ to the second batch.

\paragraph{Computing importance scores via batch-local predictive models.}
Separately for each batch $Q \in \mathcal{Q}$ and subgroup $\ell \in \hat{L}_Q$, we fit a ``batch-local'' predictive model, denoted as $\hat{\psi}^{(Q,\ell)}$, for $Y$ given $(X, \tilde{X}, Z)$ as follows.
This model can be trained using information from the cloaked dataset $\tilde{\mathcal{D}}(\mathbf{V})$ and a smaller {\em batch-local} dataset $\mathcal{D}^{(Q,\ell)} = ([\mathbf{X}_j, \tilde{\mathbf{X}}_j]_{j \in Q}, \mathbf{Y}, \mathbf{Z})^{(Q,\ell)}$, which includes only observations $(X_j, \tilde{X}_j, Y, Z)$ for all $j \in Q$ and all individuals $i \in [n]$ with $\nu_j(Z^i) = \ell$.
Concretely, we implement $\hat{\psi}^{(Q,\ell)}$ using the lasso, trained with $\mathbf{Y}^{(Q,\ell)}$ as the outcome and $[[\mathbf{X}_j, \tilde{\mathbf{X}}_j]_{j \in Q}, [\mathbf{X}, \tilde{\mathbf{X}}]_{\mathrm{swap}(\mathbf{V},-Q)}, \mathbf{Z}]^{(Q,\ell)}$ as the predictor matrix, where $[\mathbf{X}, \tilde{\mathbf{X}}]_{\mathrm{swap}(\mathbf{V},-Q)}$ denotes the matrix $[\mathbf{X}, \tilde{\mathbf{X}}]_{\mathrm{swap}(\mathbf{V})}$ after removing the columns indexed by $j$ and $p+j$ for all $j \in Q$.
The sparsity parameter of this lasso model is tuned as usual, via cross-validation.

Then, to obtain local importance scores satisfying~\eqref{eq:swap-tau}, it is sufficient to compute $(T_{j,\ell}, \tilde{T}_{j,\ell})$, for each $j \in Q$, in such a way that swapping $X_j$ and $\tilde{X}_j$ in $\mathcal{D}^{(Q,\ell)}$ would result in $T_{j,\ell}$ being swapped with $\tilde{T}_{j,\ell}$; this is analogous to Proposition~\ref{prop:sskf_local_scores-simple} in Appendix~\ref{app:proofs}.
Such equivariance is easily achieved by setting $T_{j,\ell} = |\hat{\beta}_{j, \ell}|$ and $\tilde{T}_{j,\ell} = |\hat{\beta}_{j + |Q|, \ell}|$, for each $j \in Q$, where $\hat{\beta}_{j, \ell}$ and $\hat{\beta}_{j + |Q|, \ell}$ respectively denote the (scaled) estimated regression coefficients of variable $j$ and its knockoff within the batch-local lasso model $\hat{\psi}^{(Q,\ell)}$.

Note that, unlike the procedure in Appendix~\ref{sec:powerful-statistics}, when computing local scores within a batch $(Q, \ell)$, this approach does not utilize any additional (non-local) information from the cloaked dataset $\tilde{\mathcal{D}}(\mathbf{V})$ beyond what is captured by $[[\mathbf{X}_j, \tilde{\mathbf{X}}_j]_{j \in Q}, [\mathbf{X}, \tilde{\mathbf{X}}]_{\mathrm{swap}(\mathbf{V},-Q)}, \mathbf{Z}]^{(Q,\ell)}$.
Specifically, the ``prior'' regularization weights described in Appendix~\ref{sec:powerful-statistics} are not used, effectively setting the hyper-parameter $\smash{\xi^{(Q,\ell)} = 0}$.
The motivation for this choice is two-fold. First, since the number of covariates $m$ is expected to be very small compared to both the number $p$ of variables and the sample size $n$, the effective sample size of $[[\mathbf{X}_j, \tilde{\mathbf{X}}_j]_{j \in Q}, [\mathbf{X}, \tilde{\mathbf{X}}]_{\mathrm{swap}(\mathbf{V},-Q)}, \mathbf{Z}]^{(Q,\ell)}$ is already relatively large. Second, having fewer hyper-parameters to tune via cross-validation is helpful to further reduce computational costs.

Finally, after evaluating all local importance scores, we proceed as described in Section~\ref*{sec:knockoffs-testing}, with test statistics $W_{j, \ell} = T_{j,\ell} - \tilde{T}_{j,\ell}$ for each $Q \in \mathcal{Q}$, $j \in Q$ and $\ell \in \hat{L}_Q$.

\subsubsection{Accounting for Strong Linkage Disequilibrium} \label{subsec:adaptation-gwas-LD}

Due to the presence of strong local dependencies among genetic variants on the same chromosome, a phenomenon known as {\em linkage disequilibrium}, it is often intrinsically very challenging to precisely identify individual variants that influence a complex trait. Instead, significant genetic associations are typically detected at the level of broader genetic loci, which include several SNPs in strong linkage disequilibrium.

To address this complication in GWAS data, \citet{sesia2020multi} and \citet{sesia2021false} applied the global knockoff filter of \citet{candes2018} at different levels of resolution. Specifically, they partitioned the genome into contiguous groups of variants in strong linkage disequilibrium and tested for conditional independence with the phenotype at the level of groups of SNPs. These groups of SNPs are then used as the units of inference rather than individual SNPs. 

Specifically, let $\mathcal{B}$ be a partition of the genome into contiguous groups of SNPs $b \in \mathcal{B}$. The equivalent environment-level hypothesis at the group level, see Equation~(\ref{eq:null-hyp-loc-env}), is
\begin{align*} 
  \bar{\mathcal{H}}_{b,\ell}(\nu) : Y^{(b,\ell)} \indep X^{(b,\ell)}_b \mid X^{(b,\ell)}_{-b}, Z^{(b,\ell)}\;\;\;\; b \in \mathcal{B}, \ell \in [L_b].
\end{align*}
This hypothesis is analogous to the one in Equation~(\ref{eq:null-hyp-loc-env}), except that inference is at the level of groups of SNPS, rather than single SNPs.

Similarly, the exchangeability is required to hold for groups of SNPs, see also Equation~(\ref{eq:knock_cond_1}):
\begin{align*} 
  \big[ \mathbf{X}, \tilde{\mathbf{X}} \big]_{\mathrm{swap}(b)} \mid \mathbf{Z}
  \; \overset{d}{=} \; \big[ \mathbf{X}, \tilde{\mathbf{X}} \big] \mid \mathbf{Z}, \qquad \forall b \in \mathcal{B}.
\end{align*}

\citet{sesia2021false} construct valid knockoffs for groups of SNPs using hidden Markov models, see \citet{sesia2021false} for further details. They construct groups of SNPs using complete-linkage hierarchical clustering. We can seamlessly integrate the genetic group level strategy with our aLKF method. To test the hypothesis described above, starting with a pre-defined partition $\mathcal{B}$ of the genome into contiguous groups of SNPs $b \in \mathcal{B}$ at the desired resolution, along with corresponding knockoff genotypes, as described in \citet{sesia2020multi} and \citet{sesia2021false}, we can apply our method as outlined so far, with four key modifications:

\begin{enumerate}
\item The data cloaking step using the random matrix $\mathbf{V}$ is performed at the level of groups of SNPs, assigning the same values of $V_{ij}$ to all SNPs $j$ within a group of SNPs $b \in \mathcal{B}$. This reflects the idea that the units of inference are now groups of SNPs and not single SNPs.

\item Each group of SNPs $b$ is assigned a single partition function $\hat{\nu}_b$, rather than having distinct partition functions for individual SNPs.

\item While pre-screening nominally selects a subset of individual variants, since inference is then performed at the level of groups of SNPs, the batch-local predictive models are fitted using not only the pre-screened variants but also all other variants in the same genetic group.

\item For each combination of $b$ and $\ell \in \hat{L}_b$, a single pair of local scores $(T_{b,\ell}, \tilde{T}_{b,\ell})$ is computed by summing the single-SNP scores corresponding to the individual SNPs within that group.
Then, the corresponding test statistic is $W_{b,\ell} = T_{b,\ell} - \tilde{T}_{b,\ell}$.

\end{enumerate}

Finally, the discoveries obtained with our method will be interpreted as highlighting groups of SNPs displaying significant associations with the outcome within certain subgroups determined by specific covariate values.

The choice of SNP group size reflects a well-known trade-off between resolution and power: smaller groups enable more precise localization of associations but tend to reduce power due to strong correlations among nearby variants \citep{sesia2020multi, sesia2021false, gablenz2025catch}. \citet{sesia2021false} introduce a hierarchy of LD-based groupings to support multi-resolution analyses, and \citet{gablenz2025catch} propose a method to aggregate results across resolutions while controlling the FDR. To maintain focus and clarity, we adopt a single resolution in this paper, with a median SNP group width of 3 kilobases (kb)—a scale that captures strong linkage disequilibrium while allowing variation in signal strength. Importantly, the group size does not affect the theoretical validity of our method. 

\clearpage

\section{Mathematical Proofs} \label{app:proofs}

\begin{proposition} \label{prop:sskf_local_scores-simple}
Consider a vector of local importance scores ${[\mathbf{T},\tilde{\mathbf{T}}]}$ which can be written as the output of a suitable (randomized) function $\boldsymbol{\tau}$ applied to $\mathcal{D}$ and $\tilde{\mathcal{D}}(\mathbf{V})$; that is, $[\mathbf{T},\tilde{\mathbf{T}}] = \boldsymbol{\tau} \big( \mathcal{D}, \tilde{\mathcal{D}}(\mathbf{V}) \big) = \big[ \bm{t} \big( \mathcal{D}, \tilde{\mathcal{D}}(\mathbf{V}) \big), \tilde{\bm{t}} \big( \mathcal{D}, \tilde{\mathcal{D}}(\mathbf{V}) \big) \big]$, using the same notation as in~\eqref{eq:def-tau}.
Suppose that, for any pair $(j,\ell)$, the scores $(T_{j,\ell},\smash{\tilde{T}_{j,\ell}})$ are a function of a local model $\hat{\psi}^{(j,\ell)}$, trained looking only at $\tilde{\mathcal{D}}(\mathbf{V})$ and $\mathcal{D}^{(j,\ell)} = (\mathbf{X}_j, \tilde{\mathbf{X}}_j, \mathbf{Y}, \mathbf{Z})^{(j,\ell)}$, such that swapping $X_j$ and $\tilde{X}_j$ in $\mathcal{D}^{(j,\ell)}$ results in $T_{j,\ell}$ being swapped with $\tilde{T}_{j,\ell}$; i.e., we refer to this as the ``local symmetry'' property:
\begin{align} \label{eq:local-symmetry}
  T_{j,\ell}\left( (\mathbf{X}_j, \tilde{\mathbf{X}}_j, \mathbf{Y}, \mathbf{Z})^{(j,\ell)}, \tilde{\mathcal{D}}(\mathbf{V}) \right) 
  & = \tilde{T}_{j,\ell}\left( (\tilde{\mathbf{X}}_j, \mathbf{X}_j, \mathbf{Y}, \mathbf{Z})^{(j,\ell)}, \tilde{\mathcal{D}}(\mathbf{V}) \right).
\end{align}
Notably,~\eqref{eq:local-symmetry} is equivalent to the symmetry required, at the whole-sample level, for the importance scores utilized by \citet{candes2018}.

Then, $[\mathbf{T},\tilde{\mathbf{T}}]$ must satisfy the equivariance property defined in~\eqref{eq:swap-tau}; that is, for any $\mathcal{S} \subseteq [\smash{L}]$, whose elements uniquely identify pairs $(j,\ell)$ for $j \in [p]$ and $\smash{ \ell \in[{L}_j]}$,
\begin{align*}
  \boldsymbol{\tau}\big( \mathcal{D}_{\mathrm{swap}(\mathcal{S})}, \tilde{\mathcal{D}}(\mathbf{V}) \big)
  =
  \big[
  \bm{t}\big( \mathcal{D}, \tilde{\mathcal{D}}(\mathbf{V}) \big),
  \tilde{\bm{t}}\big( \mathcal{D}, \tilde{\mathcal{D}}(\mathbf{V}) \big) \big]_{\mathrm{swap}(\mathcal{S})}.
\end{align*}
\end{proposition}

\begin{proof}[Proof of Proposition~\ref{prop:sskf_local_scores-simple}]
This proof is quite straightforward.
Consider how ${[\mathbf{T},\tilde{\mathbf{T}}]}$ would change if the input data set $\mathcal{D}$ were replaced by $\mathcal{D}_{\mathrm{swap}(\mathcal{S})}$, while keeping the cloaked data set $\tilde{\mathcal{D}}(\mathbf{V})$ unchanged.

\begin{itemize}

\item For any $(j,\ell) \notin \mathcal{S}$, the local model $\hat{\psi}^{(j,\ell)}$ is invariant because both $\tilde{\mathcal{D}}(\mathbf{V})$ and $\mathcal{D}^{(j,\ell)}$ are invariant, as explained above. Therefore, in this case $\smash{T_{j,\ell}}$ and $\smash{\tilde{T}_{j,\ell}}$ are invariant.

\item For any $(j,\ell) \in \mathcal{S}$, the local model $\hat{\psi}^{(j,\ell)}$ remains invariant up to swapping the importance score of $X_j$ with that of $\tilde{X}_j$, because $\tilde{\mathcal{D}}(\mathbf{V})$ is invariant while $X_j$ and $\tilde{X}_j$ are swapped in $\mathcal{D}^{(j,\ell)}$, by~\eqref{eq:local-symmetry}. Therefore, in this case $\smash{T_{j,\ell}}$ and $\smash{\tilde{T}_{j,\ell}}$ are swapped.
\end{itemize}

This concludes the proof that 
\begin{align*}
  \boldsymbol{\tau}\big( \mathcal{D}_{\mathrm{swap}(\mathcal{S})}, \tilde{\mathcal{D}}(\mathbf{V}) \big)
  =
  \big[
  \bm{t}\big( \mathcal{D}, \tilde{\mathcal{D}}(\mathbf{V}) \big),
  \tilde{\bm{t}}\big( \mathcal{D}, \tilde{\mathcal{D}}(\mathbf{V}) \big) \big]_{\mathrm{swap}(\mathcal{S})}.
\end{align*}

\end{proof}

\begin{proof}[Proof of Theorem~\ref*{thm:coin-flip-fixed}]

Let $\smash{\bar{\mathbf{U}} \in \{\pm 1\}^{L}}$ be a random vector with independent entries such that: $\smash{\bar{U}_{j,\ell}= \pm 1}$ with probability $1/2$ if $\smash{\mathcal{H}_{j,\ell}(\nu)}$ in~\eqref{eq:null-hyp-loc} is true and ${\bar{U}_{j,\ell} = +1}$ otherwise, for all $j \in [p]$ and ${\ell \in[L_j]}$. 
To prove Theorem~\ref*{thm:coin-flip-fixed}, it suffices to prove the {\em flip-sign} property
\begin{align} \label{eq:coin-flip-fixed}
\mathbf{W}  \; \oset{d}{=}\; \mathbf{W} \odot \bar{\mathbf{U}} ,
\end{align}
since from there the result follows immediately as in~\cite{candes2018}.

To prove~\eqref{eq:coin-flip-fixed}, we begin by recalling some useful notation.
For each $j \in [p]$, $ {L}_j(\nu) = \max_{z \in \mathcal{Z}} \nu_j(z) \in \mathbb{N} $ represents the number of disjoint subgroups induced by $ \nu_j $.
Define $\mathcal{S}(\bar{\mathbf{U}}) := \{(j,\ell) : j \in [p], \ell \in [L_j], \bar{U}_{j,\ell} = -1 \} \subseteq [L]$, where  $ L = \sum_{j=1}^{p} {L}_j $.
Recall also that $\smash{[\mathbf{X}, \tilde{\mathbf{X}}]_{\mathrm{swap}(\mathcal{S}(\bar{\mathbf{U}}))}}$ is defined as the matrix obtained from $\smash{[\mathbf{X}, \tilde{\mathbf{X}}]}$ after swapping the sub-column $\smash{\mathbf{X}^{(\ell)}_j}$, which contains all observations of $X_j$ in the subgroup $g_{j,\ell}$, with the corresponding knockoffs, for all $(j,\ell) \in \mathcal{S}(\bar{\mathbf{U}})$.

To emphasize that $\mathbf{W}$ is a function of $\mathcal{D}$ and $\tilde{\mathcal{D}}(\mathbf{V})$, we write
 $\mathbf{W} = \mathbf{W}(\mathcal{D},\tilde{\mathcal{D}}(\mathbf{V}))$. 
Specifically, recall that $\mathbf{W}$ is obtained by applying an anti-symmetric combination function to $[\mathbf{T},\tilde{\mathbf{T}}]$ in~\eqref{eq:def-tau}.
Further, Proposition~\ref{prop:sskf_local_scores-simple} implies the importance scores ${[\mathbf{T},\tilde{\mathbf{T}}]}$ computed by Algorithm~\ref{alg:sskf_second_phase} satisfy~\eqref{eq:swap-tau}.
Therefore, it follows from \eqref{eq:swap-tau} that
\begin{align*}
  \mathbf{W}(\mathcal{D}_{\mathrm{swap}(\mathcal{S}(\bar{\mathbf{U}}))},\tilde{\mathcal{D}}(\mathbf{V}))
  =
  \bar{\mathbf{U}} \odot \mathbf{W}(\mathcal{D},\tilde{\mathcal{D}}(\mathbf{V}))
\end{align*}
almost surely with respect to $\bar{\mathbf{U}},\mathbf{V}$, and $\mathcal{D}$, where $\mathbf{W}(\mathcal{D}_{\mathrm{swap}(\mathcal{S}(\bar{\mathbf{U}}))},\tilde{\mathcal{D}}(\mathbf{V}))$ represents the imaginary vector of test statistics which would be obtained by applying our method to the modified data set $\mathcal{D}_{\mathrm{swap}(\mathcal{S}(\bar{\mathbf{U}}))} = ([\mathbf{X}, \tilde{\mathbf{X}}]_{\mathrm{swap}(\mathcal{S}(\bar{\mathbf{U}}))},\mathbf{Y},\mathbf{Z})$ instead of $\mathcal{D} = (\smash{[\mathbf{X}, \tilde{\mathbf{X}}],\mathbf{Y},\mathbf{Z}})$.

Finally, the proof of~\eqref{eq:coin-flip-fixed} is completed by showing that it follows from Lemma~\ref{eq:lemma-1} that
\begin{align} \label{eq:swap-d}
  ([\mathbf{X}, \tilde{\mathbf{X}}],\mathbf{Y},\mathbf{Z}, \tilde{\mathcal{D}}(\mathbf{V}))
  \; \oset{d}{=}\;
  ([\mathbf{X}, \tilde{\mathbf{X}}]_{\mathrm{swap}(S(\bar{\mathbf{U}}))},\mathbf{Y},\mathbf{Z}, \tilde{\mathcal{D}}(\mathbf{V})),
\end{align}
for any subset $S \subseteq [L]$ of true null hypotheses $\mathcal{H}_{j,\ell}(\nu)$ in~\eqref{eq:null-hyp-loc},
which in turn implies 
$$ 
\mathbf{W}\left( \mathcal{D},\tilde{\mathcal{D}}(\mathbf{V}) \right)
\;\oset{d}{=}\;
\mathbf{W}\left( \mathcal{D}_{\mathrm{swap}(\mathcal{S}(\bar{\mathbf{U}}))},\tilde{\mathcal{D}}(\mathbf{V}) \right).
$$
To see why Lemma~\ref{eq:lemma-1} implies~\eqref{eq:swap-d}, note that we can define a matrix $\mathbf{U} \in \{\pm 1\}^{n \times p}$ such that, for each $i \in [n]$ and $j \in [p]$, $U_{i,j} = \bar{U}_{j,l}$ where $\ell = \nu_{j}(Z^i)$.
In particular, this means that $[\mathbf{X}, \tilde{\mathbf{X}}]_{\mathrm{swap}(S(\bar{\mathbf{U}}))} = [\mathbf{X}, \tilde{\mathbf{X}}]_{\mathrm{swap}(\mathbf{U})}$.
Further, it follows from the definition of $\bar{\mathbf{U}}$ and $\mathcal{H}_{j,\ell}(\nu)$ in~\eqref{eq:null-hyp-loc} that almost surely $U_{i,j}= 1$ if $\mathcal{H}^i_{j}$ in~\eqref{eq:null-hyp-ind} is {\em not true}.
Therefore, we can directly apply Lemma~\ref{eq:lemma-1} with this choice of $\mathbf{U}$ and obtain~\eqref{eq:swap-d}.

% Since different individuals are assumed to be independent of one another, it suffices to prove that a stronger version of~\eqref{eq:swap-d} holds for a single row of the data.
% More precisely, it is enough to prove that, conditional on $Z$,
% \begin{align} \label{eq:swap-d-single}
%   \left( (X,\tilde{X})_{\mathrm{swap}(s)}, Y\right) \mid Z
%   \; \oset{d}{=}\;
%   \left( (X,\tilde{X}), Y \right) \mid Z,
% \end{align}
% where $s \subseteq [p]$ is any subset of variables for which the null hypotheses $\mathcal{H}_{j,\ell}(\nu)$, for $ \ell = \nu_j(Z)$, is true.
% With this goal in mind, recall that $Y \indep \tilde{X} \mid X, Z$, by design of the knockoffs. 
% Further, $Y^{(\ell)} \indep X^{(\ell)}_j \mid X^{(\ell)}_{-j}, Z^{(\ell)}$ for all $j \in s$.
%Therefore,~\eqref{eq:swap-d-single} follows from exactly the same argument as in the proof of Lemma 3.2 in~\cite{candes2018}.

\end{proof}

\begin{lemma}\label{eq:lemma-1}
Let $\mathbf{V} \in \{\pm 1\}^{n \times p}$ be i.i.d.~Bernoulli random variables, independent of everything else.
Under the model in Equation~\eqref{eq:indep-model}, consider any fixed matrix $\mathbf{U} \in \{\pm 1\}^{n \times p}$ such that, for each $i \in [n]$ and $j \in [p]$,  $U_{i,j}= 1$ if $\mathcal{H}^i_{j}$ in~\eqref{eq:null-hyp-ind} is {\em not true}.
Let $[\mathbf{X},\tilde{\mathbf{X}}]_{\mathrm{swap}(\mathbf{U})} \in \mathbb{R}^{n \times 2p}$ be the concatenation of $\mathbf{X}$ and $\tilde{\mathbf{X}}$, with the $i$-th observation of $X_j$ swapped with its knockoff if and only if $U_{ij} = -1$.
Then,
\begin{align*}
  \left( [\mathbf{X}, \tilde{\mathbf{X}}]_{\mathrm{swap}(\mathbf{U})}, [\mathbf{X}, \tilde{\mathbf{X}}]_{\mathrm{swap}(\mathbf{V})},\mathbf{Y} \right) \mid \mathbf{Z}
  \quad \oset{d}{=}\quad
  \left( [\mathbf{X}, \tilde{\mathbf{X}}], [\mathbf{X}, \tilde{\mathbf{X}}]_{\mathrm{swap}(\mathbf{V})},\mathbf{Y}\right) \mid \mathbf{Z},
\end{align*}
which also implies
\begin{align*}
  [\mathbf{X}, \tilde{\mathbf{X}}]_{\mathrm{swap}(\mathbf{U})} \mid \left( [\mathbf{X}, \tilde{\mathbf{X}}]_{\mathrm{swap}(\mathbf{V})},\mathbf{Y},\mathbf{Z}\right)
  \quad \oset{d}{=} \quad
  [\mathbf{X}, \tilde{\mathbf{X}}] \mid \left( [\mathbf{X}, \tilde{\mathbf{X}}]_{\mathrm{swap}(\mathbf{V})},\mathbf{Y},\mathbf{Z}\right).
\end{align*}
\end{lemma}

\begin{proof}[Proof of Lemma~\ref{eq:lemma-1}]

By independence, it suffices to prove this for a single row of the data matrices, which corresponds to an individual $i \in [n]$ with a true conditional distribution $p^{(i)}_{Y\mid X,Y}$.
Keeping the individual's index $i$ implicit to simplify the notation, we aim to prove
\begin{align*}
  \left( (X,\tilde{X})_{\mathrm{swap}(u)}, Y , (X,\tilde{X})_{\mathrm{swap}(v)} \right) \mid Z
  \quad \oset{d}{=} \quad \left( (X,\tilde{X}), Y , (X,\tilde{X})_{\mathrm{swap}(v)} \right) \mid Z,
\end{align*}
where $u \in \{-1,1\}^p$ satisfies $u_j=1$ for all variables $j \in [p]$ that are true nulls for the $i$-th individual, and $v$ denotes the $i$-th row of $\mathbf{V}$.

By construction of the knockoffs $\tilde{X}$, we know that
\begin{align} \label{eq:lemma-1-eq-1}
(X,\tilde{X})_{\mathrm{swap}(u)} \mid Z \; \oset{d}{=} \; (X,\tilde{X}) \mid Z.
\end{align}
Further, for any fixed swap $s$ and random swap $v$ it holds that $v \odot s \, \oset{d}{=} \, v$, which implies
\begin{align}  \label{eq:lemma-1-eq-2}
  (X,\tilde{X})_{\mathrm{swap}(v)} \mid (X,\tilde{X})_{\mathrm{swap}(u)}, Z \; \oset{d}{=} \; (X,\tilde{X})_{\mathrm{swap}(v)} \mid (X,\tilde{X}), Z.
\end{align}
Combining~\eqref{eq:lemma-1-eq-1} and~\eqref{eq:lemma-1-eq-2} yields:
\begin{align}  \label{eq:lemma-1-eq-3}
  \left( (X,\tilde{X})_{\mathrm{swap}(u)}, (X,\tilde{X})_{\mathrm{swap}(v)}, \right) \mid Z
  \quad \oset{d}{=} \quad 
  \left( (X,\tilde{X}),  (X,\tilde{X})_{\mathrm{swap}(v)}\right) \mid Z.
\end{align}
Now, recall that $Y \indep \tilde{X} \mid X, Z$. Further, $Y \indep X_{s} \mid X_{-s}, Z$ for all $X_{s}$ swapped by $s$ because the latter only involves null variables.
Therefore, by the same argument as in the proof of Lemma 3.2 in~\cite{candes2018},
\begin{align}  \label{eq:lemma-1-eq-4}
  Y \mid (X,\tilde{X})_{\mathrm{swap}(u)}, (X,\tilde{X})_{\mathrm{swap}(v)},Z \; \oset{d}{=} \; Y \mid (X,\tilde{X}), (X,\tilde{X})_{\mathrm{swap}(v)}, Z.
\end{align}
Finally, combining~\eqref{eq:lemma-1-eq-3} with~\eqref{eq:lemma-1-eq-4} gives the desired result.
\end{proof}

\begin{proof}[Proof of Theorem~\ref*{thm:coin-flip}]

Consider a random vector $\bar{\mathbf{U}} \in \{\pm 1\}^{\hat{L}}$ with independent entries such that: $\bar{U}_{j,\ell}= \pm 1$ with probability $1/2$ if $\smash{\mathcal{H}_{j,\ell}(\hat{\nu})}$ in~\eqref{eq:null-hyp-loc} is true and $\bar{U}_{j,\ell} = +1$ otherwise, for all $j \in [p]$ and ${\ell \in[\hat{L}_j]}$. 
Then, it suffices to prove the flip-sign property $\mathbf{W} \mid \hat{\nu} \; \oset{d}{=}\; \mathbf{W} \odot \bar{\mathbf{U}} \mid \hat{\nu}$, since the FDR result follows from there directly as in~\cite{candes2018}.

Recall that, by construction, both $\hat{\nu}$ and $\smash{\hat{L}}$ are determined by the cloaked data set $\tilde{\mathcal{D}}(\mathbf{V}) = (\smash{[\mathbf{X}, \tilde{\mathbf{X}}]_{\mathrm{swap}(\mathbf{V})}, \mathbf{Y},\mathbf{Z}})$, where $\mathbf{V} \in \{0,1\}^{n \times p}$ are i.i.d.~Bernoulli random variables, as well as (possibly) by some completely independent random noise (e.g., due to a cross-validation procedure).
Therefore, we focus on proving the following stronger flip-sign property: $\mathbf{W} \mid \tilde{\mathcal{D}} \; \oset{d}{=}\; \mathbf{W} \odot \bar{\mathbf{U}} \mid \tilde{\mathcal{D}}$, which allows us to treat $\hat{\nu}$ and $\smash{\hat{L}}$ as fixed.

Recall also that the test statistics $\mathbf{W}$ are determined by applying a (possibly randomized) function $\boldsymbol{\tau}$ to the knockoff-augmented data set $\mathcal{D} = ([\mathbf{X}, \tilde{\mathbf{X}}],\mathbf{Y},\mathbf{Z})$ and $\tilde{\mathcal{D}}(\mathbf{V})$.
Since the possible source of randomness in $\boldsymbol{\tau}$ is also independent of everything else, we may ignore it without loss of generality, to simplify the notation as much as possible.

Define $\mathcal{S}(\bar{\mathbf{U}}) := \{(j,\ell) : j \in [p], \ell \in [L_j], \bar{U}_{j,\ell} = -1 \} \subseteq [L]$, where  $ \hat{L} = \sum_{j=1}^{p} \hat{L}_j $.
Recall also that $\smash{[\mathbf{X}, \tilde{\mathbf{X}}]_{\mathrm{swap}(\mathcal{S}(\bar{\mathbf{U}}))}}$ is defined as the matrix obtained from $\smash{[\mathbf{X}, \tilde{\mathbf{X}}]}$ after swapping the sub-column $\smash{\mathbf{X}^{(\ell)}_j}$, which contains all observations of $X_j$ in the subgroup $g_{j,\ell}$, with the corresponding knockoffs, for all $(j,\ell) \in \mathcal{S}(\bar{\mathbf{U}})$.
With this notation, it follows from \eqref{eq:swap-tau} that
\begin{align} \label{eq:theorem-2-1}
  \mathbf{W}([\mathbf{X}, \tilde{\mathbf{X}}]_{\mathrm{swap}(\mathcal{S}(\bar{\mathbf{U}}))},\mathbf{Y},\mathbf{Z}, \tilde{\mathcal{D}}(\mathbf{V}))
  =
  \bar{\mathbf{U}} \odot \mathbf{W}([\mathbf{X}, \tilde{\mathbf{X}}],\mathbf{Y},\mathbf{Z}, \tilde{\mathcal{D}}(\mathbf{V})).
\end{align}

Next, note that we can equivalently write (almost surely) $[\mathbf{X}, \tilde{\mathbf{X}}]_{\mathrm{swap}(\bar{\mathbf{U}})} = [\mathbf{X}, \tilde{\mathbf{X}}]_{\mathrm{swap}(\mathbf{U})}$, where $\mathbf{U} \in \{\pm 1\}^{n \times p}$ is a matrix defined such that, for each $i \in [n]$ and $j \in [p]$,  $U_{i,j}= \bar{U}_{j,\ell}$ for $\ell = \hat{\nu}_j(Z_i)$.
In particular, this definition of $\mathbf{U}$ implies that: (1) $\mathbf{U}$ may be treated as fixed conditional on $\tilde{\mathcal{D}}(\mathbf{V})$, similar to $\bar{\mathbf{U}}$; and (2) $U_{i,j} = 1$ whenever $\mathcal{H}^i_{j}$ in~\eqref{eq:null-hyp-ind} is not true.
Therefore, we can apply Lemma~\ref{eq:lemma-1}, obtaining that, conditional on $\tilde{\mathcal{D}}(\mathbf{V})$ and $\mathbf{U},\bar{\mathbf{U}}$,
\begin{align*}
  [\mathbf{X}, \tilde{\mathbf{X}}]_{\mathrm{swap}(\bar{\mathbf{U}})} 
  \; \oset{d}{=}\;
  [\mathbf{X}, \tilde{\mathbf{X}}],
\end{align*}
and thus, conditional on $\tilde{\mathcal{D}}(\mathbf{V})$ and $\mathbf{U},\bar{\mathbf{U}}$,
\begin{align} \label{eq:theorem-2-2}
  \mathbf{W}([\mathbf{X}, \tilde{\mathbf{X}}]_{\mathrm{swap}(\bar{\mathbf{U}})},\mathbf{Y},\mathbf{Z}, \tilde{\mathcal{D}}(\mathbf{V})) 
  \; \oset{d}{=}\;
  \mathbf{W}([\mathbf{X}, \tilde{\mathbf{X}}],\mathbf{Y},\mathbf{Z}, \tilde{\mathcal{D}}(\mathbf{V}) ).
\end{align}

Finally, combining Equations~\eqref{eq:theorem-2-1} and~\eqref{eq:theorem-2-2} leads to the desired result:
\begin{align*}
  \mathbf{W} \odot \bar{\mathbf{U}} \mid \tilde{\mathcal{D}}(\mathbf{V}), \hat{\nu}
  \quad \oset{d}{=} \quad
  \mathbf{W} \mid \tilde{\mathcal{D}}(\mathbf{V}), \hat{\nu}.
\end{align*}

\end{proof}

\clearpage

\section{Numerical Experiments} \label{app:experiments}

\subsection{Additional Details on the Experiments from Section~\ref*{sec:experiments}} 

We provide here more details on the joint distribution of the variables utilized in the numerical Section~\ref*{sec:experiments}, and on the corresponding construction of knockoffs.

The first 20 covariates are independently sampled from a Bernoulli(0.5) distribution:
\begin{align*}
  Z_{j} \overset{\text{i.i.d.}}{\sim} \text{Bernoulli}(0.5), \qquad \forall j \in \{1,\ldots,20\}.
\end{align*}
The remaining 60 covariates are generated by applying the normal CDF transform to a Gaussian random vector with an order-one autoregressive covariance matrix and correlation parameter 0.5. That is,
\begin{align*}
  & Z_{j} = \Phi(Z_j^{\text{latent}}-1), \qquad \forall j \in \{21,\ldots,80\}, \\
  & (Z_{21}^{\text{latent}},\ldots,Z_{80}^{\text{latent}}) \sim \text{AR}(0.5).
\end{align*}

Conditional on the covariates $Z$, the 20 explanatory variables are generated independent of each other as follows:
\begin{align*}
  X_j \mid (Z_{1},\ldots,Z_{20}) \sim \text{Bernoulli}(Z_{20+j}), \qquad \forall j \in \{1,\ldots,20\}.
\end{align*}
Therefore, the knockoffs $\tilde{X}$ are also generated independent of each other as follows:
\begin{align*}
  \tilde{X}_j \mid (Z_{1},\ldots,Z_{20}) \sim \text{Bernoulli}(Z_{20+j}), \qquad \forall j \in \{1,\ldots,20\}.
\end{align*}

\subsection{Experiments with Data from a Randomized Experiment} \label{app:randomized-experiment}

\subsubsection{Description of the Data}

We conduct additional numerical experiments that complement those presented in Section~\ref*{sec:empirical}, focusing on a setting where both the number of explanatory variables ($p$) and the number of covariates ($m$) are relatively small.
These experiments are based on a real data set from a 2-week long randomized experiment involving 80,000 blood donors in China \citep{sun2019mobile}, which was designed to investigate the effectiveness of different donation incentives.
In this experiment, 80,000 individuals were randomly divided into 7 groups: a control group of size 14,000 and 6 treated groups of size 11,000 each.
The treated groups received a text message with a reminder to donate, while the control group received nothing.
Treated groups 2--5 received further encouragement to donate: the promise of an individual monetary reward (equivalent to \$5.00--\$8.30, depending on the amount of blood donated) for group 2; a suggestion to bring a friend for group 3, both the individual monetary reward and the friend suggestion for group 4; a suggestion to bring a friend and the promise of a group reward (equivalent to the individual reward) for group 5; a suggestion to bring a friend, the promise of a group reward (as for the previous group), and the promise of a small extra gift for group 6.
See Table~\ref{table:data-blood-groups} for a summary of this experimental design.
In the end, 797 individuals made a donation.
These data also contain covariates including age, sex, weight, blood type, marital status, education level, occupation, residency status (local or non-local), and time since the last blood donation.
Missing values (less than 4\%) are imputed with the corresponding sample median (or mode, if categorical).

\begin{table}[H]
  \caption{Randomized treatment-control assignments in a field experiment conducted by~\cite{sun2019mobile} to investigate the effectiveness of different incentives for blood donors.} \vspace{0.5em}
  \label{table:data-blood-groups}
  \centering
  \begin{tabular}{ccccccc}
    \toprule
    \multirow{2}{*}{Assignment} & \multirow{2}{*}{Size} & \multicolumn{5}{c}{Treatments $(X_1,\ldots,X_5)$} \\
    \cmidrule(l{2pt}r{2pt}){3-7}
    & & Reminder & Indiv.~reward & Friend & Group reward & Group gift \\
    \midrule
    Control group     & 14k & 0 & 0 & 0 & 0 & 0 \\
    Treated group 1 & 11k & 1 & 0 & 0 & 0 & 0 \\
    Treated group 2 & 11k & 1 & 1 & 0 & 0 & 0 \\
    Treated group 3 & 11k & 1 & 0 & 1 & 0 & 0 \\
    Treated group 3 & 11k & 1 & 1 & 1 & 0 & 0 \\
    Treated group 4 & 11k & 1 & 1 & 1 & 0 & 0 \\
    Treated group 5 & 11k & 0 & 1 & 1 & 1 & 0 \\
    Treated group 6 & 11k & 1 & 1 & 1 & 0 & 1 \\
    \bottomrule
  \end{tabular}
\end{table}

\subsubsection{Construction of Knockoffs}

To simplify the task of generating knockoffs for these data, we approximate the joint distribution of the binary treatments $X_1,\ldots,X_5$ by imagining that different individuals were assigned to one of the six possible groups listed in Table~\ref{table:data-blood-groups} independently of one another, with probability proportional to the observed group size. 
Although the treatment group sizes were in truth fixed a priori, the simplification introduced by this model is useful because it leads to a manageable joint treatment distribution for which exact knockoffs can be generated by the Metropolized algorithm of~\cite{bates2020metropolized}.
% For example, note that our model implies $\P{(X_1,\ldots,X_5) = (0,0,0,0,0)} = 14/80$, while other joint probabilities can be similarly computed from Table~\ref{table:data-blood-groups}.
The approximate validity of these knockoffs is confirmed empirically by the first and second-order goodness-of-fit diagnostics reported in Tables~\ref{tab:knockoffs-1}--\ref{tab:knockoffs-2}.
In principle, it would be possible to generate knockoffs without the above independence approximation using an algorithm similar to that of \citet{sesia2021false}, but that would be much more technically involved and seems unnecessary.

%\subsection{Knockoffs for blood donation treatments}  \label{app:analysis-knocoffs}

\begin{table}[H]  \centering
  \caption{Goodness-of-fit-diagnostics for knockoff treatments, for the blood donation data. The means and standard deviations (S.D.)~of the knockoffs approximately match those of the corresponding true treatments.} \vspace{0.5em}
  \label{tab:knockoffs-1}
\begin{tabular}{rrrrrrrrrrr}
\toprule
\multicolumn{1}{c}{ } & \multicolumn{5}{c}{Treatments} & \multicolumn{5}{c}{Knockoffs} \\
\cmidrule(l{3pt}r{3pt}){2-6} \cmidrule(l{3pt}r{3pt}){7-11}
  & $X_1$ & $X_2$ & $X_3$ & $X_4$ & $X_5$ & $\tilde{X}_1$ & $\tilde{X}_2$ & $\tilde{X}_3$ & $\tilde{X}_4$ & $\tilde{X}_5$\\
\midrule
Mean & 0.825 & 0.412 & 0.550 & 0.138 & 0.138 & 0.822 & 0.411 & 0.549 & 0.139 & 0.140\\
S.D. & 0.380 & 0.492 & 0.497 & 0.344 & 0.344 & 0.382 & 0.492 & 0.498 & 0.346 & 0.347\\
\bottomrule
\end{tabular}
\end{table}

\begin{table}[H] \centering
  \caption{Goodness-of-fit-diagnostics for knockoff treatments, for the blood donation data. The pairwise correlations between different treatments approximately match those of the corresponding knockoffs, as well as those between treatments and knockoffs. Smaller values of the diagonal correlation terms in the upper-right block of the correlation matrix (in red) tend to correspond to higher power.} \vspace{0.5em}
  \label{tab:knockoffs-2}
\begin{tabular}{lrrrrrrrrrr}
\toprule
\multicolumn{1}{c}{ } & \multicolumn{5}{c}{Treatments} & \multicolumn{5}{c}{Knockoffs} \\
\cmidrule(l{3pt}r{3pt}){2-6} \cmidrule(l{3pt}r{3pt}){7-11}
  & $X_1$ & $X_2$ & $X_3$ & $X_4$ & $X_5$ & $\tilde{X}_1$ & $\tilde{X}_2$ & $\tilde{X}_3$ & $\tilde{X}_4$ & $\tilde{X}_5$\\
\midrule
$X_1$ & 1 & 0.386 & 0.509 & 0.184 & 0.184 & \textcolor{red}{0.701} & 0.385 & 0.508 & 0.185 & 0.186\\
$X_2$ & 0.386 & 1 & 0.197 & -0.335 & 0.477 & 0.389 & \textcolor{red}{0.607} & 0.197 & -0.337 & 0.481\\
$X_3$ & 0.509 & 0.197 & 1 & 0.361 & 0.361 & 0.514 & 0.197 & \textcolor{red}{0.612} & 0.364 & 0.365\\
$X_4$ & 0.184 & -0.335 & 0.361 & 1 & -0.159 & 0.186 & -0.334 & 0.362 & \textcolor{red}{0.622} & -0.161\\
$X_5$ & 0.184 & 0.477 & 0.361 & -0.159 & 1 & 0.186 & 0.478 & 0.362 & -0.161 & \textcolor{red}{0.639}\\
\addlinespace
$\tilde{X}_1$ & \textcolor{white}{0.701} & \textcolor{white}{0.389} & \textcolor{white}{0.514} & \textcolor{white}{0.186} & \textcolor{white}{0.186} & 1 & 0.371 & 0.496 & 0.187 & 0.187\\
$\tilde{X}_2$ & \textcolor{white}{0.385} & \textcolor{white}{0.607} & \textcolor{white}{0.197} & \textcolor{white}{-0.334} & \textcolor{white}{0.478} & 0.371 & 1 & 0.189 & -0.319 & 0.464\\
$\tilde{X}_3$ & \textcolor{white}{0.508} & \textcolor{white}{0.197} & \textcolor{white}{0.612} & \textcolor{white}{0.362} & \textcolor{white}{0.362} & 0.496 & 0.189 & 1 & 0.347 & 0.347\\
$\tilde{X}_4$ & \textcolor{white}{0.185} & \textcolor{white}{-0.337} & \textcolor{white}{0.364} & \textcolor{white}{0.622} & \textcolor{white}{-0.161} & 0.187 & -0.319 & 0.347 & 1 & -0.162\\
$\tilde{X}_5$ & \textcolor{white}{0.186} & \textcolor{white}{0.481} & \textcolor{white}{0.365} & \textcolor{white}{-0.161} & \textcolor{white}{0.639} & 0.187 & 0.464 & 0.347 & -0.162 & 1\\
\bottomrule
\end{tabular}
\end{table}

\subsubsection{Simulation of Synthetic Outcomes}

The donation outcomes from this randomized experiment are extremely imbalanced (the positive outcomes are approximately 1\% of the total), and therefore we cannot expect any conditional testing method searching for local associations to have sufficient power to make interesting discoveries.
Therefore, to make the analysis of these data more meaningful, we conduct a semi-synthetic study in which the true donation events are replaced by simulated outcomes from an imaginary causal model conditional on the treatments and covariates. 
This setup remains quite realistic, as it involves real variables, but it reduces the outcome imbalance and increases the effective sample size. 
Consequently, the analysis will be more informative regarding the relatively effectiveness of the our methodology compared to other approaches when applied in situations where the available data are sufficient to make discoveries.  
Further, as the ground truth is known exactly in such a controlled setting, we can have access to useful diagnostics, including the type-I errors.

For each of the 80,000 individuals, an imaginary donation outcome is simulated from a logistic model based on the $m$ real covariates $Z \in \mathbb{R}^m$ and the treatment variables $X \in \mathbb{R}^5$; i.e., $\smash{ \text{logit}\left(\mathbb{P}[Y = 1 \mid X, Z] \right) = \varphi(X,Z) }$, where:
\begin{align} \label{eq:blood-causal-model}
\begin{split}
\varphi(X,Z)
= & -c + \\
  & - b \cdot \I{\text{male}} + b \cdot \I{\text{married}} + b \cdot \I{\text{resident}} - b \cdot \I{\text{age} < 25} +  \\
  & - b \cdot \I{\text{student}} - b \cdot \I{\text{education} < 16} + b \cdot \I{\text{Rh}^-} - b \cdot \I{\text{blood type} \neq \text{O}} +  \\
  & + a \cdot X_1 \left[ (1-\I{\text{resident}}) + (1-\I{\text{donation within 12 months}}) \right]  + \\
  & + a \cdot X_2 \left[ \I{\text{student}} + \I{\text{student}} \cdot \I{\text{male}} \right]  + \\
  & + a \cdot X_3 \left[ (1-\I{\text{male}}) + (1-\I{\text{student}}) \right]  + \\
  & + a \cdot X_4 \left[ \I{\text{student}}) + \I{\text{male}} \right]  + \\
  & + a \cdot X_5 \left[ (1-\I{\text{education} <16}) \right].
\end{split}
\end{align}
Above, we set $a=0.4$, $b=0.2$, while $c$ is such that half of the individuals on average receive a simulated $Y=1$.
Therefore, the first treatment (text reminder) is effective for individuals who did not donate recently and for those who are non-residents. The individual reward is effective for students, and twice as much for those who are male. The friends request is effective for female or unmarried individuals. The group reward is effective for males or students. The small gift is effective for individuals with fewer than 16 years of education. 

\subsubsection{Results}

The goal of our analysis is to discover which treatments are effective within which subgroups, as powerfully and precisely as possible.
The adaptive local knockoff filter is applied as in the experiments of Section~\ref{sec:empirical}, defining the subgroup-specific hypotheses based on the top two candidate covariates selected by a (logistic) lasso model fitted on the cloaked data.

\begin{figure}[!htb]
  \centering
  \includegraphics[width=0.95\textwidth]{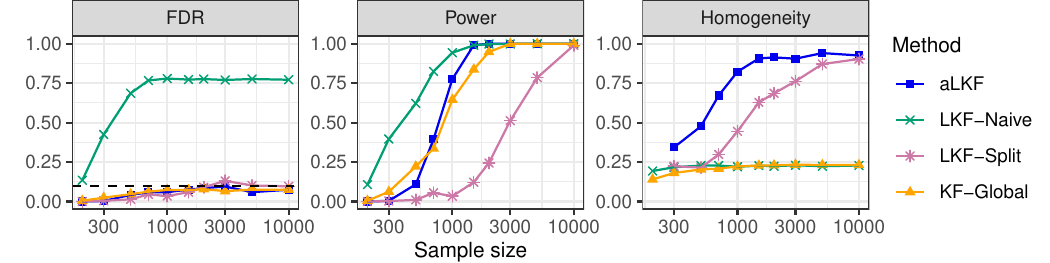}
  \caption{Performance of the adaptive local knockoff filter and benchmarks in semi-synthetic simulations based on data from a randomized blood donation experiment. Other details are as in Figure~\ref*{fig:experiment-heterogeneous-1}.}
  \label{fig:experiment-blood}
\end{figure}

Figure~\ref{fig:experiment-blood} summarizes the results of 100 experiments with independent knockoffs and outcomes, varying the sample size. 
Each time, a random subset of individuals is analyzed.
The performance of the adaptive local knockoff filter is quantified in terms of false discovery rate and power, as well as in terms of the homogeneity of the selected hypotheses, as in Section~\ref{sec:experiments}. The same benchmarks as in Section~\ref{sec:experiments} are also considered here.
The adaptive local knockoff filter controls the false discovery rate and is more powerful than data splitting. 
As expected, the naive local approach does not control the false discovery rate, while the global knockoff filter can only discover less informative population-wide associations.

\clearpage

\section{Details on the UK Biobank Applications} \label{app:gwas}

\paragraph{Background on data and pre-processing.} 
This study utilizes UK Biobank data (applications 27837 and 74923), including unrelated individuals with complete genotype and phenotype information. Observations recommended for exclusion by the UK Biobank due to insufficient data quality (data field 22010) are omitted. We use ancestries and knockoff variables generated by \citet{sesia2021false}. Covariates in all cross-validated lasso models include age (data field 21003), squared age, sex (if applicable, data field 22001), and the top five genetic principal components. The outcome of interest in Section~\ref*{sec:real-data-genetic} is Waist-Hip Ratio (waist circumference $/$ hip circumference), based on data fields 48 (waist circumference) and 49 (hip circumference).

\paragraph{Background on inference at the level of groups of SNPs.}
As discussed in Appendix~\ref{subsec:adaptation-gwas-LD} and~\ref{subsec:adaptation-gwas-LD}, \citet{sesia2021false} partition the individual SNPs into contiguous groups at seven levels of resolution based on complete linkage hierarchical clustering. The group sizes range from single SNPs to 425-kilobase (kb) wide groups. For the real data analysis we choose the resolution with median group width 81 kb and for the simulation we choose a resolution with median group width 3 kb. In the simulation, for simplicity, we only choose a single variant within each of these groups to be nonzero. Note that the presence of a single nonzero SNP within a group renders the conditional independence hypothesis relative to that group of SNPs to be false as well. 

\paragraph{Computational details.}
The GWAS data analyses in Sections~\ref*{subsec:simulation-genetic} and~\ref*{sec:real-data-genetic} are conducted in R/4.2.0 and use the ``bigstatsr'' and ``bigsnpr'' packages \citep{prive2018efficient} for handling large-scale genetic data. For data manipulation, transformation and visualization we use the ``tidyverse'' package \citep{tidyverse}. Computational details on the construction of knockoff variables based on Hidden Markov Models (HMM) can be found in \citet{sesia2021false}. Their software to construct these knockoffs is available at 
\url{https://msesia.github.io/knockoffgwas/}. For reference, the construction of the HMM knockoffs in \citet{sesia2021false} took $\approx$ 4 d using 10 cores and 80 GB of memory to generate knockoffs on chromosome 1 for the UK Biobank data ($\approx$ 1 million haplotype sequences, 600,000 SNPs, and 600,000 IBD (identical-by-descent segments)). In this paper, we run all jobs on a computing cluster and we parallelize when possible. For the simulation in Section 3.2, we submit separate jobs for each amplitude, sparsity, proportion of main effects. Runtime for 100 iterations varies by the simulation parameters and ranges from around 2 hours to 12 hours, with a median runtime of around 5 hours. We run all experiments on the computing cluster with 1 core and 25GB RAM. For the real data analysis in section 3.3, we use 20 CPU cores with a max of 200 GB. We describe the additional computational steps for GWAS data in detail in Section \ref{app:implementation}. The pre-screening step takes around 2 hours, creating the interactions and running the lasso takes around 5 hours, and the lasso on the screened data takes around 3 hours. We submit different jobs to run the local models within each batch, which take around 5 hours. In total, the entire process takes around 15 hours. Our batch scripts containing additional details available at 
\if1\blind
\url{https://github.com/msesia/i-modelx}.
\else
\url{https://github.com/[masked-url]}.
}
\fi

\section{Additional Simulations with the UK Biobank Data}\label{app:add-sims}

\subsection{Varying the Proportion of Global Effects}

\begin{figure}[H]
  \centering
  \includegraphics[width=16cm]{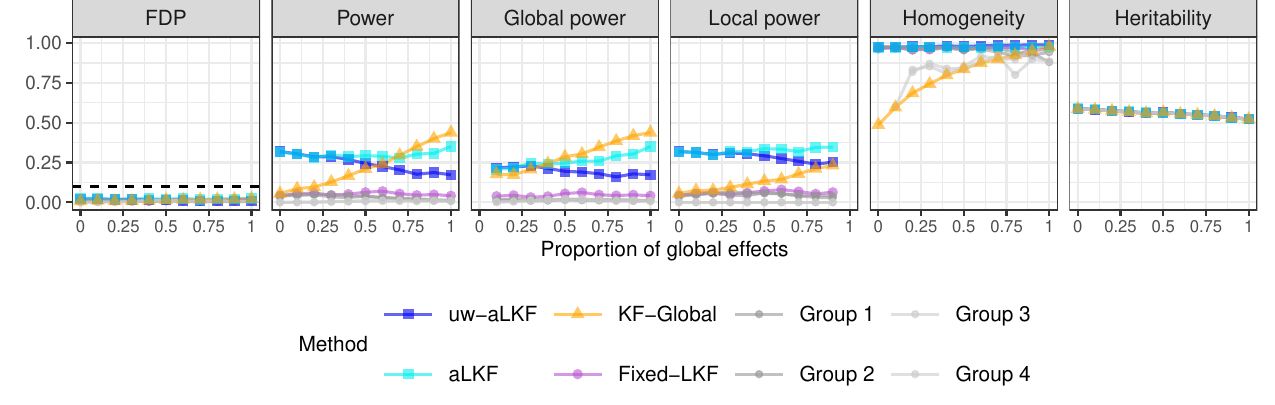}\\[-0.5em]
  \caption{Performance of the local knockoff filter (aLKF) and benchmark methods on real genotype data with a simulated outcome while varying the proportion of global effects for a fixed signal amplitude of 15. Group 1 indicates the subgroup $Z_1=0,Z_2=0$, group 2 indicates $Z_1=0,Z_2=1$, group 3 indicates $Z_1=1,Z_2=0$, and group 4 indicates $Z_1=1,Z_2=1$. The nominal false discovery rate (FDR) level is 0.1. 1\% of SNPs are nonzero. 100 iterations. }
  \label{fig:experiment-genetic-1-appendix}
\end{figure}

All simulations in this section follow the same general setting as described in Section~\ref*{subsec:simulation-genetic}. In Figure~\ref{fig:experiment-genetic-1-appendix}, we fix a signal amplitude of 15, which approximately corresponds to a heritability of 0.5, and vary the proportion of global effects. Below we compare the performance of the {\em aLKF} with benchmark methods as described in Section~\ref*{subsec:simulation-genetic}, as well as the separate subgroups that the {\em Fixed-LKF} is based on. Additionally, we include an ``unweighted'' version of the LKF, where the multiplicative factor $c \in (0, 1]$ applied to the lasso penalty is equal to $1$ for all coefficients ({\em uw-aLKF}).

Figure~\ref{fig:experiment-genetic-1-appendix} shows that all methods control the FDR. As expected, for lower proportions of global effects, and, in turn, higher proportions of local effects, the {\em aLKF} methods are more powerful than the global approach. 

For high proportions of main effects, {\em KF-Global} has highest power as expected. However, our objective extends beyond merely identifying main effects - we aim to uncover local effects as well. While {\em KF-Global} is able to also reject some of the variants that are locally nonzero, it is not able to pinpoint to the relevant subgroup. Moreover, while the informativeness metric (homogeneity) is close to one for the {\em aLKF} methods, it is only increasing to one for higher proportions of global effects for {\em KF-Global}.

The {\em aLKF} is able to recuperate some of the effects lost to the {\em uw-aLKF} with higher proportions of main effects. Overall, the {\em aLKF} is able to interpolate between the {\em uw-aLKF} for lower proportions of global effects and the {\em KF-Global} for higher proportions of global effects.

Figure~\ref{fig:experiment-genetic-1-diff-sparsity} below shows the results of the same simulation setting for signal sparsity levels of 0.5\% and 2\%. The results are qualitatively similar.

\begin{figure}[H]
  \centering
  \includegraphics[width=16cm]{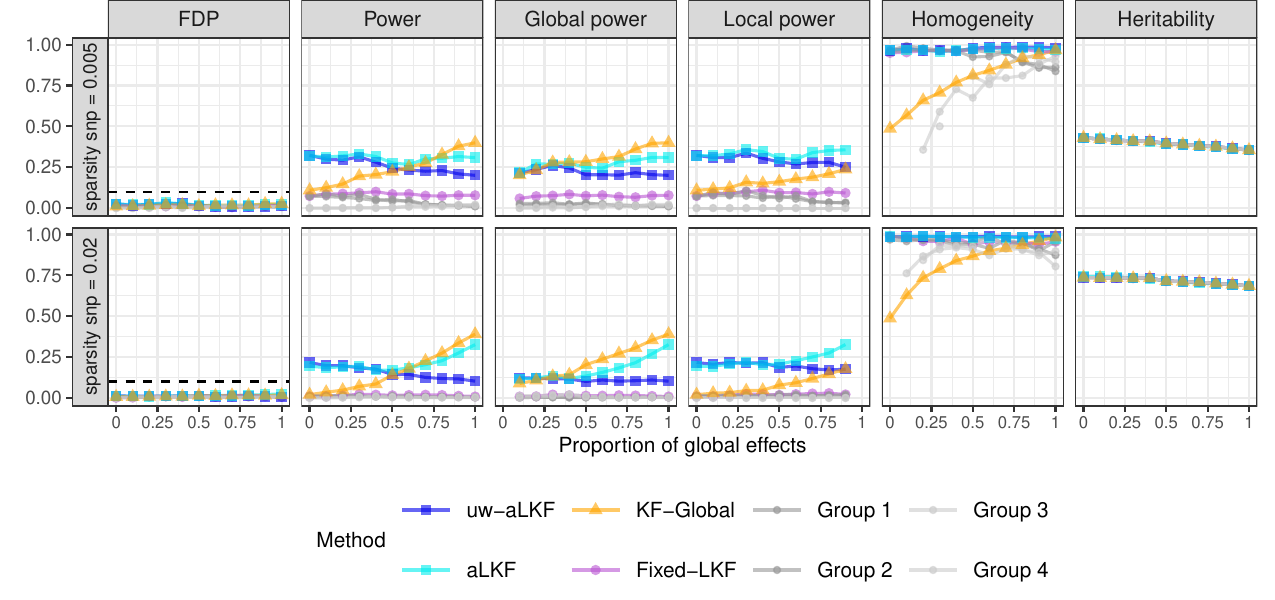}\\[-0.5em]
  \caption{Simulation setting as in Figure~\ref{fig:experiment-genetic-1-appendix} with different levels of sparsity. Sparsity as indicated in row headings.}
  \label{fig:experiment-genetic-1-diff-sparsity}
\end{figure}

The next two figures below repeat the same simulation setting, however, the simulations use $Z_1$ as the only covariate. Again, the results are qualitatively similar as discussed above.

\begin{figure}[H]
  \centering
  \includegraphics[width=16cm]{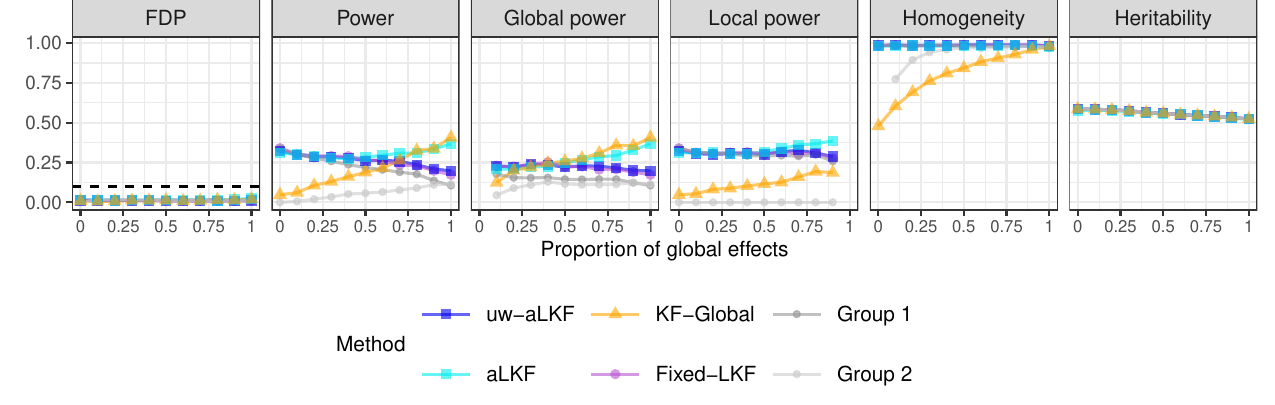}\\[-0.5em]
  \caption{Same general simulation setting as \ref{fig:experiment-genetic-1-appendix}, however, only $Z_1$ is used as a covariate. Group 1 indicates the subgroup $Z_1=0$ and group 2 indicates $Z_1=1$.}
  \label{fig:experiment-genetic-1cov}
\end{figure}

\begin{figure}[H]
  \centering
  \includegraphics[width=16cm]{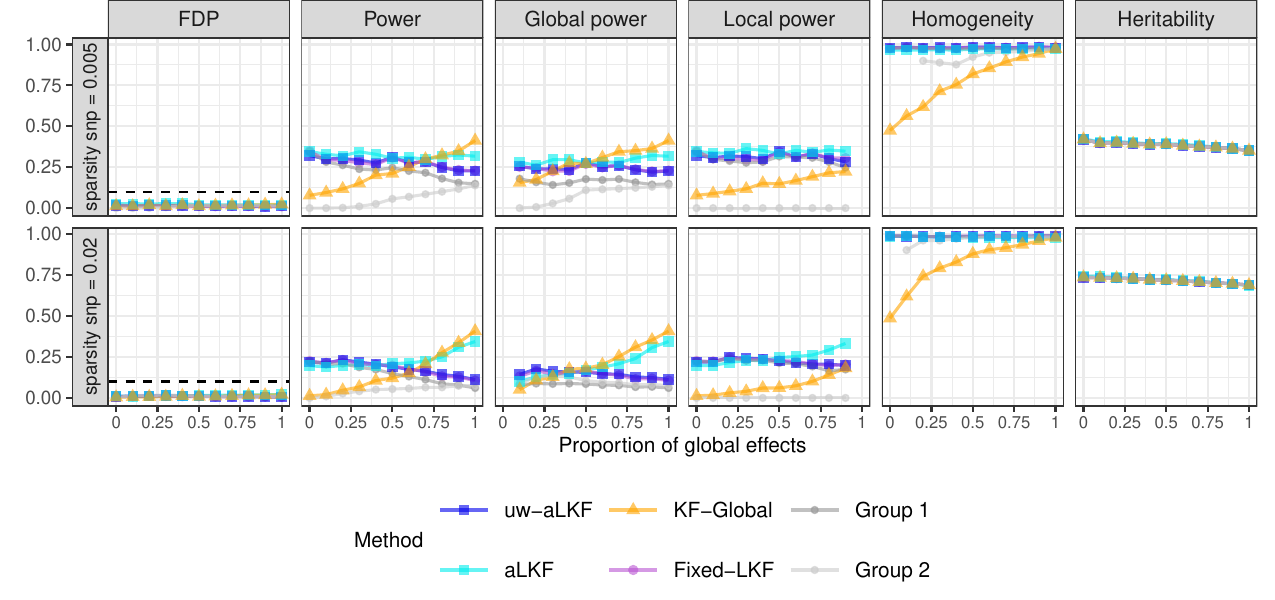}\\[-0.5em]
  \caption{Same simulation setting as ~\ref{fig:experiment-genetic-1cov} with different levels of sparsity.}
  \label{fig:experiment-genetic-1-diff-sparsity-1cov}
\end{figure}

\subsection{Varying the Signal Strength}

\begin{figure}[H]
  \centering
  \includegraphics[width=16cm]{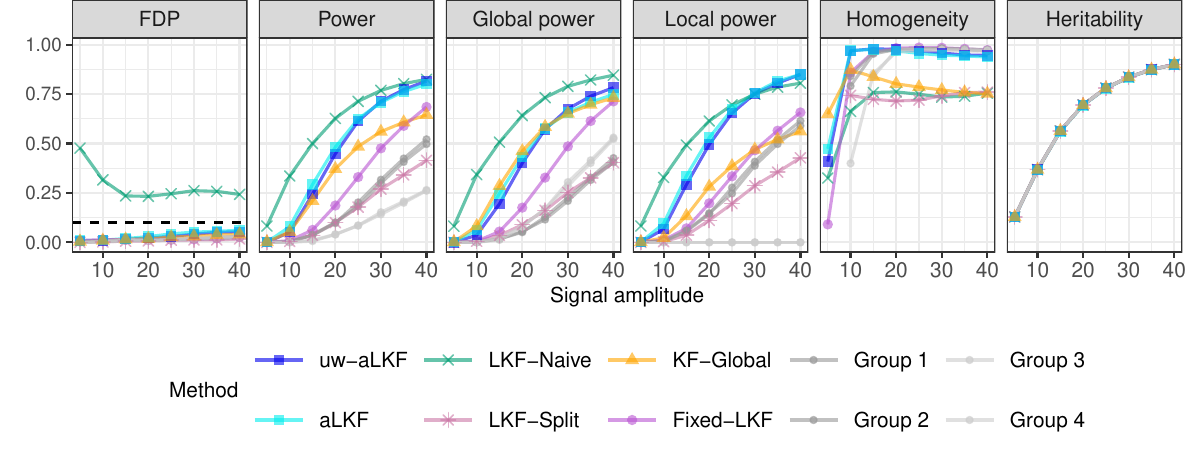}
  \caption{Identical to Figure~\ref*{fig:experiment-genetic-2}, with the difference being the inclusion of the {\em uw-aLKF}, groups 1 to 4 and heritability.}
  \label{fig:experiment-genetic-2-appendix}
\end{figure}

We include additional simulations in the spirit of Figure~\ref*{fig:experiment-genetic-2} in Section~\ref*{subsec:simulation-genetic}: we vary signal amplitude for a fixed proportion of global effects of 50\%. Figure~\ref{fig:experiment-genetic-2-appendix} is identical to Figure~\ref*{fig:experiment-genetic-2}, but includes the {\em uw-aLKF} and the separate groups that the {\em Fixed-LKF} is based on.

%%%%% vary amplidute %%%%%%%

Since there are only local effects in the group $Z_1 = 0$, the subgroups $Z_1=0,Z_2=0$, $Z_1=0,Z_2=1$ have higher power compared to the groups $Z_1=1,Z_2=0$,  $Z_1=1,Z_2=1$. The next figure displays results for sparsity levels of 0.5\% and 2\%. The results are consistent with those discussed earlier.

\begin{figure}[H]
  \centering
  \includegraphics[width=16cm]{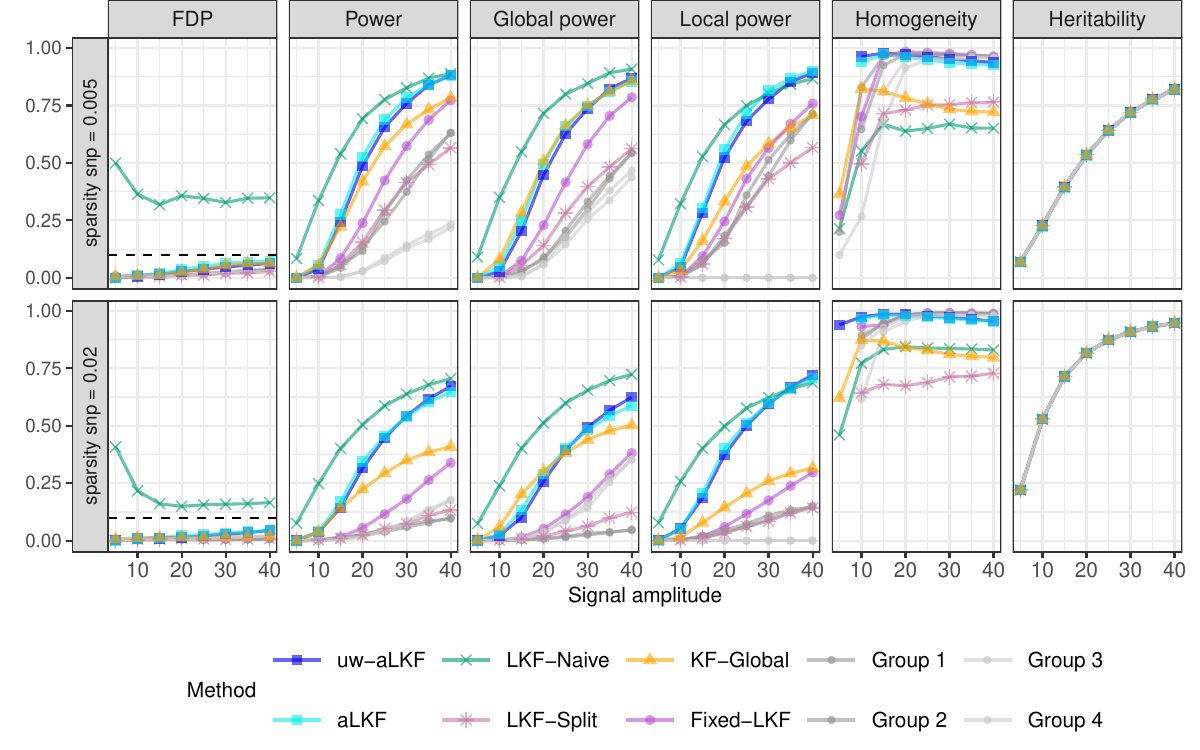}
  \caption{Same simulation setting as in Figure~\ref{fig:experiment-genetic-2-appendix}, except that different levels of sparsity are used, as indicated in row headings.}
  \label{fig:experiment-genetic-2-appendix-diff-sparsity}
\end{figure}

The next two figures repeat the simulations above, but use $Z_1$ as the only covariate. 

\begin{figure}[H]
  \centering
  \includegraphics[width=16cm]{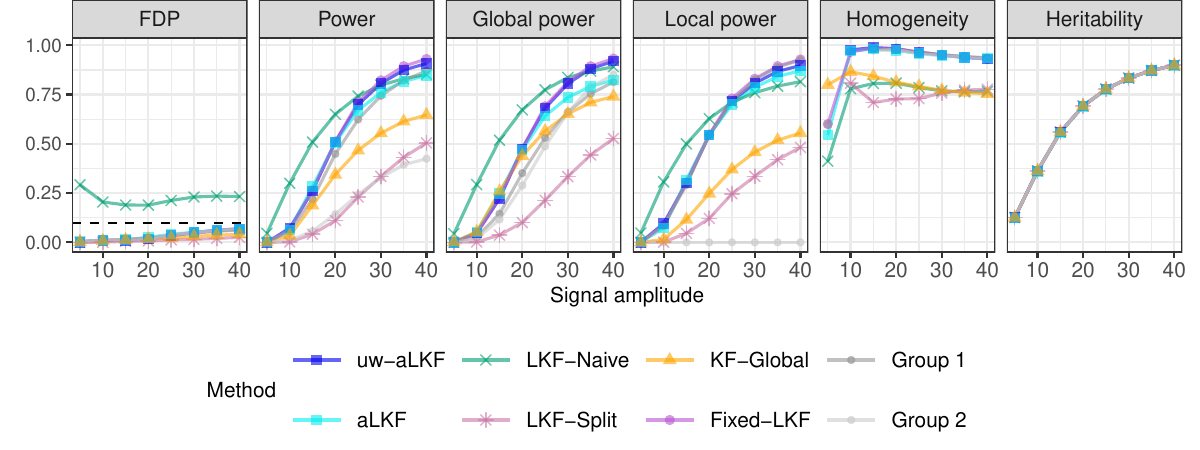}
  \caption{Same simulation setting as in \ref{fig:experiment-genetic-2-appendix}, however, only $Z_1$ is used as a covariate. Group 1 indicates the subgroup $Z_1=0$ and group 2 indicates $Z_1=1$.}
  \label{fig:experiment-genetic-2-appendix-1cov}
\end{figure}

\begin{figure}[H]
  \centering
  \includegraphics[width=16cm]{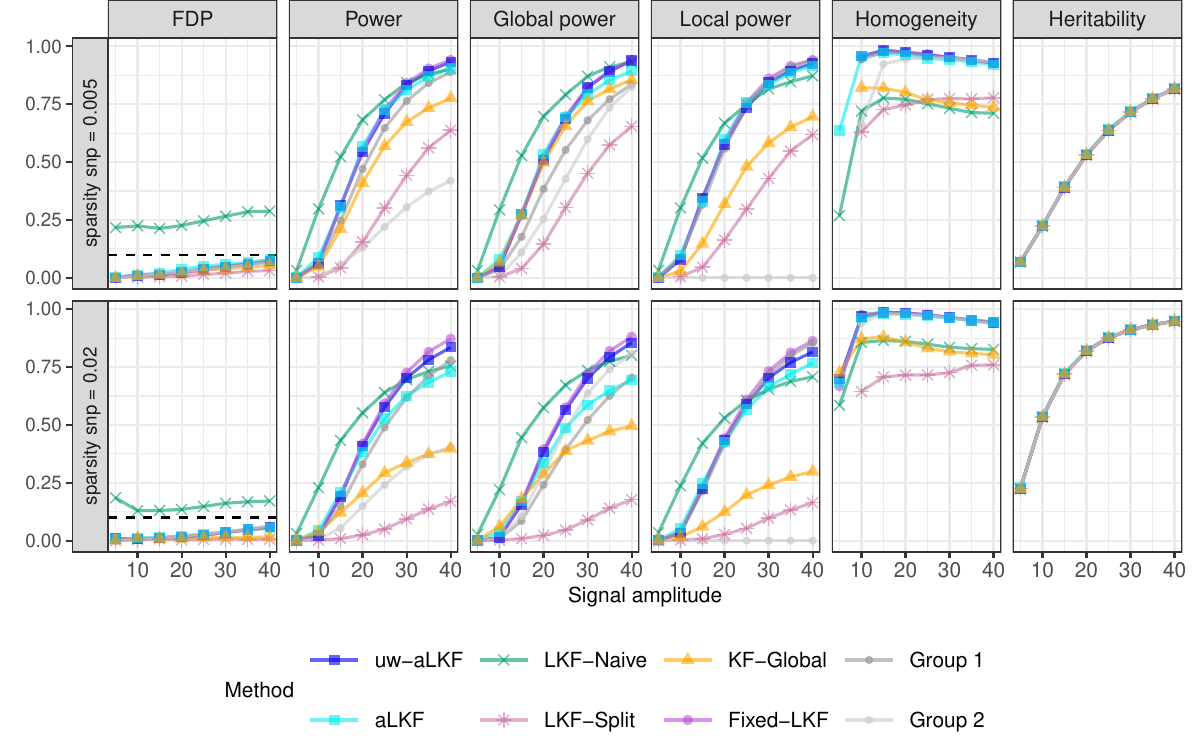}
  \caption{Same simulation setting as in Figure~\ref{fig:experiment-genetic-2-appendix-1cov}, except that different levels of sparsity are used, as indicated in row headings.}
  \label{fig:experiment-genetic-2-appendix-1cov-diff-sparsity}
\end{figure}

\section{Additional Results from the Analysis of WHR}\label{app:add-whr}

\begin{figure}[H]
  \centering
  \includegraphics[width=16cm]{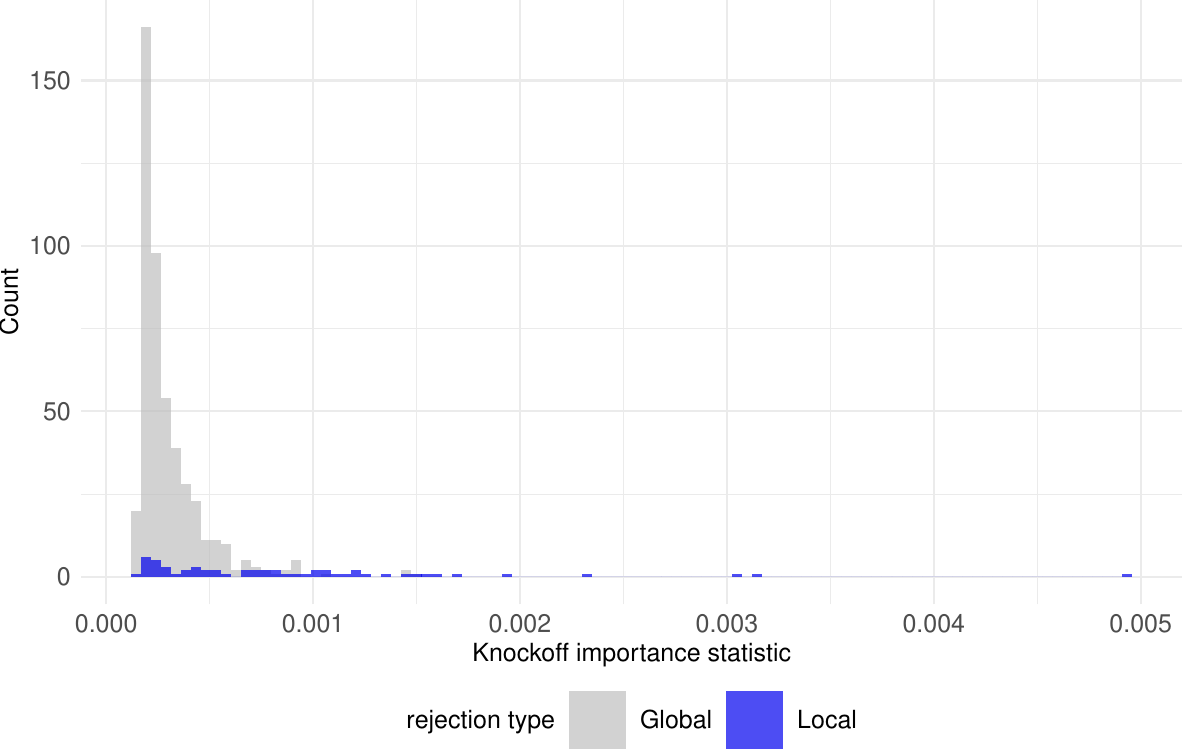}\\[-0.5em]
  \caption{Histogram of the knockoff importance statistics for rejected hypotheses. Local rejections are indicated in blue and global rejections in gray.}
  \label{fig:ukb-hist-rej-w}
\end{figure}

% venn diagram comparison 
\begin{figure}[H]
  \centering
  \includegraphics[width=8cm]{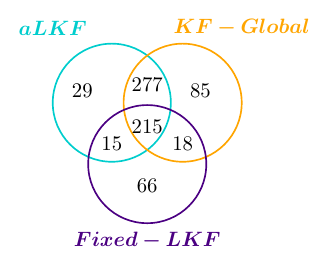}\\[-0.5em]
  \caption{Partition agnostic Venn diagram comparing the rejections of the {\em aLKF}, {\em KF-Global} and the {\em Fixed-LKF}  (considering only unique loci). There are 536 unique loci rejected for the {\em aLKF} (541 total rejections, 10 are rejected in both males and females, i.e. 536 unique loci). The totals for the {\em aLKF} in this Venn diagram add up to 536 ($277 + 215 + 29 + 15 = 536$). Similarly, there are 314 unique loci rejections for the {\em Fixed-LKF} method ($215 + 15 + 18 + 66 = 314$). {\em KF-Global} only makes unique loci rejections, therefore the total based on this Venn diagram is 595 as well ($277 + 215 + 18 + 85 = 595$).}
  \label{fig:venn-gwas}
\end{figure}

\begin{table}[h!]
\centering
 \footnotesize
\begin{tabular}{@{}l l|cccc||c@{}}
\hline
& & \multicolumn{4}{c||}{\textbf{{\em Fixed-LKF}}} & \\ \cline{3-6}
& & Female only & Male only & Both Male and Female & Not detected &  \\ \hline
\multirow{5}{*}{\rotatebox{90}{\textbf{{\em aLKF}}}} 
& Entire population & 127 & 36 & 24 & 297 & 484 \\
& Female only & 35 & 0 & 0 & 6 & 41 \\
& Male only & 0 & 4 & 0 & 2 & 6 \\
& Both Male and Female & 2 & 1 & 1 & 1 & 5 \\
& Not detected & 68 & 16 & 0 &  & (84) \\ \hline \hline
&  & 232 & 57 & 25 & (306) & 314;536 \\ \hline
\end{tabular}
\caption{Comparison of the rejections from the {\em aLKF} to those from the {\em Fixed-LKF} method for unique loci. For the {\em aLKF}, 10 loci are rejected twice. There are are 541 total rejections for the {\em aLKF} in total, which means that $541 - 10/2 = 536$ are unique. The total number of {\em aLKF} rejections in this table adds up to 536 ($484 + 41 + 6 + 5 = 536$). Similarly, 314 of the {\em Fixed-LKF} rejections are unique. The total number of unique rejections (314; 536) do not include the totals for ``Not detected'' loci. The numbers in this table also match with those in the Venn diagram in Figure~\ref{fig:venn-gwas}.}
\label{tab:alkf_vs_fixed}
\end{table}

\begin{landscape}

\begin{center}

\begin{ThreePartTable}
                  
	              \begin{TableNotes}
                                    \item Local rejections for WHR on the British unrelated population, sorted by the knockoff importance statistic $W$. FDR level: $\alpha = 0.1$. 
                                    \item ``SNP lead'' is the individual SNP within the rejected group of variants with the largest knockoff importance statistic. ``BP lead'' is the basepair (BP) of ``SNP lead''. The minimum and maximum BP for the group are given by ``min BP'' and ``max BP''. ``Partition'' indicates the partition of the rejection, i.e. whether the local rejection was made for males or females. The column ``rej. global'' indicates whether the group was rejected by KF-Global. If ``rej. global'' is blank, then the group was rejected by KF-Global, if it is ``no'', then it was not rejected.
                                    \item ``Gene'' indicates the gene(s) that the leading SNP or group overall is closest to. These genes have been found using the NHGRI-EBI Catalog \citep{sollis2023nhgri}, FAVOR \citep{zhou2023favor}, and the CVDKP \citep{costanzo2023cardiovascular}. ``Previous discovery'' indicates whether each gene has been (locally) associated with WHR or a closely related phenotype, WHR adjusted for BMI, before. The entry ``local diff.'' means that the gene has been found to have differential effects by sex in previous research \citep{funkhouser2020deciphering, heid2010meta, shungin2015new, randall2013sex, winkler2015influence}. In contrast, the entry ``global'' means that previous studies have reported a (significant) association between the gene and WHR, or WHR adjusted for BMI, based on the NHGRI-EBI Catalog and/or the CVDKP  \citep{sollis2023nhgri, costanzo2023cardiovascular}. If the entry ``previous discovery'' is missing, then the gene was reported to be associated with WHR (or WHR adjusted for BMI) according to the CVDKP  \citep{costanzo2023cardiovascular}, but not significantly.

								\end{TableNotes}
		
								  \scriptsize
								\begin{longtable}{@{\extracolsep{2pt}} lllllllllllllll} 
								\caption{Validation and annotation of local discoveries} 
        \label{tab:gwas-local-annotation}
									\\[-1.8ex]\hline 
\hline \\[-1.8ex] 
CHR & Group & W & SNP lead & BP lead & min BP & max BP & partition & rej. vanilla & gene & previous discovery \\ 
\hline \\[-1.8ex] 
$6$ & $735$ & $0.00496$ & rs72961013 & $127,529,780$ & $126,504,799$ & $127,530,850$ & female &  & RSPO3 & local diff. \\ 
$6$ & $735$ & $0.00121$ & rs72961013 & $127,529,780$ & $126,504,799$ & $127,530,850$ & male &  & RSPO3 & local diff. \\ 
$2$ & $973$ & $0.00314$ & rs79953491 & $165,555,539$ & $165,501,849$ & $165,829,017$ & female &  & COBLL1 & local diff. \\ 
$6$ & $371$ & $0.00305$ & rs4711750 & $43,757,082$ & $43,757,082$ & $43,781,321$ & female &  & VEGFA & local diff. \\ 
$3$ & $807$ & $0.00234$ & rs9833879 & $129,333,910$ & $129,097,692$ & $129,637,348$ & female &  & PLXND1 & local diff. \\ 
$3$ & $496$ & $0.00193$ & rs4132228 & $64,708,114$ & $64,692,129$ & $64,862,311$ & female &  & ADAMTS9-AS2 & local diff. \\ 
$1$ & $997$ & $0.00170$ & rs6427245 & $170,303,073$ & $170,171,258$ & $170,345,383$ & female &  & GORAB-AS1, LINC01681 & local diff. \\ 
$9$ & $595$ & $0.00159$ & rs10991417 & $107,688,629$ & $107,676,940$ & $107,743,189$ & female &  & ABCA1 & local diff. \\ 
$5$ & $365$ & $0.00155$ & rs3936510 & $55,860,866$ & $55,828,080$ & $55,880,301$ & female &  & MAP3K1, C5orf67 & local diff. \\ 
$12$ & $824$ & $0.00151$ & rs4765127 & $124,460,167$ & $124,256,619$ & $124,499,584$ & female &  & CCDC92, RFLNA, ZNF664 & local diff. \\ 
$4$ & $557$ & $0.00146$ & rs1377290 & $89,719,368$ & $89,468,618$ & $89,744,242$ & female &  & FAM13A & local diff. \\ 
$1$ & $1,006$ & $0.00135$ & rs9286854 & $172,366,806$ & $172,302,544$ & $172,463,933$ & female &  & DNM3 & global \\ 
$1$ & $1,283$ & $0.00126$ & rs4846567 & $219,750,717$ & $219,622,746$ & $219,979,718$ & female &  & ZC3H11B, LYPLAL1-AS1 & local diff. \\ 
$12$ & $258$ & $0.00119$ & rs718314 & $26,453,283$ & $26,440,451$ & $26,532,354$ & female &  & ITPR2 & local diff. \\ 
$11$ & $402$ & $0.00114$ & rs35169799 & $64,031,241$ & $63,858,714$ & $64,219,679$ & female &  & PLCB3 & global \\ 
$3$ & $164$ & $0.00110$ & rs4684859 & $12,498,401$ & $12,477,055$ & $12,859,500$ & female &  & PPARG,TSEN2, CAND2 & local diff. \\ 
$3$ & $164$ & $0.00027$ & rs2305398 & $12,856,856$ & $12,477,055$ & $12,859,500$ & male &  & PPARG,TSEN2, CAND2 & local diff. \\ 
$11$ & $393$ & $0.00107$ & rs72927213 & $62,351,198$ & $62,309,586$ & $62,649,463$ & female &  & TUT1 & global \\ 
$6$ & $598$ & $0.00106$ & rs2503097 & $100,606,568$ & $100,599,675$ & $100,637,297$ & female &  & MCHR2-AS1,SIM1,LOC728012 & local diff. \\ 
$5$ & $678$ & $0.00103$ & rs17145265 & $118,698,998$ & $118,676,043$ & $118,969,960$ & female & no & TNFAIP8 & local diff. \\ 
$7$ & $244$ & $0.00100$ & rs1055144 & $25,871,109$ & $25,852,328$ & $25,938,593$ & female &  & LINC03007,MIR148A & local diff. \\ 
$2$ & $945$ & $0.00099$ & rs55920843 & $158,412,701$ & $158,390,468$ & $158,487,634$ & female &  & ACVR1C & global \\ 
$4$ & $245$ & $0.00090$ & rs2048508 & $26,079,994$ & $25,990,084$ & $26,123,616$ & female &  & LINC02357, RBPJ & global \\ 
$6$ & $298$ & $0.00087$ & rs114344942 & $34,207,990$ & $34,060,867$ & $34,238,514$ & female &  & HMGA1 & local diff. \\ 
$2$ & $707$ & $0.00083$ & rs17041868 & $111,894,720$ & $111,810,633$ & $112,007,943$ & female &  & BCL2L11,MIR4435-2HG & global \\ 
$3$ & $163$ & $0.00081$ & rs4279078 & $12,314,846$ & $12,029,991$ & $12,475,557$ & female &  & SYN2,PPARG & local diff. \\ 
$12$ & $820$ & $0.00079$ & rs73228032 & $123,322,415$ & $123,159,176$ & $123,387,922$ & female & no & HIP1R, HCAR1 & local diff. \\ 
$8$ & $748$ & $0.00076$ & rs6982636 & $126,479,315$ & $126,447,844$ & $126,543,098$ & female & no & TRIB1AL & global \\ 
$20$ & $292$ & $0.00075$ & rs143384 & $34,025,756$ & $33,886,799$ & $34,797,801$ & male &  & GDF5 & local diff. \\ 
$11$ & $364$ & $0.00074$ & rs12361415 & $47,474,146$ & $47,370,041$ & $49,425,463$ & male &  & CELF1,RAPSN, AGBL2 & global \\ 
$11$ & $364$ & $0.00022$ & rs12286721 & $47,701,528$ & $47,370,041$ & $49,425,463$ & female &  & CELF1,RAPSN, AGBL2 & global \\ 
$10$ & $815$ & $0.00067$ & rs2254069 & $122,875,589$ & $122,816,108$ & $123,000,561$ & female &  & WDR11,RPL19P16, MIR5694 & local diff. \\ 
$12$ & $825$ & $0.00067$ & rs12829378 & $124,516,911$ & $124,504,283$ & $124,637,099$ & female & no & RFLNA & global \\ 
$17$ & $249$ & $0.00060$ & rs16961845 & $17,432,456$ & $17,432,456$ & $17,693,891$ & female &  & PEMT & global \\ 
$14$ & $87$ & $0.00055$ & rs12432376 & $25,942,829$ & $25,920,720$ & $25,990,890$ & male & no & STXBP6 & local diff. \\ 
$1$ & $872$ & $0.00053$ & rs114056923 & $154,868,744$ & $154,824,330$ & $155,019,710$ & female & no & PMVK,KCNN3 & global \\ 
$1$ & $528$ & $0.00049$ & rs111573261 & $66,001,886$ & $65,896,561$ & $66,169,679$ & female &  & LEPR & global \\ 
$3$ & $932$ & $0.00047$ & rs62271364 & $150,050,596$ & $150,004,071$ & $150,229,958$ & female & no & TSC22D2,LOC646903, AC018545.1 & local diff. \\ 
$10$ & $608$ & $0.00046$ & rs1236816 & $89,684,214$ & $89,591,344$ & $89,850,500$ & female &  & PTEN & global \\ 
$1$ & $252$ & $0.00046$ & rs649458 & $23,712,528$ & $23,294,671$ & $23,829,648$ & female &  & TCEA3 & global \\ 
$1$ & $873$ & $0.00042$ & rs74593044 & $156,016,868$ & $155,025,361$ & $156,022,802$ & female & no & UBQLN4 & global \\ 
$1$ & $799$ & $0.00041$ & rs72694053 & $115,391,936$ & $114,931,419$ & $115,672,145$ & male &  & SYCP1 & global \\ 
$5$ & $335$ & $0.00037$ & rs72765634 & $50,799,239$ & $50,420,111$ & $50,950,593$ & male & no & HMGB1P47,RNA5SP182 & global \\ 
$3$ & $853$ & $0.00034$ & rs2306374 & $138,119,952$ & $137,784,095$ & $138,565,389$ & female & no & MRAS, ESYT3 & local diff. \\ 
$3$ & $853$ & $0.00017$ & rs808925 & $138,177,119$ & $137,784,095$ & $138,565,389$ & male & no & MRAS, ESYT3 & local diff. \\ 
$2$ & $755$ & $0.00030$ & rs954244 & $121,309,231$ & $121,305,604$ & $121,355,408$ & female & no & LINC01101, Y\_RNA & global \\ 
$2$ & $421$ & $0.00029$ & rs4953618 & $48,954,842$ & $48,954,842$ & $49,011,274$ & female & no & GTF2A1L,LHCGR,STON1-GTF2A1L & global \\ 
$2$ & $375$ & $0.00026$ & rs116520905 & $44,037,396$ & $43,937,198$ & $44,049,795$ & female & no & ABCG5, DYNC2LI1 & global \\ 
$1$ & $19$ & $0.00026$ & rs12044475 & $2,709,392$ & $2,373,987$ & $2,802,562$ & male & no & TTC34 &  \\ 
$1$ & $396$ & $0.00025$ & rs77213763 & $46,440,828$ & $45,782,786$ & $46,897,679$ & male &  & MAST2,AL358075.4,PIK3R3 &  \\ 
$1$ & $396$ & $0.00020$ & rs785493 & $46,584,859$ & $45,782,786$ & $46,897,679$ & female &  & MAST2,AL358075.4,PIK3R3 &  \\ 
$2$ & $251$ & $0.00024$ & rs73920682 & $25,578,778$ & $25,487,656$ & $25,592,436$ & female & no & DNMT3A,AC012074.1 & global \\ 
$2$ & $135$ & $0.00021$ & rs4799 & $10,933,249$ & $10,881,867$ & $10,969,662$ & female & no & PDIA6 & global \\ 
$1$ & $105$ & $0.00020$ & rs2071931 & $9,329,289$ & $9,305,445$ & $9,339,467$ & female &  & H6PD & local diff. \\ 
$1$ & $99$ & $0.00018$ & rs301805 & $8,481,016$ & $8,374,253$ & $8,982,906$ & male & no & RERE & global \\ 
$7$ & $283$ & $0.00017$ & rs12112389 & $30,936,024$ & $30,932,526$ & $30,942,744$ & female & no & MINDY4,AQP1 & global \\ 
$16$ & $669$ & $0.00017$ & rs9926861 & $85,156,713$ & $85,150,638$ & $85,172,619$ & female & no & LINC02139,FAM92B & global \\ 
\hline \\[-1.8ex] 
\insertTableNotes

								\end{longtable}	
										
\end{ThreePartTable}

\end{center}

\end{landscape}

\clearpage

\FloatBarrier
\section{The Robust Local Knockoff Filter}  \label{app:extension-ralkf}

\subsection{Testing Partial Conjunction Local Hypotheses} \label{app:methods-ralkf}

Recall that the local knockoff filter is designed to test hypotheses~\eqref{eq:null-hyp-loc} of the type
\begin{align*}
  \mathcal{H}_{j,\ell}(\nu) = \bigcap_{i \in [n] \,:\, \nu_j(Z^i) = \ell} \mathcal{H}^i_{j}, \qquad j \in [p], \ell \in [L_j],
\end{align*}
for a (possibly random) partition function $\nu$ with total width $L = \sum_{j=1}^{p} L_j$.
Here, we present an extension of our method, inspired by \citet{li2021searching}, that makes it possible to test the following family of {\em partial conjunction} \citep{friston2005conjunction,benjamini2008screening} hypotheses:
\begin{align} \label{eq:null-pici}
\mathcal{H}^{\mathrm{pc},r}_{j}(\nu) : \left| \left\{ \ell \in [L_j] : \mathcal{H}_{j,\ell}(\nu) \text{ is true} \right\} \right| > L_j-r,
\end{align}
for any fixed $r \in \{1,\ldots, L_j\}$ and any $j \in [p]$.
Intuitively, the hypothesis in~\eqref{eq:null-pici} states that the number of subgroups for variable $j$ in which $\mathcal{H}_{j,\ell}(\nu)$ is false is strictly smaller than $r$.
Therefore, rejecting~\eqref{eq:null-pici} suggests $X_j$ is associated with $Y$ in at least $r$ distinct subgroups identified by the partition function $\nu$.

If $r = 1$, the partial conjunction hypothesis becomes equivalent to a global hypothesis; in that case, rejecting $\mathcal{H}^{\mathrm{pc},r}_{j}(\nu)$ suggests the existence of least one subgroup in which $X_j$ is conditionally associated with $Y$.
By contrast, rejecting $\mathcal{H}^{\mathrm{pc},r}_{j}(\nu)$ for $r=G$ requires evidence of a conditional association between $X_j$ and $Y$ across {\em all} subgroups identified by the partition function $\nu$.

One reason why it can be interesting to test $\mathcal{H}^{\mathrm{pc},r}_{j}(\nu)$ is that this hypothesis may point towards the discovery of variables whose association with $Y$ is {\em robust} to changes in the covariate distribution~\citep{li2021searching}.
Intuitively, testing $\mathcal{H}^{\mathrm{pc},r}_{j}(\nu)$ can be seen as a way for discovering conditional associations that do {\em not} involve interactions, thus taking a point of view complementary to that of aLKF.
The $p$ partial conjunction hypotheses defined in~\eqref{eq:null-pici} are relatively easy to test while controlling the FDR, if one starts with a collection $\mathbf{W}$ of aLKF test statistics, satisfying the flip-sign property in the proof of Theorem~\ref*{thm:coin-flip}: $\mathbf{W} \mid \hat{\nu} \; \oset{d}{=}\; \mathbf{W} \odot \mathbf{U} \mid \hat{\nu}$. In fact, it suffices to apply the multi-environment knockoff filter of~\citet{li2021searching} to these statistics, as explained below.
We call this approach ``robust local knockoff filter'', or {\em Robust-aLKF}.

For each $j \in \{1,\ldots,p\}$, let $n^-_j$ count the negative signs among the $W_{j,\ell}$ statistics for all subgroups $\ell \in [L_j]$, and let $\smash{n^0_j}$ be the number of zeros among them. For simplicity, assume $\hat{\nu}$ partitions the covariate space into $L_0$ subgroups for each variable; this is without loss of generality, as it is otherwise sufficient to set the remaining undefined statistics to zero if $L_j < L_0$ for some $j \in [p]$.
Then, compute
\begin{align}  \label{eq:pici-pvalues}
\begin{split}
  p_j^{\mathrm{pc},r}
  & = \Psi\left( n^-_j -1, (L_0-r+1-n^0_j) \lor 0, \frac{1}{2} \right) + U_j \cdot \Psi'\left( n^-_j, (L_0-r+1-n^0_j) \lor 0, \frac{1}{2}\right),
\end{split}
\end{align}
where $\Psi(\cdot, m,\pi)$ is the binomial cumulative distribution function, $\Psi'(\cdot,m,\pi)$ is the corresponding probability mass, and $U_j$ is a uniform random variable on $[0,1]$ independent of everything else.
Further, define
\begin{align} \label{eq:pici-order}
  |W_j^{\mathrm{pc},r}| = \bar{w} \left( |W_{j,1}|, \dots,|W_{j,L_0}| \right),
\end{align}
for some symmetric function $\bar{w}$, such as that which multiplies the top $r$ largest entries by absolute value:
\begin{align*}
w \left( |W_{j,1}|, \dots,|W_{j,L_0}| \right) = \prod_{\ell = 1}^r  \bar{|W|}_{j,(L_0-\ell+1)}.
\end{align*}
Above, $\bar{|W|}_{j,(\ell)}$ are the order statistics for $\{|W_{j,1}|, \dots,|W_{j,L_0}|\}$.
Then, the FDR for~\eqref{eq:null-pici} can be controlled by applying the usual selective SeqStep+ sequential testing procedure of~\citet{barber2015controlling}, thanks to a relatively simple extension of Theorem~1 from~\citet{li2021searching}.
\begin{theorem} \label{th:seqstep-fdr}
Selective SeqStep+ applied to the p-values \eqref{eq:pici-pvalues} ordered by~\eqref{eq:pici-order} controls the FDR for~\eqref{eq:null-pici} if the statistics $\mathbf{W}$ satisfy the flip-sign property in the proof of Theorem~\ref*{thm:coin-flip}: $\mathbf{W} \mid \hat{\nu} \; \oset{d}{=}\; \mathbf{W} \odot \mathbf{U} \mid \hat{\nu}$.
\end{theorem}
\begin{proof}
This result follows immediately from Theorem~1 in~\citet{li2021searching} with a simple extension of Proposition~5 therein.
The original statement of Proposition~5 in~\citet{li2021searching} assumed the statistics $\mathbf{W}$ to be computed using data collected from separate and a-priori fixed experimental settings (or {\em environments}), but it is easy to see that their proof only requires the flip-sign property established by our Theorem~\ref*{thm:coin-flip}.
\end{proof}

Alternatively, the FDR for~\eqref{eq:null-pici} could be (approximately) controlled by applying to the p-values \eqref{eq:pici-pvalues} ordered by~\eqref{eq:pici-order} the accumulation test~\citep{li2017accumulation} instead of SeqStep+; this can be easily proved with an analogous extension of Theorem~2 from~\citet{li2021searching}.

\FloatBarrier

\subsection{Experiments in a ``Transfer Learning'' Setting} \label{app:experiments-discovery}

We demonstrate the empirical performance of the Robust-aLKF applied in a ``transfer learning'' setting inspired by~\cite{li2022transfer}.
The goal is complementary to that of the experiments in Section~\ref*{sec:empirical}: we wish to detect variables with a robust association across different subgroups.
This is useful for example to find which variables maintain their association with $Y$ within a future data set with covariate shift, in which the distribution of $Z$ differs from the current one but the true model for $Y \mid X,Z$ is the same~\citep{li2022transfer}.

To simulate this covariate shift scenario, we generate data from a model similar to that in Section~\ref*{sec:experiments}, but now only half of the non-null variables interact with the covariates; i.e., $\smash{\beta_j^{(i)}} = \bar{\beta}$ for half of the variables.
Further, the data dimensions are increased to ensure sufficiently many discoveries can be made. Specifically, the number of variables is $p=40$, while the number of covariates is $m=160$, of which 120 are continuous.
The covariate shift is imagined to be such that the 40 binary covariates are always equal to zero instead of following a symmetric Bernoulli distribution.
Therefore, the goal is to identify the $20$ variables whose association does not involve interactions, as the covariate shift would make the others irrelevant.
These data are analyzed as in Section~\ref*{sec:experiments}, with the only difference that the aLKF is replaced by its robust version outlined in Appendix~\ref{app:methods-ralkf}, which is designed to test~\eqref{eq:null-pici}.

\begin{figure}[H]
  \centering
  \includegraphics[width=\textwidth]{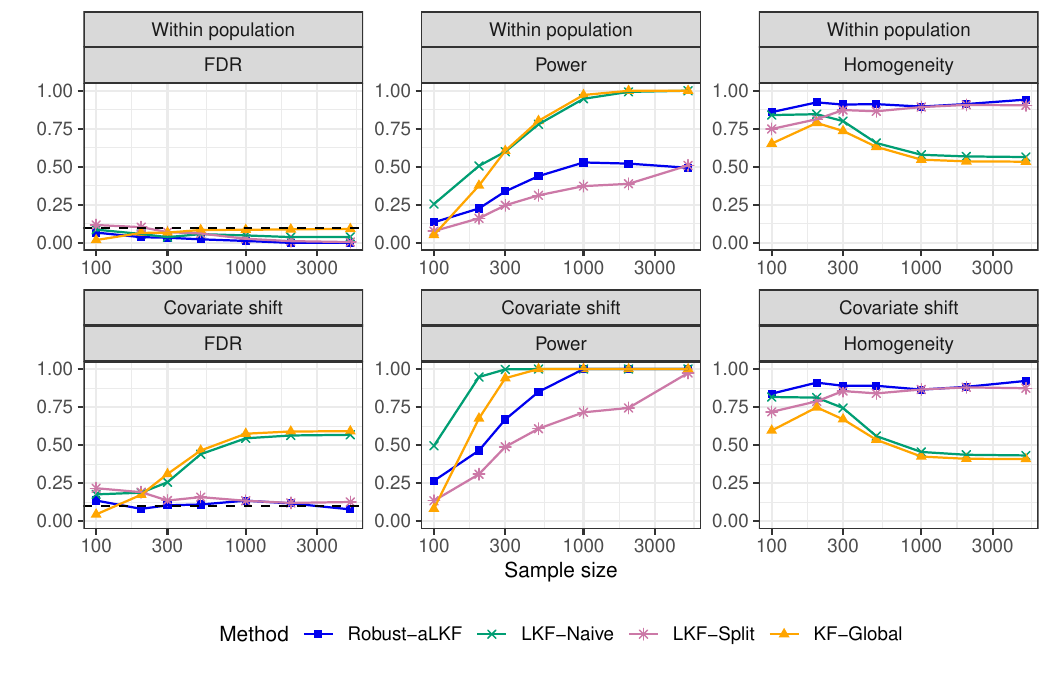}\\[-1em]
  \caption{Performance of the Robust-aLKF and alternative benchmarks for testing local conditional hypotheses under covariate shift, in numerical experiments with synthetic data. The findings summarized in the bottom panel are counted as true if and only if they report a variable whose conditional association with the outcome is robust to changes in the covariate distribution. Other details are as in Figure~\ref*{fig:experiment-heterogeneous-1}.}
  \label{fig:experiment-transfer-1}
\end{figure}

Figure~\ref{fig:experiment-transfer-1} reports on the results of 100 independent experiments, as a function of the sample size.
The performance of the robust version of the aLKF is quantified in terms of the average FDP, the power, and the homogeneity of the reported findings, separately within the training population and under covariate shift. In the latter case, only the discoveries of variables with a direct conditional association with the outcome (i.e., not mediated by any interactions) are counted as true, while the others are considered false positives.
Therefore, neither the naive benchmark nor the global knockoff filter control this notion of FDR under covariate shift because they tend to report all associated variables, including those that are non-null only thanks to interactions. By contrast, the Robust-aLKF empirically controls the FDR even under covariate shift.
The data-splitting benchmark here is a modified version of our Robust-aLKF in which half of the observations are used for partitioning the covariate space, without randomly swapping the knockoffs, and the remaining half are utilized for testing the selected hypotheses.
 The results demonstrate data-splitting is valid under covariate shift but it is not as powerful as our method, especially if the sample size is moderately large.

%%% Local Variables:
%%% mode: latex
%%% TeX-master: "supplement_jasa"
%%% End:

\end{document}

\end{document}

%%% Local Variables:
%%% mode: latex
%%% TeX-master: t
%%% End: